\def\tsecompldate{16th December, 1994}
\def\tseprepno{Imperial/TP/94-95/6}
\def\tsehepphno{{\tt astro-ph/9412059}}

\documentstyle[12pt]{article}

\def\tsetrue{T} \def\tsefalse{F} % see TeX book pp 206 and environs

\let\tseletter=\tsetrue  % select Letter style paper else A4
\let\tsepaper=\tsefalse  % select paper/proceedings version else preprint
\makeatletter
\if\tseletter\tsetrue 
\else 
\fi
\def\setb@se#1{\baselineskip=#1 \normalbaselineskip=#1}
\setb@se{12pt}
\long\def\title#1{\vspace*{11.5pc}{\pretolerance=10000\raggedright
  \setb@se{12pt}\bf #1\par}\nobreak\ignorespaces}
\long\def\author#1{\vspace{4pc}\begin{list}{\hfill}%
{\topsep=0pt\parskip=0pt\parsep=0pt\partopsep=0pt\listparindent=0pt%
\itemsep=0pt\rightmargin=0pt\labelsep=0pt\labelwidth=5pc\leftmargin=5pc}%
\item\normalsize{#1}\end{list}\vspace{14pt}}
\long\def\affil#1{\begin{list}{\hfill}
{\topsep=0pt\parskip=0pt\parsep=0pt\partopsep=0pt\listparindent=0pt
\itemsep=0pt\rightmargin=0pt\labelsep=0pt\labelwidth=5pc\leftmargin=5pc}
  \item\normalsize{\rm #1}\end{list}\vspace{7pt}}
\long\def\beginabstract{\vspace{21pt plus 7pt minus 7pt}\begin{list}{\hfill}%
{\topsep=0pt\parskip=0pt\parsep=0pt\partopsep=0pt\listparindent=0pt%
\itemsep=0pt\rightmargin=0pt\labelsep=0pt\labelwidth=5pc\leftmargin=5pc}%
\item\normalsize{\bf Abstract. }}
\long\def\endabstract{\end{list}\vspace{28pt plus14pt minus 14pt}%
\normalsize\noindent}
\def\@sect#1#2#3#4#5#6[#7]#8{\ifnum #2>\c@secnumdepth
  \def\@svsec{}\else
  \refstepcounter{#1}\edef\@svsec{\csname the#1\endcsname.\hskip 0.5em}\fi
  \@tempskipa #5\relax
   \ifdim \@tempskipa>\z@
     \begingroup #6\relax
       \@hangfrom{\hskip #3\relax\@svsec}{\interlinepenalty \@M #8\par}
     \endgroup
    \csname #1mark\endcsname{#7}\addcontentsline
      {toc}{#1}{\ifnum #2>\c@secnumdepth \else
		   \protect\numberline{\csname the#1\endcsname}\fi
		 #7}\else
     \def\@svsechd{#6\hskip #3\@svsec #8\csname #1mark\endcsname
		   {#7}\addcontentsline
			{toc}{#1}{\ifnum #2>\c@secnumdepth \else
			  \protect\numberline{\csname the#1\endcsname}\fi
		    #7}}\fi
  \@xsect{#5}}

\def\section{\@startsection {section}{1}{\z@}{-28pt plus -14pt minus
-14pt}{2.3ex plus .2ex}{\normalsize\bf}}

\def\subsection{\@startsection{subsection}{2}{\z@}{-14pt plus -8pt
minus -4pt}{1.5ex plus .2ex}{\normalsize\bf}}
\def\subsubsection{%
    \@startsection{subsubsection}{3}{\z@}{-14pt plus
    -8pt minus -4pt}{-1.5ex plus -.2ex}{\normalsize\bf}}
\def\paragraph{\@startsection
     {paragraph}{4}{\z@}{3.25ex plus 1ex minus
.2ex}{-1em}{\normalsize\bf}}
\def\subparagraph{\@startsection
     {subparagraph}{4}{\parindent}{3.25ex plus 1ex minus
     .2ex}{-1em}{\normalsize\bf}}

\def\caption{\refstepcounter\@captype
\@dblarg{\@caption\@captype}}
\long\def\@caption#1[#2]#3{\par\addcontentsline{\csname
 ext@#1\endcsname}{#1}{\protect\numberline{\csname
 the#1\endcsname}{\ignorespaces #2}}\begingroup
   \@parboxrestore
   \hspace*{28pt}
   \parbox{394pt}{\@makecaption{{\bf\csname
	    fnum@#1\endcsname}}{\ignorespaces #3}}\par
\vspace{7pt}\endgroup}

\long\def\@makecaption#1#2{
   \vskip 14pt
   \setbox\@tempboxa\hbox{#1{\bf.} #2}
   \ifdim \wd\@tempboxa >\hsize   % IF longer than one line:
       #1{\bf.} #2\par            %   THEN set as ordinary
>paragraph.
     \else                        %   ELSE  set flush left.
       \hbox to\hsize{\box\@tempboxa\hfil}
   \fi}

\newcommand{\boldarrayrulewidth}{1pt} % Width of bold rule
\def\bhline{\noalign{\ifnum0=`}\fi\hrule \@height
		    \boldarrayrulewidth \futurelet
		    \@tempa\@xhline}

\def\@xhline{\ifx\@tempa\hline\vskip \doublerulesep\fi
      \ifnum0=`{\fi}}

\def\thebibliography#1{\section*{REFERENCES\@mkboth
  {REFERENCES}{REFERENCES}}\list
  {\hfil[\arabic{enumi}]}{\itemsep=0pt\labelsep=7pt\itemindent=-14pt
    \settowidth\labelwidth{[#1]}
    \leftmargin\labelwidth
    \advance\leftmargin\labelsep
    \advance\leftmargin -\itemindent
    \usecounter{enumi}}\setb@se{12pt}\small
    \def\newblock{\hskip .11em plus .33em minus .07em}
    \sloppy\clubpenalty4000\widowpenalty4000
    \sfcode`\.=1000\relax}
\def\references{\section*{REFERENCES\@mkboth
{REFERENCES}{REFERENCES}}\list{}{\itemsep=0pt\labelsep=0pt\itemindent=-28pt
\labelwidth=0pt\leftmargin=28pt}\setb@se{12pt}\small
\def\newblock{\hskip .11em plus .33em minus .07em}
\sloppy\clubpenalty4000\widowpenalty4000
\sfcode`\.=1000\relax}
\makeatother

\let\tsenoteon=\tsefalse   % tse note system on
\let\tsedevon=\tsefalse  % select development options on or off
\if\tsetrue\tsedevon \newcommand{\tsedevelop}[1]{{#1}}
\else \newcommand{\tsedevelop}[1]{{}}
\fi

\newcommand{\tnote}[1]{\if\tsetrue\tsenoteon \footnote{#1} \fi}
\if\tsetrue\tsenoteon{
\typeout{--- Tim Footnotes Included ---}
}\else 
\fi

\newcommand{\ttitle}[1]{{\em #1}}

\newcommand{\half}{\frac{1}{2}}
\newcommand{\bea}{\begin{eqnarray}}
\newcommand{\eea}{\end{eqnarray}}
\newcommand{\beq}{\begin{equation}}
\newcommand{\eeq}{\end{equation}}
\newcommand{\nnel}{\nonumber \\ {}}
\newcommand{\bpi}{\begin{picture}}
\newcommand{\epi}{\end{picture}}

\newcommand{\quarter}{\frac{1}{4}}
\newcommand{\phiv}{\phi_{v}}
\newcommand{\phibar}{\bar{\phi}}
\newcommand{\modphi}{|\phi|}

\newcommand{\dash}[1]{{#1}^\prime{}}
\newcommand{\tdash}{\dash{t}}

\newcommand{\thetadash}{\dash{\theta}}
\newcommand{\veck}{\vec{k}}
\newcommand{\vecp}{\vec{p}}
\newcommand{\vecx}{\vec{x}}
\newcommand{\vecxdash}{\dash{\vec{x}}}
\newcommand{\kbar}{\bar{k}}

\newcommand\ba{\begin{array}}
\newcommand\ea{\end{array}}
\newcommand\ben{\begin{equation}}
\newcommand\een{\end{equation}}

\def\fun#1#2{\lower3.6pt\vbox{\baselineskip0pt\lineskip.9pt
\ialign{$\mathsurround=0pt#1\hfill##\hfil$\crcr#2\crcr\sim\crcr}}}

\def\math{\mathsurround 0pt}
\def\oversim#1#2{\lower.5pt\vbox{\baselineskip0pt \lineskip-.5pt

\ialign{$\math#1\hfil##\hfil$\crcr#2\crcr{\scriptstyle\sim}\crcr}}}

\def\({\left(} \def\){\right)}
\def\[{\left[} \def\]{\right]}

\def\half{{\mathchoice{{\textstyle{1\over 2}}}{1\over 2}{1\over 2}{1
\over 2}}}
\def\unit#1{\ifinner \;
            \else \quad \fi
            {\rm #1}}

\setbox1=\hbox{$\mathchar'173$}
\def\dbar{\raise0.3em\copy1\mskip-12mud}
\def\deltabar{\raise0.3em\copy1\mskip-12mu\delta}

\begin{document}

\typeout{--- Title page start ---}
\if\tsepaper\tsefalse
 \begin{flushright}
\tsecompldate \\
\tsedevelop{ (LaTeX-ed on \today ) \\}
\tseprepno \\
\tsehepphno
%\hskip 1 cm \>{next preprint number}\\
\end{flushright}
\fi

\title{THE PRODUCTION OF STRINGS AND MONOPOLES AT PHASE TRANSITIONS
\if\tsepaper\tsefalse\footnote{Invited lectures given by R.J.R.\
to the Nato Advanced Study Institute and Euroconference on
Formation and Interaction of Topological Defects,  Isaac Newton
Institute, Cambridge (UK), September, 1994.
Available through anonymous ftp from
{\tt ftp://euclid.tp.ph.ic.ac.uk/papers/93-4\_09.tex}
or on WWW at {\tt http://euclid.tp.ph.ic.ac.uk/Papers/index.html} }\fi
}
\author{R.J. Rivers
\if\tsepaper\tsefalse \footnote{E-mail: {\tt R.Rivers@IC.AC.UK}}\fi
and T.S. Evans
\if\tsepaper\tsefalse\footnote{E-mail: {\tt T.Evans@IC.AC.UK}}\fi
}

\affil{Blackett Laboratory,\if\tsepaper\tsefalse\footnote{WWW:
{\tt http://euclid.tp.ph.ic.ac.uk/}}\fi
 Imperial College, \\
Prince Consort Road, London SW7 2BZ  U.K.}

\beginabstract
We shall show that the density of defects produced at a
second-order phase transition is determined by the correlation
length of the fields.  This is true both for defects appearing in
the Ginzburg regime and for defects produced at a quench, when the
Ginzburg regime is irrelevant.
\endabstract

%See pp.175
%\renewcommand{\thefootnote}{\arabic{footnote}}
%\setcounter{footnote}{0}

%\npagepub

\typeout{--- Main Text Start ---}

\section{ INTRODUCTION}

These notes are based on lectures given by one of us (R.J.R) at the
Nato Advanced Study Institute and Euroconference on {\it Formation
and Interactions of Topological Defects} held at the Isaac Newton
Institute, Cambridge (UK) in September, 1994.  They aim to provide a
preliminary discussion of how strings (vortices) and monopoles can
be produced at the phase transitions of relativistic quantum fields.

The applications that we have in mind are to the early universe,
where it has been argued (see Kibble, these proceedings and
elsewhere \cite{TK}) that cosmic strings produced at the era of Grand
Unification (i.e. at energy scales $10^{15}$ - $10^{16}$ GeV) can
provide the seeds for the  large-scale structure formation in the
universe that we see today.  There are many attractive features to
this idea, and we shall not recount them here.  However, it should be
remembered that the isotropy of the universe suggests that it has
passed through a period of rapid inflation.  Inflation
in itself generates large-scale structure but, of greater importance
in this context, it was
originally introduced to dilute the undesirably high monopole density that is
almost inevitable  in unified theories.
Some ingenuity is required
for strings not to suffer a similar fate.  Should cosmic strings
turn out to be nothing more than an elaborate fancy we are consoled by the
observation that they are, in many ways, the relativistic
counterparts to the vortices in superfluids and superconductors.  So
much so, that vortex production \cite{HLM} in superfluid $^4 He$ has been
invoked as simulating cosmology in the laboratory (see Zurek, these
proceedings, and elsewhere \cite{Z} ).  But for their relativistic nature, our
methods are equally applicable to these more homely materials.
[By homely, we mean terrestrial rather than non-exotic.  See
Salomaa, these proceedings, for a description of some of the very
exotic defects of superfluid $^{3}He$, for example].
However, we shall not change to a non-relativistic gear here.

In practice, our calculations are, as yet, too primitive to be able
to address the details of early  universe cosmology directly, even
if we had wished.
However, our main conclusion is general, and does not require a
cosmological backcloth.  It is that defects play an important role
in any second-order transition at which they can be produced,
appearing initially in
essentially the maximum numbers compatible with retaining their individuality
as diffuse entities.
Specifically, the density
of defects, after
their production at the phase transition, is determined largely by the
distance $\xi (t)$ over which the fields are correlated, and
which also characterises defect size.  Very
crudely, we predict one monopole per correlation volume
$v=O(\xi^{3}(t))$ and one string passing through each correlation
area $a=O(\xi^{2}(t))$.  This applies equally to
\begin{enumerate}
\item defects produced from large fluctuations in the Ginzburg regime
near a second-order transition and to
\item defects produced from long wavelength spinodal decomposition
after a quench.
\end{enumerate}
In the first case this result was anticipated qualitatively by Kibble,
on making
reasonable assumptions about domain formation.  It has now become
sufficiently part of the folklore that the results in the second case
will be no surprise, even though circumstances are different there.
However, our conclusions are reached as a result of quantum
calculations rather than semiclassical arguments which,
for gauge theories, have led to some confusion.
Of course,  there are several caveats (e.g. weak couplings, short to
intermediate times)  as will be seen.  Whether these would be
satisfied in realistic many-body systems is dubious, but our
calculations can be improved upon in principle.  In practice we have
yet to do so, and the results presented here are chosen, in part,
for their simple analytic nature.

These lecture notes essentially fall into two parts, corresponding
to cases i) and ii) above.  The first is
concerned with phase transitions as viewed from the platform of
equilibrium  thermal field theory.  To be concrete and simple we
restrict ourselves to the transitions of global and local $O(N)$
theories.  In $D=3$ spatial dimensions an $O(N)$ theory permits strings
when $N=2$ and monopoles when $N=3$. We examine the presence of both
global and local defects in
the Ginzburg regime close to a second-order transition, characterised
by large fluctuations.  However, while confirming our basic
ideas about fluctuations we are unable to provide a satisfactory
mechanism as to how defects appearing in this way can
 persist as the system cools.
The second part of the lectures
attempts to circumvent this problem by avoiding any discussion of
equilibrium theory, adopting a
non-equilibrium approach from the start for global $O(N)$ theories.
Specifically, we consider the  production of
defects  as a consequence of spinodal decomposition at the onset of a
transition, whose freezing in is much less problematical.
Even then, a quantitative description has yet to be given.  We
conclude with some tentative steps in this direction.

Our introduction to equilibrium and non-equilibrium quantum
field theory is, or
should be , well-known.  Some of the further material is taken from
unpublished lecture notes (T.S.E) and from published work (R.J.R) with Mark
Hindmarsh \cite{HR}.  The work on defect production is more recent, as yet
unpublished (although in a preliminary form some of the conclusions
were presented \cite{RR} (R.J.R) at the recent Nato Advanced Workshop on {\it
Electroweak Theory and the Early Universe},
held in Sintra, Portugal,  March,
1994).  However, the main results of the latter part have already
been submitted (R.J.R. and A. Gill) \cite{GR}.

In addition we have drawn heavily on the recent series of papers by
Dan Boyanovsky, Hector de Vega and co-authors \cite{B,boyanovsky},
which provide an excellent introduction to behaviour out of
equilibrium, and to the much earlier (equilibrium) work of Halperin
\cite{halperin}  concerning global defects.  We are indebted to
Mark Hindmarsh for helpful comments on the latter.  The reader
looking for background material to amplify some of our more cursory
comments should find the Proceedings of the {\it Third Thermal
Fields Workshop},  Banff (Canada),  August 1993, most helpful
\cite{Banff}.

\section{ THE PHASE TRANSITIONS OF RELATIVISTIC QUANTUM FIELDS}

The defects in which we are interested are not fundamental entities
like superstrings.  They are diffuse field configurations formed at
phase transitions in the early universe, which survive because of
their topological stability. To understand how phase transitions
occur we need to recapulate the rudiments of thermal field theory
\cite{TFT,Ka}.

\subsection{ What is Thermal Field Theory?}

Thermal field theory is a combination of two theories.  The first
is relativistic quantum field theory, used to describe the behaviour
of elementary particles, when only a few such particles are involved
e.g. in $e^{+}-e^{-}$ collisions at LEP.  The specific attribute
that special relativity brings is the annihilation and creation of
particles, the conversion of rest-mass to energy.  This is combined
with the fundamental
ingredient of quantum physics, quantum fluctuations as encoded by
Heisenberg's uncertainty relation
\beq
\Delta E\Delta t \geq\hbar.
\eeq

\begin{figure}[htb]
\begin{center}
\setlength{\unitlength}{2pt}%
\begin{picture}(50,11)(-11,-3)
\thicklines
\put(-11,0){\vector(1,0){6}}
\put(-5,0){\line(1,0){5}}
%\put(22,0){\vector(1,0){11}}
\put(22,0){\vector(1,0){8}}
\put(30,0){\line(1,0){3}}
\put(0,0){\circle*{3}}
\put(22,0){\circle*{3}}
% Internal lines
\put(2,0){\oval(4,4)[tl]}
\put(2,4){\oval(4,4)[br]}
\multiput(4,4)(-0.1,0.2){5}{\line(1,0){0.1}}
\multiput(3.5,5)(-0.1,0.1){5}{\line(1,0){0.1}}
\multiput(3,5.5)(-0.1,0.2){5}{\line(1,0){0.1}}
\put(4.5,6.5){\oval(4,4)[tl]}
\multiput(4.5,8.5)(0.2,-0.1){5}{\line(1,0){0.1}}
\multiput(5.5,8)(0.1,-0.1){5}{\line(1,0){0.1}}
\multiput(6,7.5)(0.2,-0.1){5}{\line(1,0){0.1}}
\put(7,9){\oval(4,4)[br]}
\put(11,9){\oval(4,4)[t]}
\put(15,9){\oval(4,4)[bl]}
\multiput(15,7)(0.2,0.1){5}{\line(1,0){0.1}}
\multiput(16,7.5)(0.1,0.1){5}{\line(1,0){0.1}}
\multiput(16.5,8)(0.2,0.1){5}{\line(1,0){0.1}}
\put(17.5,6.5){\oval(4,4)[tr]}
\multiput(19.5,6.5)(-0.1,-0.2){5}{\line(1,0){0.1}}
\multiput(19,5.5)(-0.1,-0.1){5}{\line(1,0){0.1}}
\multiput(18.5,5)(-0.1,-0.2){5}{\line(1,0){0.1}}
\put(20,4){\oval(4,4)[bl]}
\put(20,0){\oval(4,4)[tr]}
\put(11,0){\oval(22,22)[b]}
\put(22,0){\line(0,-1){1}}
\end{picture}
\begin{picture}(50,11)(-11,-3)
\thicklines
\put(-11,0){\vector(1,0){6}}
\put(-5,0){\line(1,0){5}}
%\put(22,0){\vector(1,0){11}}
\put(22,0){\vector(1,0){8}}
\put(30,0){\line(1,0){3}}
\put(0,0){\circle*{3}}
\put(22,0){\circle*{3}}
% Internal lines
\put(2,0){\oval(4,4)[tl]}
\put(2,4){\oval(4,4)[br]}
\multiput(4,4)(-0.1,0.2){5}{\line(1,0){0.1}}
\multiput(3.5,5)(-0.1,0.1){5}{\line(1,0){0.1}}
\multiput(3,5.5)(-0.1,0.2){5}{\line(1,0){0.1}}
\put(4.5,6.5){\oval(4,4)[tl]}
\multiput(4.5,8.5)(0.2,-0.1){5}{\line(1,0){0.1}}
\multiput(5.5,8)(0.1,-0.1){5}{\line(1,0){0.1}}
\multiput(6,7.5)(0.2,-0.1){5}{\line(1,0){0.1}}
\put(7,9){\oval(4,4)[br]}
\put(11,9){\oval(4,4)[t]}
\put(15,9){\oval(4,4)[bl]}
\multiput(15,7)(0.2,0.1){5}{\line(1,0){0.1}}
\multiput(16,7.5)(0.1,0.1){5}{\line(1,0){0.1}}
\multiput(16.5,8)(0.2,0.1){5}{\line(1,0){0.1}}
\put(17.5,6.5){\oval(4,4)[tr]}
\multiput(19.5,6.5)(-0.1,-0.2){5}{\line(1,0){0.1}}
\multiput(19,5.5)(-0.1,-0.1){5}{\line(1,0){0.1}}
\multiput(18.5,5)(-0.1,-0.2){5}{\line(1,0){0.1}}
\put(20,4){\oval(4,4)[bl]}
\put(20,0){\oval(4,4)[tr]}
\put(0,0){\vector(1,-2){5}}
\put(22,0){\line(-1,-2){5}}
\put(8,-16){\makebox(0,0)[lb]{$n_f$}}
\put(22,0){\line(0,-1){1}}
\end{picture}
\begin{picture}(50,11)(-11,-3)
\thicklines
\put(-11,0){\vector(1,0){6}}
\put(-5,0){\line(1,0){5}}
%\put(22,0){\vector(1,0){11}}
\put(22,0){\vector(1,0){8}}
\put(30,0){\line(1,0){3}}
\put(0,0){\circle*{3}}
\put(22,0){\circle*{3}}
% Internal lines
\put(2,0){\oval(4,4)[tl]}
\put(2,4){\oval(4,4)[br]}
\multiput(4,4)(-0.1,0.2){5}{\line(1,0){0.1}}
\multiput(3.5,5)(-0.1,0.1){5}{\line(1,0){0.1}}
\multiput(3,5.5)(-0.1,0.2){5}{\line(1,0){0.1}}
\put(4.5,6.5){\oval(4,4)[tl]}
\multiput(4.5,8.5)(0.2,-0.1){5}{\line(1,0){0.1}}
\multiput(5.5,8)(0.1,-0.1){5}{\line(1,0){0.1}}
\multiput(6,7.5)(0.2,-0.1){5}{\line(1,0){0.1}}
\put(8,10){\makebox(0,0)[lb]{$n_b$}}
%\put(7,9){\oval(4,4)[br]}
%\put(11,9){\oval(4,4)[t]}
%\put(15,9){\oval(4,4)[bl]}
\multiput(15,7)(0.2,0.1){5}{\line(1,0){0.1}}
\multiput(16,7.5)(0.1,0.1){5}{\line(1,0){0.1}}
\multiput(16.5,8)(0.2,0.1){5}{\line(1,0){0.1}}
\put(17.5,6.5){\oval(4,4)[tr]}
\multiput(19.5,6.5)(-0.1,-0.2){5}{\line(1,0){0.1}}
\multiput(19,5.5)(-0.1,-0.1){5}{\line(1,0){0.1}}
\multiput(18.5,5)(-0.1,-0.2){5}{\line(1,0){0.1}}
\put(20,4){\oval(4,4)[bl]}
\put(20,0){\oval(4,4)[tr]}
\put(11,0){\oval(22,22)[b]}
\put(22,0){\line(0,-1){1}}
\end{picture}
\begin{picture}(50,11)(-11,-3)
\thicklines
\put(-11,0){\vector(1,0){6}}
\put(-5,0){\line(1,0){5}}
\put(22,0){\vector(1,0){8}}
\put(30,0){\line(1,0){3}}
\put(0,0){\circle*{3}}
\put(22,0){\circle*{3}}
% Internal electron lines
\put(0,0){\vector(1,-2){5}}
\put(22,0){\line(-1,-2){5}}
\put(8,-16){\makebox(0,0)[lb]{$n_f$}}
% Now do the photon lines
\put(2,0){\oval(4,4)[tl]}
\put(2,4){\oval(4,4)[br]}
\multiput(4,4)(-0.1,0.2){5}{\line(1,0){0.1}}
\multiput(3.5,5)(-0.1,0.1){5}{\line(1,0){0.1}}
\multiput(3,5.5)(-0.1,0.2){5}{\line(1,0){0.1}}
\put(4.5,6.5){\oval(4,4)[tl]}
\multiput(4.5,8.5)(0.2,-0.1){5}{\line(1,0){0.1}}
\multiput(5.5,8)(0.1,-0.1){5}{\line(1,0){0.1}}
\multiput(6,7.5)(0.2,-0.1){5}{\line(1,0){0.1}}
\put(8,10){\makebox(0,0)[lb]{$n_b$}}
%\put(7,9){\oval(4,4)[br]}
%\put(11,9){\oval(4,4)[t]}
%\put(15,9){\oval(4,4)[bl]}
\multiput(15,7)(0.2,0.1){5}{\line(1,0){0.1}}
\multiput(16,7.5)(0.1,0.1){5}{\line(1,0){0.1}}
\multiput(16.5,8)(0.2,0.1){5}{\line(1,0){0.1}}
\put(17.5,6.5){\oval(4,4)[tr]}
\multiput(19.5,6.5)(-0.1,-0.2){5}{\line(1,0){0.1}}
\multiput(19,5.5)(-0.1,-0.1){5}{\line(1,0){0.1}}
\multiput(18.5,5)(-0.1,-0.2){5}{\line(1,0){0.1}}
\put(20,4){\oval(4,4)[bl]}
\put(20,0){\oval(4,4)[tr]}
%\put(11,0){\oval(22,22)[b]}
%\put(22,0){\line(0,-1){1}}
\end{picture}
\end{center}
%\vspace{1cm}
\caption{Different types of process which occur when in a heat bath.}
\label{fese}
\end{figure}
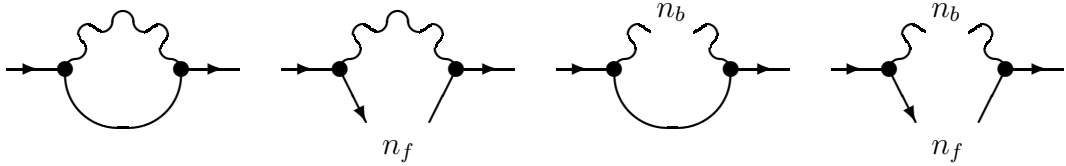

One of the great steps forward in the development of quantum field
theory was the use of  Feynman diagrams
to represent the effects of quantum fluctuations on physical
processes.  For  instance, for the case of an electron travelling in
vacuo, in quantum electrodynamics we must include corrections such
as the first diagram in Fig. \ref{fese}
% *** Thermal and quantum fluctuations ************************
\typeout{*** electron self-energy diagrams}
The internal lines represent interactions between the electron
(the external legs) and possible vacuum fluctuations involving the
emission and absorption of photons (the wavy line).  The latter are
virtual, as are the intermediate electrons,
existing only for the short time allowed by Heisenberg's
uncertainty principle.  The picture we have is of an electron
travelling through  a vacuum, empty of real particles but full of
virtual shortlived particles, which interact with it.

The second ingredient in thermal field theory is thermodynamics/
statistical mechanics, used to study many-body problems.
The key idea is that a few thermodynamic or bulk properties are
sufficient to characterise the essential physics.  In doing this,
the precise initial state of the system is assumed unknown, the
choice of any particular state only specified by a probability.
This statistical uncertainty is quite different from the quantum
uncertainty discussed above.  In particular, we shall typically
assume a Boltzmann distribution, $e^{-\beta E}$, to describe the
probability of being in a state of energy $E$.  In units in which
Boltzmann's constant $k_{B} =1$, $\beta$ is the inverse temperature
$T^{-1}$ of the system.  When combined with
quantum physics this leads to fluctuations in the number of real
particles, as described by the usual Bose-Einstein or Fermi-Dirac
number distributions
\beq
n_{b,f} = \frac{1}{e^{\beta E}\pm 1}
\eeq
(with $\mp$ taken according as the particles are bosons/fermions).
A useful picture is to think of doing experiments in a background,
the heat bath or reservoir, which is full of real particles, whose
precise number is unknown.

This combination of relativistic quantum field theory and
statistical mechanics, which constitutes thermal field theory,
describes quite a different physical situation from that of normal
quantum field theory.  Both quantum and statistical fluctuations
have to be accounted for simultaneously and, as outlined above, they
are dissimilar in their nature.  Although it is not surprising that
thermal field theory has some quite different properties from
standard quantum field theory, what is amazing is that the two can
be described in a very similar way.

As an example of thermal physics, return to the electron, now taken
to be propagating through a QED plasma.  The quantum fluctuations
mentioned above (the first diagram in Fig. \ref{fese}) are always
present, whereas the remaining diagrams in the Figure do not appear
in ordinary quantum field theory, involving interactions with the
real particles present in the plasma or heat bath.  The last diagram
in Fig.\ref{fese} represents such an interaction with two real
particles, while the middle two diagrams  represent the effects on
the electron's propagation of a mixture of quantum (virtual) and
statistical fluctuations.  All of these will change the electron's
inertial mass, and the latter lead to the dissipation of its energy.

\subsection{ Why Are There Phase Transitions?}

The diagrams of Fig. \ref{fese} suggest that a particle in a heat-bath
behaves as if its momentum-space propagator were \cite{DJ}
\beq
G(k) = \frac{i}{k^{2} - m^{2} + i\epsilon} + 2\pi\delta (k^{2} - m^{2})
n_{b}(\omega).
\eeq
The first term is the usual virtual particle exchange, present for
all four-momenta $k$, whereas the second term describes real particles
in the heatbath, present only when $k^{2} = m^{2}$
(in units in which $c=1$).
We have taken spinless bosons as an example.
It will be shown later that the situation is rather more
complicated, but $G(k)$ of (2.3) is all that is necessary for one-loop
diagrams, which are sufficient to show the presence of phase
transitions.

Consider the theory of a real scalar field $\phi$, given by the action
\beq
S[\phi] = \int d^{4}x [\frac{1}{2}(\partial_{\mu}\phi)(\partial^{\mu}\phi)
-\frac{1}{2} m_{0}^{2}\phi^{2} - \frac{1}{8} g_{0}^{2}\phi^{4}]
\eeq
in thermal equilibrium at temperature $T$.  The presence of the
heatbath affects the inertial properties of particles fired into it.
 The effective mass $m^{2}(T)$ of  $\phi$ -field quanta is represented
diagramatically (to one loop) by
% *** Self-energies ****************************************************
\typeout{*** Higgs Self-energies}
\bea
\bpi(40,26)(28,13)
\put(0,16){\line(1,0){16}}
\put(48,16){\line(1,0){16}}
\put(32,16){\circle{32}}
\put(16,16){\circle*{3}}
\put(48,16){\circle*{3}}
\put(32,17){\makebox(0,0)[b]{\footnotesize 1-loop}}
\put(32,13){\makebox(0,0)[t]{\mbox{\footnotesize 1PI}}}
\epi
&=&
\bpi(50,11)(-11,-3)
\put(-11,0){\line(1,0){44}}
\put(11,0){\circle*{3}}
\put(11,0){\circle{5}}
\epi
\bpi(0,40)
\epi
+
\bpi(44,11)(-11,-3)
\put(-5,-11){\line(1,0){32}}
\put(11,-11){\circle*{3}}
\put(11,1){\circle{22}}
\epi
%+
%\bpi(44,11)(-11,-3)
%\put(-11,0){\line(1,0){11}}
%\put(22,0){\line(1,0){11}}
%\put(-1,0){\circle*{3}}
%\put(23,0){\circle*{3}}
%\put(11,0){\circle{22}}
%\epi
%+
%\bpi(44,11)(-11,-5.5)
%\put(0,-19){\line(1,0){22}}
%\put(11,-19){\line(0,1){8}}
%\put(11,-19){\circle*{3}}
%\put(11,-11){\circle*{3}}
%\put(11,1){\circle{22}}
%\epi
\eea
where the solid line now denotes the full propagator (3),
and the first term is $m_{0}^{2}$, read off from
(4).  [The fact that $m^{2}(T)$ is independent of momentum is a
reflection of the relative simplicity of the one-loop diagrams in
$g^{2}\phi^{4}$ theory, in contrast to those of QED.]
The second term of (5) is ultraviolet divergent, but this causes
no problem, since the bare mass $m_{0}$  has no physical meaning,  To
$O(\hbar)$, the physical $(mass)^{2}$, denoted by $m^{2}$,
is defined as the sum of the first two
terms.  The end result is that the heat-bath induces a
temperature-dependent term to the effective mass of the form
\bea
m^{2}(T)&=& m^{2} + \frac{3}{2}g^{2}\hbar\int \dbar^{4}k 2\pi
\delta (k^{2} - m^{2})n_{b}(\omega)\nonumber\\
&=& m^{2} + \frac{3}{2}g^{2}\hbar\int\frac {\dbar^{3}k}{\omega
(k)}\; \frac{1}{e^{\beta\omega (k)} -1}
\eea
where $\dbar^{D}k = (2\pi)^{-D} d^{D}k$,
and $\omega (k)=\sqrt{\veck^{2} +m^{2}}$.
Since the Boltzmann factor
cuts off the momentum integration at $|\veck| = O(T)$  the second term in
(6) is ultraviolet finite.  Its temperature-dependence (for $T\gg m$)
can be read off from its ultraviolet divergence, had $n_{b}$   been absent,
as $O(g^{2} T^{2})$.  Specifically, up to logarithms in the final term
\beq
m^{2}(T) = m^{2} + \frac{1}{8}\hbar
g^{2}(T^{2} - \frac{3}{\pi} Tm + O(m^{2})).
\eeq
[At one loop there is no coupling constant renormalisation and we
have renamed $g_{0}$  as $g$].  That is, the effect of the heatbath is to
increase the $(mass)^{2}$  value of the quanta.  In particular, massless
bosons put in a heatbath acquire mass.  With qualifications this is
true for photons, as will be seen later,

Suppose now that the cold $(mass)^{2}$ parameter is {\it negative}, $m^{2}  =
-\half g^{2}\eta^{2}$
, say, some $\eta$  .  That is, on redefining the parameters to accomodate
the one-loop quantum fluctuations, the action $S[\phi ]$ for $\phi$   is
\beq
S[\phi] = \int d^{4}x [\frac{1}{2} (\partial_{\mu}\phi)(\partial^{\mu}\phi)
- \frac{1}{8} g^{2}(\phi^{2}-\eta^{2})^{2}],
\eeq
the familiar double-well potential, with vacuum values  $\phi =\pm\eta$
. The choice of one of these minima as the ground state breaks the
reflection invariance
$\phi\rightarrow -\phi$ of the action (or Hamiltonian) and
this $Z_{2}$    symmetry is said to be spontaneously broken.  The upturned
quadratic potential $-\quarter g^{2}\eta^{2}\phi^{2}$
near the origin characterises a region
of instability for field fluctuations of small amplitude.  For
high $T$, when only the first term in the brackets of (1.8) need be
taken, $ m^{2} (T)$ becomes positive, however negative $m^{2}$  may be.
That is,
the $Z_{2}$   symmetry is restored \cite{KL,DJ,W}.
The effective $(mass)^{2}$ for
small field amplitudes is then most
transparently written as
\beq
m^{2}(T) = m^{2}\left( 1-\frac{T^{2}}{T_{c}^{2}}\right),
\eeq
where $T_{c}^{2} = 4\eta^{2}$
and we are working in units in which $\hbar =1$. In this
approximation $T_{c}$  defines the critical temperature of a second-order
transition, above which the symmetry is restored, and below which it
is broken.  At the transition the correlation length $\xi (T) = |m(T)|^{-1}$
diverges as
\beq
\xi (T)\propto |T - T_{c}|^{-\gamma}, \gamma = \half.
\eeq
This mean-field result can be improved by a renormalisation-group
analysis
%(which gives $\gamma = ?$)
but, at the qualitative level at which
we are working, it is good enough.

\subsection{ Phase Transitions of O(N) Theories.}

The theories in which we shall be interested are particular examples
of the $O(N)$ extension of the theory (4), in which N scalar fields
$\phi_{a}$ ($a =1,2,...N$)
transform as the fundamental vector representation.
If we are only concerned with invariance under {\it global} $O(N)$
transformations, the action (4) is elevated to
\beq
S[\phi] = \int d^{4}x [\frac{1}{2}
(\partial_{\mu}\phi_{a})(\partial^{\mu}\phi_{a})
- \frac{1}{8} g^{2}(\phi_{a}^{2}-\eta^{2})^{2}],
\eeq
where D = 3 is the number of spatial dimensions and summation over
the $O(N)$ label $a$ is assumed.
For $\eta^{2}>0$ the $O(N)$ symmetry is broken
to $O(N-1)$, the vacuum manifold ${\cal M}  =O(N)/O(N-1)$
being equal to $S^{N-1}$,
the $(N-1)$-sphere.  On expanding about any point on the sphere we find
a single Higgs boson with mass $m_{H} = g\eta$
and $(N-1)$ massless Goldstone
bosons.  If we heat the system to a plasma at temperature $ T$ we would
expect that, as for the single field, at some temperature $T_{c} = O(\eta )$
the symmetry would be
restored.  This is indeed the case,  Some care is needed to
guarantee the masslessness of the Goldstone modes at all $T\leq T_{c}$
\cite{Ka} but
basically we proceed as in (6) in determining the effective mass
$m(T)$ of the small-field potential.  The only difference is that the
loop diagram of (5) is extended to  all N scalar fields.
% *** Self-energies ****************************************************
%\typeout{*** Higgs Self-energies}
%\bea
%\bpi(40,26)(28,13)
%\put(0,16){\line(1,0){16}}
%\put(48,16){\line(1,0){16}}
%\put(32,16){\circle{32}}
%\put(16,16){\circle*{3}}
%\put(48,16){\circle*{3}}
%\put(32,17){\makebox(0,0)[b]{\footnotesize 1-loop}}
%\put(32,13){\makebox(0,0)[t]{\mbox{\footnotesize 1PI}}}
%\epi
%&=&
%\bpi(44,11)(-11,-3)
%\put(-5,-11){\line(1,0){32}}
%\put(11,-11){\circle*{3}}
%\put(11,1){\circle{22}}
%\epi
%+
%\bpi(44,11)(-11,-3)
%\put(-11,0){\line(1,0){11}}
%\put(22,0){\line(1,0){11}}
%\put(-1,0){\circle*{3}}
%\put(23,0){\circle*{3}}
%\put(11,0){\circle{22}}
%\epi
%+
%\bpi(50,11)(-11,-3)
%\put(-11,0){\line(1,0){44}}
%\put(11,0){\circle*{3}}
%\put(11,0){\circle{5}}
%\epi
%\bpi(0,40)
%\epi
%+
%\bpi(44,11)(-11,-5.5)
%\put(0,-19){\line(1,0){22}}
%\put(11,-19){\line(0,1){8}}
%\put(11,-19){\circle*{3}}
%\put(11,-11){\circle*{3}}
%\put(11,1){\circle{22}}
%\epi
%\eea
%for fixed external leg label $a$, rather than the single diagram of
%(6).
The end result is that, ignoring terms relatively $O(\eta /T_{c})$,
the one-loop calculation gives \cite{DJ}
\bea
m^{2}(T) &=& -\frac{1}{2} g^{2}\eta^{2} + (N+2)g^{2}\frac{T^{2}}{24}\\
&=& m^{2}\left( 1 - \frac{T^{2}}{T^{2}_{c}}\right)
\eea
with $T_{c}^{2}  = 12\eta^{2}  / (N+2)$.  For $T<T_{c}$ the symmetry
is broken, with global minima at $|\phi| = \eta (T)$, where
$\eta^{2}(T) = \eta^{2} (1 - T^{2}/T_{c}^{2})$.  The effective Higgs
mass $m_{H}$, that measures the curvature in the radial field at these
minima, is $m_{H}(T) = g\eta (T)$ or,equivalently, is given by
$m_{H}^{2}(T) = -2m^{2}(T)$.
As before, there is a second-order
transition (with the
same index $\gamma =\half$  at one loop).

Global invariance sits uneasily in contemporary particle physics,
for which there is no evidence for Goldstone bosons.
To extend the theory to be invariant under {\it local} $O(N)$
transformations, it is necessary to introduce
$\half N(N-1)$ gauge fields $A_{\mu}$
 transforming as the adjoint representation of $O(N)$.  The
Lagrangian density can be written as
\beq
{\cal L} = \half D_{\mu}\phi D^{\mu}\phi -
\quarter Tr F_{\mu\nu}F^{\mu\nu}
-\frac{1}{8} g^{2}(\phi^{2} - \eta^{2})^{2}.
\eeq
Here the covariant derivative of $\phi$  is
\beq
D_{\mu}\phi = \partial_{\mu}\phi + eA_{\mu}\phi,
\eeq
and
\beq
F_{\mu\nu} = \partial_{\mu} A_{\nu} - \partial_{\nu} A_{\mu}
+ e[A_{\mu}, A_{\nu}].
\eeq
The $(N-1)$ Goldstone modes of the global theory now transmute into
longitudinal modes of the vector fields, enabling $(N-1)$ gauge fields
to acquire mass $m_{v} = e\eta$,
while the Higgs mass is $m_{H}  =g\eta$,   as before.
The remaining gauge fields stay massless.
In calculating $m^{2}(T)$ we now have to include gauge-field
(wavy line) one-loop diagrams
% *** Self-energies ****************************************************
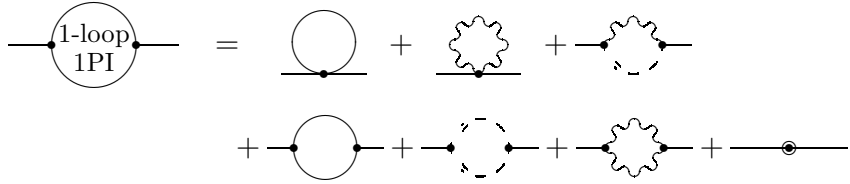
\typeout{*** Abelian Higgs Self-energies}
\bea
\bpi(40,26)(28,13)
\put(0,16){\line(1,0){16}}
\put(48,16){\line(1,0){16}}
\put(32,16){\circle{32}}
\put(16,16){\circle*{3}}
\put(48,16){\circle*{3}}
\put(32,17){\makebox(0,0)[b]{\footnotesize 1-loop}}
\put(32,13){\makebox(0,0)[t]{\mbox{\footnotesize 1PI}}}
\epi
&=&
\bpi(44,11)(-11,-3)
\put(-5,-11){\line(1,0){32}}
\put(11,-11){\circle*{3}}
\put(11,1){\circle{22}}
\epi
+
\bpi(44,11)(-11,-3)
\put(-5,-11){\line(1,0){32}}
\put(11,-11){\circle*{3}}
\put(2,0){\oval(4,4)[tl]}
\put(2,4){\oval(4,4)[br]}
\multiput(4,4)(-0.1,0.2){5}{\line(1,0){0.1}}
\multiput(3.5,5)(-0.1,0.1){5}{\line(1,0){0.1}}
\multiput(3,5.5)(-0.1,0.2){5}{\line(1,0){0.1}}
\put(4.5,6.5){\oval(4,4)[tl]}
\multiput(4.5,8.5)(0.2,-0.1){5}{\line(1,0){0.1}}
\multiput(5.5,8)(0.1,-0.1){5}{\line(1,0){0.1}}
\multiput(6,7.5)(0.2,-0.1){5}{\line(1,0){0.1}}
\put(7,9){\oval(4,4)[br]}
\put(11,9){\oval(4,4)[t]}
\put(15,9){\oval(4,4)[bl]}
\multiput(15,7)(0.2,0.1){5}{\line(1,0){0.1}}
\multiput(16,7.5)(0.1,0.1){5}{\line(1,0){0.1}}
\multiput(16.5,8)(0.2,0.1){5}{\line(1,0){0.1}}
\put(17.5,6.5){\oval(4,4)[tr]}
\multiput(19.5,6.5)(-0.1,-0.2){5}{\line(1,0){0.1}}
\multiput(19,5.5)(-0.1,-0.1){5}{\line(1,0){0.1}}
\multiput(18.5,5)(-0.1,-0.2){5}{\line(1,0){0.1}}
\put(20,4){\oval(4,4)[bl]}
\put(20,0){\oval(4,4)[tr]}
\put(2,0){\oval(4,4)[bl]}
\put(2,-4){\oval(4,4)[tr]}
\multiput(4,-4)(-0.1,-0.2){5}{\line(1,0){0.1}}
\multiput(3.5,-5)(-0.1,-0.1){5}{\line(1,0){0.1}}
\multiput(3,-5.5)(-0.1,-0.2){5}{\line(1,0){0.1}}
\put(4.5,-6.5){\oval(4,4)[bl]}
\multiput(4.5,-8.5)(0.2,0.1){5}{\line(1,0){0.1}}
\multiput(5.5,-8)(0.1,0.1){5}{\line(1,0){0.1}}
\multiput(6,-7.5)(0.2,0.1){5}{\line(1,0){0.1}}
\put(7,-9){\oval(4,4)[tr]}
\put(11,-9){\oval(4,4)[b]}
\put(15,-9){\oval(4,4)[tl]}
\multiput(15,-7)(0.2,-0.1){5}{\line(1,0){0.1}}
\multiput(16,-7.5)(0.1,-0.1){5}{\line(1,0){0.1}}
\multiput(16.5,-8)(0.2,-0.1){5}{\line(1,0){0.1}}
\put(17.5,-6.5){\oval(4,4)[br]}
\multiput(19.5,-6.5)(-0.1,0.2){5}{\line(1,0){0.1}}
\multiput(19,-5.5)(-0.1,0.1){5}{\line(1,0){0.1}}
\multiput(18.5,-5)(-0.1,0.2){5}{\line(1,0){0.1}}
\put(20,-4){\oval(4,4)[tl]}
\put(20,0){\oval(4,4)[br]}
\epi
+
\bpi(44,11)(-11,-3)
\put(-11,0){\line(1,0){11}}
\put(22,0){\line(1,0){11}}
\put(0,0){\circle*{3}}
\put(22,0){\circle*{3}}
\put(2,0){\oval(4,4)[tl]}
\put(2,4){\oval(4,4)[br]}
\multiput(4,4)(-0.1,0.2){5}{\line(1,0){0.1}}
\multiput(3.5,5)(-0.1,0.1){5}{\line(1,0){0.1}}
\multiput(3,5.5)(-0.1,0.2){5}{\line(1,0){0.1}}
\put(4.5,6.5){\oval(4,4)[tl]}
\multiput(4.5,8.5)(0.2,-0.1){5}{\line(1,0){0.1}}
\multiput(5.5,8)(0.1,-0.1){5}{\line(1,0){0.1}}
\multiput(6,7.5)(0.2,-0.1){5}{\line(1,0){0.1}}
\put(7,9){\oval(4,4)[br]}
\put(11,9){\oval(4,4)[t]}
\put(15,9){\oval(4,4)[bl]}
\multiput(15,7)(0.2,0.1){5}{\line(1,0){0.1}}
\multiput(16,7.5)(0.1,0.1){5}{\line(1,0){0.1}}
\multiput(16.5,8)(0.2,0.1){5}{\line(1,0){0.1}}
\put(17.5,6.5){\oval(4,4)[tr]}
\multiput(19.5,6.5)(-0.1,-0.2){5}{\line(1,0){0.1}}
\multiput(19,5.5)(-0.1,-0.1){5}{\line(1,0){0.1}}
\multiput(18.5,5)(-0.1,-0.2){5}{\line(1,0){0.1}}
\put(20,4){\oval(4,4)[bl]}
\put(20,0){\oval(4,4)[tr]}
\put(0,0){\line(0,-1){1}}
\multiput(0,-1)(0.04,-0.2){5}{\line(1,0){0.1}}
\multiput(5,-7.1)(-0.1,0.15){6}{\line(1,0){0.1}}
\multiput(5,-7.1)(0.1,-0.1){14}{\line(1,0){0.1}}
\multiput(3.9,-8.5)(0.15,-0.1){6}{\line(1,0){0.1}}
\multiput(10,-11)(-0.2,0.04){5}{\line(1,0){0.1}}
\put(10,-11){\line(1,0){2}}
\multiput(12,-11)(0.2,0.04){5}{\line(1,0){0.1}}
\multiput(19.5,-7.1)(0.1,0.15){6}{\line(1,0){0.1}}
\multiput(19.5,-7.1)(-0.1,-0.1){14}{\line(1,0){0.1}}
\multiput(18.1,-8.5)(-0.15,-0.1){6}{\line(1,0){0.1}}
\multiput(22,-1)(-0.04,-0.2){5}{\line(1,0){0.1}}
\put(22,0){\line(0,-1){1}}
\epi
\nonumber\\
&&
\bpi(-14,35)
\epi
+
\bpi(44,11)(-11,-3)
\put(-11,0){\line(1,0){11}}
\put(22,0){\line(1,0){11}}
\put(-1,0){\circle*{3}}
\put(23,0){\circle*{3}}
\put(11,0){\circle{22}}
\epi
+
\bpi(44,11)(-11,-3)
\put(-11,0){\line(1,0){11}}
\put(22,0){\line(1,0){11}}
\put(0,0){\circle*{3}}
\put(22,0){\circle*{3}}
\put(0,0){\line(0,1){1}}
\multiput(0,1)(0.04,0.2){5}{\line(1,0){0.1}}
\multiput(5,7.1)(-0.1,-0.15){6}{\line(1,0){0.1}}
\multiput(5,7.1)(0.1,0.1){14}{\line(1,0){0.1}}
\multiput(3.9,8.5)(0.15,0.1){6}{\line(1,0){0.1}}
\multiput(10,11)(-0.2,-0.04){5}{\line(1,0){0.1}}
\put(10,11){\line(1,0){2}}
\multiput(12,11)(0.2,-0.04){5}{\line(1,0){0.1}}
\multiput(19.5,7.1)(0.1,-0.15){6}{\line(1,0){0.1}}
\multiput(19.5,7.1)(-0.1,0.1){14}{\line(1,0){0.1}}
\multiput(18.1,8.5)(-0.15,0.1){6}{\line(1,0){0.1}}
\multiput(22,1)(-0.04,0.2){5}{\line(1,0){0.1}}
\put(22,0){\line(0,1){1}}
\put(0,0){\line(0,-1){1}}
\multiput(0,-1)(0.04,-0.2){5}{\line(1,0){0.1}}
\multiput(5,-7.1)(-0.1,0.15){6}{\line(1,0){0.1}}
\multiput(5,-7.1)(0.1,-0.1){14}{\line(1,0){0.1}}
\multiput(3.9,-8.5)(0.15,-0.1){6}{\line(1,0){0.1}}
\multiput(10,-11)(-0.2,0.04){5}{\line(1,0){0.1}}
\put(10,-11){\line(1,0){2}}
\multiput(12,-11)(0.2,0.04){5}{\line(1,0){0.1}}
\multiput(19.5,-7.1)(0.1,0.15){6}{\line(1,0){0.1}}
\multiput(19.5,-7.1)(-0.1,-0.1){14}{\line(1,0){0.1}}
\multiput(18.1,-8.5)(-0.15,-0.1){6}{\line(1,0){0.1}}
\multiput(22,-1)(-0.04,-0.2){5}{\line(1,0){0.1}}
\put(22,0){\line(0,-1){1}}
\epi
+
\bpi(44,11)(-11,-3)
\put(-11,0){\line(1,0){11}}
\put(22,0){\line(1,0){11}}
\put(0,0){\circle*{3}}
\put(22,0){\circle*{3}}
\put(2,0){\oval(4,4)[tl]}
\put(2,4){\oval(4,4)[br]}
\multiput(4,4)(-0.1,0.2){5}{\line(1,0){0.1}}
\multiput(3.5,5)(-0.1,0.1){5}{\line(1,0){0.1}}
\multiput(3,5.5)(-0.1,0.2){5}{\line(1,0){0.1}}
\put(4.5,6.5){\oval(4,4)[tl]}
\multiput(4.5,8.5)(0.2,-0.1){5}{\line(1,0){0.1}}
\multiput(5.5,8)(0.1,-0.1){5}{\line(1,0){0.1}}
\multiput(6,7.5)(0.2,-0.1){5}{\line(1,0){0.1}}
\put(7,9){\oval(4,4)[br]}
\put(11,9){\oval(4,4)[t]}
\put(15,9){\oval(4,4)[bl]}
\multiput(15,7)(0.2,0.1){5}{\line(1,0){0.1}}
\multiput(16,7.5)(0.1,0.1){5}{\line(1,0){0.1}}
\multiput(16.5,8)(0.2,0.1){5}{\line(1,0){0.1}}
\put(17.5,6.5){\oval(4,4)[tr]}
\multiput(19.5,6.5)(-0.1,-0.2){5}{\line(1,0){0.1}}
\multiput(19,5.5)(-0.1,-0.1){5}{\line(1,0){0.1}}
\multiput(18.5,5)(-0.1,-0.2){5}{\line(1,0){0.1}}
\put(20,4){\oval(4,4)[bl]}
\put(20,0){\oval(4,4)[tr]}
\put(2,0){\oval(4,4)[bl]}
\put(2,-4){\oval(4,4)[tr]}
\multiput(4,-4)(-0.1,-0.2){5}{\line(1,0){0.1}}
\multiput(3.5,-5)(-0.1,-0.1){5}{\line(1,0){0.1}}
\multiput(3,-5.5)(-0.1,-0.2){5}{\line(1,0){0.1}}
\put(4.5,-6.5){\oval(4,4)[bl]}
\multiput(4.5,-8.5)(0.2,0.1){5}{\line(1,0){0.1}}
\multiput(5.5,-8)(0.1,0.1){5}{\line(1,0){0.1}}
\multiput(6,-7.5)(0.2,0.1){5}{\line(1,0){0.1}}
\put(7,-9){\oval(4,4)[tr]}
\put(11,-9){\oval(4,4)[b]}
\put(15,-9){\oval(4,4)[tl]}
\multiput(15,-7)(0.2,-0.1){5}{\line(1,0){0.1}}
\multiput(16,-7.5)(0.1,-0.1){5}{\line(1,0){0.1}}
\multiput(16.5,-8)(0.2,-0.1){5}{\line(1,0){0.1}}
\put(17.5,-6.5){\oval(4,4)[br]}
\multiput(19.5,-6.5)(-0.1,0.2){5}{\line(1,0){0.1}}
\multiput(19,-5.5)(-0.1,0.1){5}{\line(1,0){0.1}}
\multiput(18.5,-5)(-0.1,0.2){5}{\line(1,0){0.1}}
\put(20,-4){\oval(4,4)[tl]}
\put(20,0){\oval(4,4)[br]}
\epi
+
\bpi(50,11)(-11,-3)
\put(-11,0){\line(1,0){44}}
\put(11,0){\circle*{3}}
\put(11,0){\circle{5}}
\epi
\nonumber
\eea
The precise diagrams involved depend on the gauge (these are
appropriate for the Landau limit of the covariant gauges  $\zeta=0$)
and some of these diagrams do not
give $O(T^2)$ contributions.  Further, we have not included tadpole
corrections so that we are assuming that we have shifted the vacuum
expectation value $\eta \rightarrow \eta(T)$.\tnote{This may not be
compatible with this result.  Maybe everthing is included here.
Must check Tom's notation and Dolan-Jackiw.}
At leading order in $T/\eta$ , the $(N-1)$ massive gauge modes cause
the effective $m^{2} (T)$ of the global scalar theory
to be changed from (12) to
\beq
m^{2}(T) = -\half g^{2}\eta^{2} +
[6(N-1)e^{2} + (N+2)g^{2}]\frac{T^{2}}{24},
\eeq
lowering the temperature at which $m^{2} (T)$ vanishes to
\beq
T_{c}^{2} = \frac{12g^{2}\eta^{2}}{6(N-1)e^{2} + (N+2)g^{2}}.
\eeq

However, a second-order transition at
temperature $T_{c}$ is no longer guaranteed.
The approximation of retaining only the $O(T^{2})$
term in (17) may not be valid if the ratio $e/g$ is sufficiently
large,  What we have calculated here is the
$\half m^{2}(T)\phi_{a}^{2}$    contribution
to the scalar sector effective potential,  Had we retained terms of
relative order $m_{v}/T$, the one-loop contribution of the gauge field is
not so much $(N-1)e^{2}T^{2}/4$, as seems from (17).  More accurately,
it is
\beq
m^{2}(T) = 6(N-1)e^{2}\left( \frac{T^{2}}{24} -
\frac{\pi}{8} Tm_{V}(\phi)\right),
\eeq
where $m_{V}(\phi) = e|\phi|$    becomes $m_{V}$ when $|\phi| =\eta$.
This term, linear in $|\phi|$,
induces a cubic term $O(e^{2}T|\phi|^{3})$ in the effective
potential \cite{LL}.

Such terms, if strong enough (i.e. $e/g$ large enough) can turn the
second-order transition into a first-order transition at a
temperature that is still essentially $T_{c}$  of (18).  Some caution is
necessary.  Eq. (19) is too simplistic,  As will be seen later, the
effect of the plasma is to induce electric and (for $N> 1$) magnetic
screening masses to the gauge fields, and $m_{V}$  (no longer a single
mass) is not simply proportional to $|\phi|$ \cite{LL}.  Further, magnetic
screening is intrinsically non-perturbative in $e$.  Nonetheless, for
$e/g\gg 1$, we do expect a first-order transition.  Defect formation
at a strong first-order transition is different from that at a
second-order transition, proceeding largely by bubble nucleation.
We shall not consider such a possibility and only assume
second-order transitions (inevitable for the global theory, as we
have seen).  In fact, this may not be a great restriction.  The
satisfactory evolution of the early universe string network
requires a mechanism for strings to chop one another up and for the
fragment loops to decay.  A strong first-order theory would have
strings with very different intercommutativity properties from
second-order strings (small $e/g$), and it
could well be that they would have undesirable consequences.

\section{ FIELD FLUCTUATION PROBABILITIES IN THERMAL EQUILIBRIUM}

Topological defects like strings and monopoles can be characterised
by non-local field configurations (e.g. magnetic flux through a
surface) or by local configurations (e.g. field zeros). In either
case, the likelihood of their appearance at phase transitions can
be determined if we know the probabilities for arbitrary field
configurations.

We begin by calculating the configuration probabilities of $O(N)$
scalar fields in thermal equilibrium at temperature $T$. To do so it
is  more convenient to adopt the imaginary-time approach to thermal
field theory than the real-time approach (with its heatbath
populated by real particles whose propagation we follow) adopted earlier.

\subsection{ The Imaginary-Time Formalism.}

The idea is simple \cite{TFT,Ka}.
First consider the theory of the single scalar
field $\phi$ of (4).  Let $H[\pi ,\phi]$ be the Hamiltonian derived
 from $S[\phi]$ (with $\pi = \dot{\phi}$).  Then, at temperature $T =
\beta^{-1}$, the partition function $Z$
\beq
Z = tr\rho = tr e^{-\beta {\hat H}}
\eeq
can be written as
\beq
Z = \sum_{n} <\Phi_{n},t_{0}|e^{-\beta{\hat H}}|\Phi_{n},t_{0}>
\eeq
where, in evaluating $tr\rho$ we have chosen a basis of eigenstates
of $\hat{\phi}$ at time $t_{0}$. i.e.
\beq
{\hat \phi}(t_{0},\vecx)|\Phi_{n},t_{0}> =
\Phi_{n}(\vecx)|\Phi_{n},t_{0}>.
\eeq
For simplicity, the $\Phi_{n}$ have been taken to be denumerable
(e.g. by the imposition of periodic boundaries).

Already, the interplay between thermal and quantum fluctuations is
apparent.  The diagonal matrix element
$ <\Phi_{n},t_{0}|e^{-\beta{\hat H}}|\Phi_{n},t_{0}>$
permits {\it two}
interpretations:-
\beq
 <\Phi_{n},t_{0}|e^{-\beta{\hat H}}|\Phi_{n},t_{0}> =
p_{t_{0}}[\Phi_{n}],
\eeq
the (relative) {\it probability} that the field takes the value
$\Phi_{n}(\vecx)$ at time $t_{0}$, or
\beq
<\Phi_{n},t_{0}|e^{-\beta{\hat H}}|\Phi_{n},t_{0}> =
<\Phi_{n},t_{0}-i\beta\hbar|\Phi_{n},t_{0}>,
\eeq
the {\it probability amplitude} that the field has value $\Phi_{n}$
at time $t_{0}-i\beta$ if it had value $\Phi_{n}$
at time $t_{0}$.  We continue to work in units in which $\hbar =1$.

Since the system is in thermal equilibrium both interpretations are
independent of $t_{0}$.  The first (statistical mechanical) is just
a reiteration of the probabilities of the Boltzmann distribution and
gives us the quantity $p[\Phi]$ we wish to calculate.  However, our
ability to interpret this probability as a probability {\it
amplitude} permits the quantum mechanical
path integral realisation
\beq
p_{t_{0}}[\Phi] =
\int_{\phi(t_{0},\vecx) = \Phi (\vecx)} {\cal D}\phi\; e^{iS[\phi]},
\eeq
where the sum is restricted to fields periodic in imaginary time,
$\phi(t_{0},\vecx) =\phi(t_{0}-i\beta,\vecx)$, period $\beta$.

The periodicity of the field in Euclidean (imaginary) time is most
simply implemented by integrating in a straight line from $t_{0}$
to $t_{0}-i\beta$ as in Fig. 2.
% *** ITF curve ******************************************
\typeout{*** ITF Curve}
\setlength{\unitlength}{2pt}%
\begin{figure}[htb]
\begin{center}
\setlength{\unitlength}{0.5pt}
\begin{picture}(237,360)(25,455)
\put(140,625){\makebox(0,0)[lb]{\Large $C_I$}}
\thicklines
\put(135,735){\circle{10}}
\put(135,485){\circle{10}}
\put(135,625){\line( 0,-1){140}}
\put(135,735){\vector( 0,-1){110}}
\put(25,735){\vector( 1, 0){200}} % re axis
\put(40,455){\vector( 0, 1){360}} % im axis
%\put(35,485){\line( 1, 0){ 10}}
\put(42,785){\makebox(0,0)[lb]{\Large $\Im m(t)$}}
\put(42,745){\makebox(0,0)[lb]{\Large $0$}}
\put(140,480){\makebox(0,0)[lb]{\Large $t_0-i \beta$}}
\put(135,745){\makebox(0,0)[lb]{\Large $t_0$}}
\put(195,705){\makebox(0,0)[lb]{\Large $\Re e(t)$}}
\end{picture}
\end{center}
%\vspace{1cm}
\caption{ITF curve.}
\label{fitf}
\end{figure}
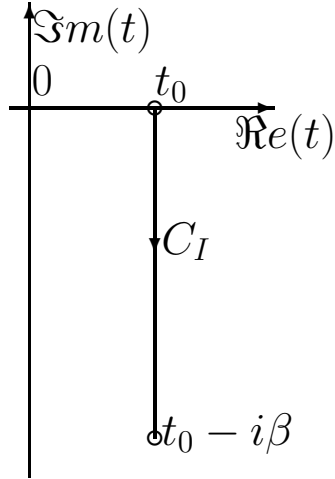
%**************************************************************

$p_{t_{0}}[\Phi]$ of (25) then becomes
\beq
p_{t_{0}}[\Phi] =
\int_{\phi(t_{0}) = \Phi} {\cal D}\phi\; e^{-S_{E}[\phi]},
\eeq
where
\beq
S_{E}[\phi] = \int_{0}^{\beta} d\tau\int d^{3}x
[\frac{1}{2} \dot{\phi}^{2}
+\frac{1}{2}(\nabla\phi)^{2}
+\frac{1}{2} m_{0}^{2}\phi^{2} + \frac{1}{8} g^{2}\phi^{4}]
\eeq
for fields periodic in $\tau$ with period $\beta$ (position labels
have been omitted, and the dot denotes differentiation with respect
to $\tau$).  This periodicity is enforced by Fourier expanding
$\phi(\tau,\vecx)$ as
\beq
\phi(\tau,\vecx) = \sum_{n} \phi_{n}(\vecx)e^{2\pi in\tau/\beta}
\eeq
whereby the quadratic part of $S_{E}[\phi]$      becomes
\beq
S_{E}[\phi]_{qu} = \beta\sum_{n}\int d^{3}x
[\half (\nabla\phi_{n})^{2}
+\half m_{n}^{2}\phi^{2}_{n}]
\eeq
in which
\beq
m_{n}^{2} = m_{0}^{2} + \left(\frac{2\pi n}{\beta}\right)^{2}.
\eeq
At high $T$ the ratio $m_{n}/m_{0}$ is large.  The $n=0$    mode
is the {\it light} mode of the field,
the $n\neq 0$ modes the {\it heavy} modes.
One very successful approach to high -T equilibrium theory is to
integrate out the heavy modes, to obtain an effective
{\it three}-dimensional  $\phi_{0}$ - theory.
That is, we rely on the controlled
decoupling of massive modes to perform a dimensional reduction from
four to three dimensions, valid when
\beq
\beta\ll \xi\ll \L
\eeq
where $L$  is the system size, $\xi$   the field correlation length, and
$\beta$  the thermal wavelength.

In estimating probabilities, it is good
enough to observe that, from (25),
\bea
p_{t_{0}}[\Phi] &=&
\int\prod_{n}{\cal D}\phi_{n} \;
\delta (\sum_{n}\phi_{n}(0) -\Phi)e^{-S_{E}[\phi]}\\
&\simeq& \int {\cal D}\phi_{0}\; \delta (\phi_{0}(0)
-\Phi)\int\prod_{n}\; D\phi_{n}
e^{-S_{E}[\phi]},
\eea
since heavy mode fluctuations are small.
The $\delta$ denotes a $\delta$-functional in which $\phi_{0}(0,\vecx) =
\Phi (\vecx)$ at each point in space.
On defining the three-dimensional effective action $S_{3}[\phi_{0}]$ by
\beq
Ne^{-\beta S_{3}[\phi_{0}]} =
\int\prod_{n\neq 0}{\cal D}\phi_{n}\; e^{-S_{E}[\phi]}
\eeq
where $N$ is a normalisation constant, it then follows from (33) that
\beq
p_{t_{0}}[\Phi]\simeq Ne^{-\beta S_{3}[\Phi_{0}]}.
\eeq
[A better calculation \cite{HR} would show that
\beq
p_{t_{0}}[\Phi] = Ne^{-\beta H[\Phi_{0}]},
\eeq
where
\beq
H[\Phi] = S_{3}[\Phi] - \frac{\beta^{2}}{24}\int\left(\frac{\delta
S_{3}}{\delta\Phi}\right)^{2} + O(\beta^{4})
\eeq
but, at the level to which we are working, (35) is sufficient.]
For $S_{E}[\phi]$ of (29), the one-loop high-T behaviour of
$S_{3}[\Phi]$ (in which
only terms $O(g^{2}T^{2})$ are displayed) is
\beq
S_{3}[\Phi] = \int d^{3}x
[\half (\nabla\Phi)^{2}
+\half m^{2}(T)\Phi^{2} + \frac{1}{8} g^{2}\Phi^{4}],
\eeq
where $m^{2}(T)$ is the effective mass (9). It is now understood as the
light-mode mass whose vanishing signals the occurrence of a phase
transition.  [By not containing $\phi_{0}$   fluctuations,$m^{2}(T)$ of (38)
differs from $m^{2}(T)$ of (9) by non-leading terms, which we ignore].
A more accurate calculation of $S_{3}$ \cite{KK}
would show it to be non-local.  The
expression (38) is sensible only for distances larger than the
thermal wave length $\beta$, at which there is an effective cutoff.

\subsection{ Gaussian Fluctuations.}

Remaining with the single field, the probability
$p_{t_{0}}[\Phi]$, that $\phi(t_{0},\vecx) = \Phi(\vecx)$,
is just a starting point.
 From it we wish to calculate the more general probabilities
$p(Q[\phi] =q)$ that some functional of
$\phi(t_{0},\vecx)$, $Q[\phi]$, equals $q$.
The quantity $q$ could be a
number, or a function $q(\vecx)$, depending on $Q$.  For simplicity, take
$Q$ as a pure functional, $q$ a number.  Generalisation is
straightforward.  If $< ...>$  denotes averaging with respect to
$p_{t_{0}}[\Phi]$, then
\bea
p(Q[\phi] =q) &=& <\delta(Q[\phi] - q)>\\
&=& \int d\alpha <e^{i\alpha (Q[\phi] - q)}>.
\eea
The cumulant expansion of (40) is, up to quadratic terms
\beq
p(Q[\phi] =q) = \int d\alpha\; e^{-i\alpha q}
e^{-\half \alpha^{2}[<QQ> - <Q>^{2}] + ...}.
\eeq
In the Gaussian approximation, in which only these terms are
retained, the $\alpha$  integration can be performed, to give
\beq
p(Q[\phi] =q) = N e^{-\half (q - <Q>)^{2}/<QQ>_{c}},
\eeq
where $<QQ>_{c} = <QQ> - <Q>^{2}$ is the connected two-$Q$
correlation function.

Our immediate interest is to determine in what circumstances field
fluctuations are large.  The simplest measure of fluctuations is the
coarse-grained field $Q[\phi] = \phi_{v}$ :-
\beq
\phi_{v} = \frac{1}{v} \int_{\vecx\in v} d^{3}x\; \phi(t_{0},\vecx)
\eeq
for an arbitrary volume $v$. [We have used the symbol $v$ to denote
both the position of the volume and its magnitude].
Introducing the window function (indicator function) $I(\vecx)$:
$I(\vecx) = 1, \vecx\in v; I(\vecx) = 0,\vecx\notin v$,
                       it follows that
\beq
<\phiv\phiv> = \frac{1}{v^{2}}\int d\vecx d\vec{\dash{x}}
I(\vecx)I(\vec{\dash{x}})G(\vecx -\vec{\dash{x}}),
\eeq
where $G(\vecx -\vec{\dash{x}})$ is the equal-time correlation
function, read off from (38) as
\beq
G(\vecx -\vec{\dash{x}})\simeq T\int \dbar^{3}k\; \frac{e^{i\veck.(\vecx
-\vec{\dash{x}})}}{\veck^{2} + m^{2}(T)}
\eeq
Suppose $m^{2}(T) > 0$ in (45).  Then, in the Gaussian approximation,
\beq
p(\phiv = \phibar) = N e^{-\half\phibar^{2}/<\phiv\phiv>}.
\eeq
The momentum integration in (45) is cut off at $|\veck| = O(T)$.  It follows
 from (45) that $\Phi_{0}(\vecx)$        is correlated over a distance
$\xi = m(T)^{-1}$.

The variance (44) can be written in terms of the Fourier
transforms
$\tilde{I} (\veck)$ of the window function (normalised to $\tilde{I}
(\vec{0}) = v$) and
\beq
G(\vec{k})\simeq\frac{T}{\vec{k}^{2} + m^{2}(T)},
\eeq
as
\beq
<\phiv\phiv> = \frac{1}{v^{2}}\int \dbar^{3} k
\; |\tilde{I}(\vec{k})|^{2} G(\vec{k}).
\eeq
Since the effect of $\tilde{I}$ is to cut off the fluctuations with
wavelengths shorter than $R$, where $v = O(R^{3})$, we can replace (48)
by
\beq
<\phiv\phiv> = \frac{1}{v^{2}}\int_{|\veck|\leq R^{-1}} \dbar^{3} k
\; G(\vec{k}).
\eeq
For
a correlation volume $v = O(\xi^{3})$, $\xi = m^{-1}(T)$, it can be shown
\cite{HR} that
\beq
<\phiv\phiv> \simeq ATm(T),
\eeq
where $A\simeq 10^{-1}$. Thus field fluctuations on the scale of
correlation volumes have {\it rms}
  value
\beq
\Delta\phi\simeq (Tm(T))^{\half}.
\eeq
[The prefector $A$ is important in demonstrating that thermal
fluctuations are, in general, not likely to supplant bubble
nucleation in first-order transitions \cite{GK} but we shall not
pursue this here.]

\subsection{ The Ginzburg Regime.}

Now suppose that $m^{2}(T) < 0$. Still for a single field,
$S_{3}[\Phi]$ can be written (cf.(38)) as
\beq
S_{3}[\Phi] = \int d^{3}x
[\frac{1}{2}(\nabla\Phi)^{2}
+\frac{1}{8} g^{2}(\Phi^{2} - \eta^{2} (T))^{2}].
\eeq
Consider the fluctuations of the `Higgs' field in one of
minima, $\Phi = \eta$  say, averaged over a correlation volume
$v = O(m_{H}^{-3}(T)) = O(g^{-3}\eta (T)^{-3})$ for the
Higgs field.  As before, $\Delta\Phi\simeq (Tg\eta (T))^{\half})$.
On going close to the critical
temperature $T_{c} = 2\eta$ a point is reached at which
$\Delta\Phi\simeq\eta (T)$     .  At this stage
correlation volumes of the field can fluctuate from the true vacuum
at $\phi =\eta (T)$   (or $\phi = -\eta(T)$) to the false vacuum
($\phi = 0$) with significant
probability.  The temperature at which this happens is the Ginzburg
temperature $T_{G}$ \cite{G} which, from above, is equally defined by
\beq
\Delta\phi(T_{G})\simeq \eta(T_{G})
\eeq
or
\beq
g_{3} = \frac{gT_{G}}{\eta(T_{G})} = \frac{g^{2}T_{G}}{m_{H}(T_{G})} \simeq 1
\eeq
or
\beq
\left(1 - \frac{T_{G}^{2}}{T_{c}^{2}}\right) = O(g^{2}).
\eeq
Equation (53) tells us how close we can get to the critical
temperature before fluctuations get too violent.  The equivalent
equation (54) says that the effective dimensionless coupling
constant of the three-dimensional theory (which, from (36) and
(38), is $g_{3}$) becomes strong.  Analysis of higher-loop diagrams
would also show that it is at $T = T_{G}$ that the one-loop approximation,
with its mean-field behaviour, breaks down.
Some care is needed on trying to define $g_{3}$  nearer to the critical
point.  As a three dimensional coupling constant it runs to an
ultraviolet fixed point \cite{Chris}
i.e. $g$ vanishes with $\eta (T)$ (or $m(T)$) at
$T = T_{c}$.
However, for our purposes we do not need to get closer to $T_{c}$   than
$T_{G}$ and hope that the one-loop result remains
qualitatively correct.

In the symmetric phase the extension to a global $O(N)$ theory is
straightforward. $ \Phi$   now
becomes an N-component column vector.  In the Gaussian
approximation, the field distribution probabilities
$p_{t_{0}}[\Phi]$
become, from (35)
\beq
p_{t_{0}}[\Phi] = N exp\{-\half\int d\vecx  d\vec{\dash{x}}
\; \Phi_{a}(\vecx) G_{ab}(\vecx -\vec{\dash{x}})\Phi_{b}(\vec{\dash{x}} \},
\eeq
where $G_{ab}$   is diagonal, as
\beq
G_{ab}(\vecx -\vec{\dash{x}}) =\delta_{ab}G(\vecx -\vec{\dash{x}}),
\eeq
with $G(\vecx -\vec{\dash{x}})$     given by (45).
Root mean square fluctuations in $|\Phi|$    over a correlation-volume are
$O((Tm(T))^{\half}$    , as before.

In the broken symmetry phase the situation is more complicated
because of the masslessness of the Goldstone modes,   In a Gaussian
approximation $G_{ab}$ is no longer proportional to $\delta_{ab}$.
However, the
$O(N)$ radial field (Higgs) fluctuations must be such as to take
$|\phi|$ in a
correlation-volume from a true vacuum at $|\phi|=\eta(T)$ to the false
vacuum at $\phi = 0$   with significant probability at a Ginzburg temperature
$T_{G}$  satisfying (55).
For {\it local} gauge theories the scalar field $\phi$  is gauge-variant and we
cannot define probabilities for arbitrary field configurations.
Rather, we shall be interested in the probabilities $p(Q[\phi]=q)$       for
gauge-invariant $Q$.  These will be developed as needed.

\section{ STRINGS AND MONOPOLES NEAR $T_{c}$ }

So far our discussion of phase transitions has made no reference to
the possible defects in the $O(N)$ theories.  However a
symmetry-breaking transition is implemented, the $O(N)$
vector field does not switch
uniformly in space (and time) from its symmetric vacuum value $\phi
= 0$   to some value $\phi = \phi_{c}, |\phi_{c}| = \eta$      on the
vacuum manifold ${\cal M} = S^{N-1}$ which characterises
the vacuum at large times.
Rather, initially we expect $\phi$  to relax towards different elements
$\phi_{c}$ of ${\cal M}$  in different regions of space.
What regions and how is the
problem that we shall attempt to solve.  Whatever, at some fairly
early stage after the transition we expect a domain structure to
have formed, such that within a single domain the $O(N)$ vector field
$\phi$ is strongly
correlated. It is between the domains that defects will be found.

\subsection{ Global Strings.}

If we take a closed path through several domains the field $\phi$
takes a closed path on ${\cal M}$. If ${\cal M}$  is not simply-connected
this path in ${\cal M}$  may be non-contractable.
In that case the path in space has trapped a
string (vortex) or strings.
However, a non-contractable path is a `large' path in field space.
For there to be a reasonable chance for such a path to be executed
the differences in the fields between adjacent domains must be
substantial.  That is, field fluctuations must be large.  The idea
that defects are born out of large fluctuations of domains was first
articulated by Kibble.

We have just seen that one circumstance in which there are large
fluctuations is close to a second-order phase transition,
as characterised by the
Ginzburg temperature.  For global theories we know that we have the
second-order transition for this to apply.
We stress that this is not the only situation
in which large fluctuations arise, and we shall turn to others
later. If we knew more of the nature of phase transitions in the
early universe we could determine whether $T_{G}$ has any relevance.
We don't know, but because it is the simplest option we examine it
first, following an argument due to Kibble. [There are several
mechanisms proposed by Kibble for the production of defects,
according to the order of the transition, the timescales, etc.. See
these proceedings.  This is the simplest.]

Specifically, \cite{TK} we have seen that the Higgs
field is correlated over a volume of
characteristic size $\xi^3$ where $\xi=m_H^{-1}(T)$.
 From our earlier discussion, the first time that one
can define such patches as the
temperature falls below $T_{c}$ is at the Ginzburg
temperature.  In each patch the
Higgs field is in the vacuum manifold but
this in itself is not enough to determine the frequency of strings,
since it is the field phases, rather
than field magnitudes, that enforce their presence.
The next step is to assume
that the field phases are correlated on the
same length scale $\xi$ (even if they are Goldstone modes).
In the Ginzburg regime the fluctuations are expected, from our
previous analysis, to be so large that each patch will have a
phase capable of  differing by a large amount $O(1)$ from that of
its neighbours.   Thus, if it is further assumed that the
phases in different correlation volumes are randomly distributed,
we get large random jumps
in phase as we cross domain boundaries.
This is all that we need to show the presence of strings in large
quantities.

Instead of having to make these assumptions, they
should be derivable for equilibrium field theory
at the Ginzburg temperature.
We shall spend the rest of this
chapter seeing to what extent this mechanism, the Kibble mechanism,
is justified in this simple form, both for strings and for monopoles.

By definition, if ${\cal M}$ is not simply-connected
then its first (fundamental)
homotopy group $\Pi_{1}({\cal M})$ is non-trivial.
Since $\Pi_{n}(S^{m})$ is trivial unless $n=m$, when $\Pi_{n}(S^{n})=Z$,
 in $D=3$ dimensions there will be $O(N)$ strings only if $N=2$ i.e.
$O(2)$ or $U(1)$.  The elements $n\in Z$ are the
winding numbers of the strings, a change in the phase of the complex
$U(1)$ field
\beq
\phi = \phi_{1} + i\phi_{2}
\eeq
of $2\pi n$ along the loop corresponding to a winding
number $n$.  For the field phase $\phi /|\phi|$   to change by $2\pi$ (or a
non-zero multiple of $2\pi$) requires that it cannot
be well-defined everywhere since $\phi$ is continuous and single-valued.
As a result $\phi$ must vanish somewhere in the
loop.
$O(2)$ or $U(1)$ strings are thus characterised by  lines of zeros of the
field doublet, i.e. they are tubes of false vacuum, for which
$\phi\simeq 0$, embedded in a true vacuum in which $|\phi|\simeq\eta$.

The properties of global $O(2)$ strings are well-documented \cite{SV}.
In the
present context of thermal equilibrium they are the local axisymmetric
{\it instanton} solutions to the field equations
\beq
\frac{\delta S_{3}[\phi]}{\delta\phi_{a}} =
-\nabla^{2}\phi_{a} + \half g^{2}\phi_{a}(\phi^{2} - \eta^{2}) = 0.
\eeq
The $\phi_{a}$ in (59) have only spatial arguments $\phi_{a}
=\phi_{a}(\vecx)$, since dimensional compactification has already
been effected in our use of $S_{3}$. More generally they are the
solutions to the four-dimensional Euler-Lagrange equations for the
action (11) i.e. {\it solitons}.  However, as written in (59), we
have traded dynamical degrees of freedom for temperature-dependence
of the parameters of the theory in a straightforward way.

In cylindrical coordinates ($r, \theta, \phi$) a single string with
winding number $n$,
as solution to (59), takes the form
\beq
\phi =\eta (T) e^{in\theta} f(r g\eta(T)),
\eeq
where
\beq
f(r)\approx c_{n}r^{n}, r\rightarrow 0 ; f(r)\approx 1 - O(r^{-2}),
r\rightarrow\infty.
\eeq
As a result, we can deduce that

a) the string thickness $a(T)$ is $O(\xi(T)) = O(m_{H}^{-1}(T)) =
O(g^{-1}\eta(T)^{-1})$, the Higgs field correlation length at
temperature $T$.

b) strings with winding number $n >1$ (which can be thought of as
lines of multiple zeros) are unstable, preferentially splitting into
strings for which $n = 1$.

c) the energy per unit length diverges logarithmically, cut off by
the presence of nearest neighbour strings in any network.
Logarithms apart, the energy/length is $O(\eta^{2}(T))$, vanishing
as we appproach the transition.

d) the net winding number of strings through a surface $S$ is
given as the line integral
\beq
N_{S} = \frac{1}{2\pi} \oint_{\partial S}
\vec{dl}.\; \vec{\partial}\alpha ,
\eeq
where we have adopted the radial/angular field decomposition
\beq
\phi = \rho e^{i\alpha}.
\eeq
In the complex field notation of (58), $N_{S}$ of (62) can be
reexpressed as
\bea
N_{S} &=& \frac{-i}{2\pi} \oint_{\partial S}
\vec{dl} .\; \frac{\phi^{\dagger}
\stackrel{\leftrightarrow}{\partial}\phi}
{|\phi|^{2}}.
\eea
More
usefully, as a surface integral, it is equivalent to
\beq
N_{S}
%&=& \frac{1}{\pi} \int_{S} \vec{dS}
%.(\underline{\partial}\phi_{1}\wedge\;\underline{\partial}\phi_{2})
%\frac{1}{|\phi|^{2}}
%\\
= \frac{-i}{2\pi} \int_{S} \vec{dS}.\frac{
(\vec{\partial}\phi^{\dagger}\wedge\;\vec{\partial}\phi)}
{|\phi|^{2}}
\eeq
or, in terms of $\rho$ and $\alpha$
\beq
N_{S}
= \frac{1}{\pi} \int_{S} \vec{dS}.\frac{
(\vec{\partial}\rho\wedge\;\vec{\partial}\alpha)}
{\rho^{2}}.
\eeq

To see that fluctuations near the phase transition
are indeed capable of creating vortices
we need to  calculate the probability that the winding number of the field
through a loop $\partial S$ bounding a surface $S$ be $n$.
In practice it is more
convenient to evaluate the related quantity
\bea
\bar{N}_{S} &=& \frac{-i}{2\pi} \oint_{\partial S}
\vec{dl} .\frac{\phi^{\dagger}
\stackrel{\leftrightarrow}{\partial}\phi}
{\eta(T)|\phi|}
\nonumber\\
&=& \frac{2}{2\pi\eta(T)} \int_{S} \vec{dS}'
.(\vec{\partial}\rho\wedge\vec{\partial}\alpha)
\eea
For a large loop $\partial S$  the difference between $N_{S}$ and
$\bar{N}_{S}$ (not integer) is vanishingly small if no vortices pass close to
$\partial S$,
and $\bar{N}_{S}$ remains a good indicator of vortex production.

In the Gaussian approximation the probability that $\bar{N}_{S}(t)$   takes the
value $n$ is
\beq
p(\bar{N}_{S} = n) = \exp\{-\half n^{2}/<\bar{N}_{S}
\bar{N}_{S}>\}.
\eeq
On decomposing the radial mode as $\rho = \eta (T) + h$ for Higgs
field $h$ and defining the Goldstone mode $g$ by $g =\eta (T)\alpha$, from
(67) it follows that
\beq
<\bar{N}_{S}\bar{N}_{S}> =
(\frac{2}{2\pi\eta^{2}(T)})^{2}
\int\int_{S} dS' dS''<(\partial h'\wedge\partial g')(\partial
h''\wedge\partial g'')>.
\eeq
The primes (doubleprimes) denote fields in the
infinitesimal areas $dS', dS''$ of $S$ respectively.
For economy of notation we have not made ths scalar products explicit.
Without loss of generality we take $S$  in the 1-2
plane, whence
\beq
<\bar{N}_{S}\bar{N}_{S}> =
(\frac{2}{2\pi\eta^{2}(T)})^{2}\int\int_{i,j = 1.2} dS' dS''
<\partial_{i} h'\partial_{i} h''\partial_{j} g'\partial_{j}
g'' -\partial_{i} h'\partial_{j} h''\partial_{j} g'\partial_{i} g''>.
\eeq
It is convenient to refine our notation
further, decomposing space-time as $x = (t,\vecx) =
(t,\vecx_{L},x_{T})$
where $\vecx_{L} = (x_{1},x_{2})$
denotes the co-ordinates of $S$, and $x_{T} = x_{3}$ the transverse
direction to $S$.
Similarly, we separate 4-momentum $p$  as $p = (E,\vecp_{L},p_{T})$.
    .

Let $G_{h}(\vecx' -\vecx'') = <h(\vecx')h(\vecx'')>$,
$G_{g}(\vecx' -\vecx'') = <g(\vecx')g(\vecx'')>$
be the Higgs field and Goldstone mode correlation functions
respectively as read off from $S_{3}$.  As a first step we ignore
correlations between Higgs and Goldstone fields.  That is, we retain
only the disconnected parts of $<\bar{N}_{S}\bar{N}_{S}>$.  Eqn. (70) then
simplifies to
\bea
<\bar{N}_{S}\bar{N}_{S}> =
(\frac{2}{2\pi\eta^{2}(T)})^{2}\int\int_{i,j = 1.2} dS' dS''&&
[<\partial_{i} h'\partial_{i} h''><\partial_{j} g'\partial_{j}
g''>\nonumber\\
&& -<\partial_{i} h'\partial_{j} h''><\partial_{j} g'\partial_{i} g''>]
\eea
which can be written as
\beq
<\bar{N}_{S}\bar{N}_{S}> =
(\frac{2}{2\pi\eta^{2}(T)})^{2}\int\int\dbar^{3} p'\dbar^{3} p''\;
G_{h}(\vec{p}')
G_{g}(\vec{p}'')
|\tilde{I} (\vec{p}_{L}'' -\vec{p}_{L}')|^{2}
[(\vecp_{L}')^{2}(\vec{p}_{L}'')^{2} - (\vec{p}_{L}' .\vec{p}_{L}'')^{2}].
\eeq
In (72) $\tilde{I} (\vec{p}_{L})$ is the Fourier transform
of the window function
$I(\vecx_{L})$ of the surface $S$ (i.e. $I(\vecx_{L}) = 1$ if
$\vecx_{L}\in S$, otherwise zero).
%The $G(\vec{p})$ are defined as
%in (3.22).

We coarse-grain in the transverse and longitudinal directions by imposing a
cut-off in three-momenta at $|p_{i}|<\Lambda = l^{-1}$,  for some $l$, as
before.  Thus $\bar{N}_{S}$   is now understood as the average value over a
closed set of correlation-volume `beads' through which $\partial S$ runs
like a necklace.

For large loops $\partial S$, $\tilde{I}
(\vec{q}_{L})\simeq\deltabar (\vec{q}_{L})$, enabling us to write
\bea
\lefteqn{<\bar{N}_{S}\bar{N}_{S}> =}
\\
&& (\frac{2}{2\pi\sigma^{2}})^{2}\int\dbar q_{T}\int\dbar^{3} p\;
G_{h}(\vec{p} +\vec{q}_{T},t)G_{g}(\vec{p},t)\int\dbar^{2} q_{L}
|\tilde{I} (\vec{q}_{L})|^{2}
[(\vecp_{L})^{2}(\vec{q}_{L})^{2} - (\vec{p}_{L} .\vec{q}_{L})^{2}]
\nonumber
\eea
By $\vec{p} +\vec{q}_{T}$ we mean $(\vec{p}_{L}, p_{T} + q_{T})$.
The dependence on the contour $\partial S$
is contained in the final integral
\bea
{\cal J} &=&\int\dbar^{2} q_{L}\;
|\tilde{I} (\vec{q}_{L})|^{2}
[(\vecp_{L})^{2}(\vec{q}_{L})^{2} - (\vec{p}_{L} .\vec{q}_{L})^{2}]
\\
&\simeq& \pi p_{L}^{2}\int^{\Lambda} dq_{L}\; q_{L}^{3}|\tilde{I} (q_{L})|^{2}
\eea
If this is evaluated for a circular loop of
radius $R$, we find
\beq
{\cal J} = p_{L}^{2}\;O(2\pi R/l),
\eeq
as we might have anticipated.  The $rms$ winding number behaves with
path length as $\Delta n \propto {\cal J}^{\half} = O(L^{\half})$
, where $L$  is the number of steps of
length $l$.  This is consistent with the coarse-grained field
phases of different
volumes $v$ being randomly distributed.

The final step is to relate the magnitude of the fluctuations in
winding number $N_{S}$ to the magnitude of the radial (Higgs) field
fluctuations and angular (Goldstone) field fluctuations.
To estimate the fluctuations we $a)$ neglect the positivity of
$\rho$ and the Jacobian from the non-linear transformation (43) and
$b)$ the non-singlevaluedness of $\alpha$.  While valid for small
fluctuations about the global minima this can only be approximate
for large fluctuations.
With these provisos, at temperature $T$ the
equilibrium propagators are time-independent,
dominantly the free-field propagators
\bea
G_{h}(\vec{p}) &=& \frac{T}{\vec{p}^{2} + m_{H}^{2}(T)}
\\
G_{g}(\vec{p}) &=& \frac{T}{\vec{p}^{2}}
\eea
If we take
$\vec{p}_{L}^{2} = 2\vec{p}^{2}/3$ in the integral
then, up to numerical factors, we can approximate (4.16) as
\beq
<\bar{N}_{S}\bar{N}_{S}> \simeq
{\cal J} \left(\frac{2}{2\pi\sigma^{2}}\right)m_{H}(T)T
\int_{|\vec{p}|<m_{H}(T)} \dbar\vec{p} G_{h}(\vec{p}),
\eeq
where we have coarse-grained to the Higgs correlation
length $\xi = m_{H}^{-1}(T)$.
[We have further assumed that the $q_{T}$ integration can be
approximated by setting $q_{T}$ to zero in the integrand.
Qualitatively this is a reasonable simplification].
The integral in (79) is essentially the integral (49).  The end
result is that \cite{RR}, after substitution,
\beq
<\bar{N}_{S}\bar{N}_{S}> =
O\left(\left(\frac{m_{H}(T)T}{\eta^{2}(T)}\right)^{2}\right),
O(L)
\eeq
where $L=Rm_{H}(T)$ is the length of the path in units of $\xi$.
Equivalently, on using our previous results for equilibrium
\bea
<\bar{N}_{S}\bar{N}_{S}>&=&
O\left(\left(\frac{<h_{v}h_{v}>}{\eta^{2}(T)}\right)^{2}\right)
O(L)
\\
&=&
O\left(\left(\frac{g^{2}T}{m(T)}\right)^{2}\right)
O(L)
\eea
%One power of $\expect{h_{v}h_{v}}$ comes from $G_{h}$,
%the other from residual factors.
 Thus, as anticipated by Kibble, the phase fluctuations are
indeed scaled by the Higgs
fluctuations.  Further, in the Ginzburg regime, when correlation-volumes
of the Higgs field can fluctuate to the false vacuum with
significant probability, the fluctuations in field phase on the same
distance scale are of order unity and, from the $O(L)$ term, are
distributed randomly.
%[It is unclear whether the inclusion of connected
%correlation functions - linking the Higgs and Goldstone fields -
%would change the results qualitatively in this regime, and is under
%examination.  The inclusion of four-point correlation functions in
%the real scalar field calculations had no dramatic effect, although
%the circumstances were somewhat different.]

Even though fluctuations are large enough
to produce non-trivial winding number there may still be some doubt that
this is directly related to strings.  However, strings intersect
sufficiently infrequently, at low density at least, for energy to be
approximately
proportionately to length, with a calculable entropy density
\beq
s_{st} = O(g\eta (T)) = O(m_{H}(T)).
\eeq
The effective string tension, whose vanishing signals string
proliferation at a
Shockley/Hagedorn phase transition, is \cite{CCER}
\beq
\sigma_{eff} = \sigma -Ts_{st},
\eeq
where $\sigma = O(\eta^{2} (T))$ is the string energy/unit length.
On inspection it is seen that $\sigma_{eff}$ also vanishes at the
same Ginzburg temperature $T_{G}$ of
(55) for which the
fluctuations  in $N_{S}$ become large.
Since strings produced from random field phases approximately
execute random walks, the majority of string lies in `infinite'
strings, rather than small loops of length $O(\xi)$. [The
probability of a self-avoiding walk not forming a loop is
approximately seventy percent].
Thus, at $T_{G}$ it costs nothing to produce
such  macroscopic defects from local fluctuations.

\subsection{ Global Monopoles.}

If, instead of taking a closed {\it path} through the domains after the
transition, we take a closed {\it surface} through them, the $O(N)$ field
vector on the closed surface executes a closed surface in ${\cal M}
= S^{N-1}$.
If, in turn, this closed surface is non-contractable, then point
defects (monopoles) can be trapped within it.  As with strings, the
formation of monopoles requires large field fluctuations.  The
noncontractability of surfaces in ${\cal M}$  requires the
second homotopy group $\Pi_{2}(S^{N-1})$ be non-trivial.  In $D=3$
dimensions this only happens when $N = 3$.
Thus, at temperature $T$, a global $O(3)$ theory possesses global
monopoles (balls of
false vacuum), for which
the simplest solution (winding number $n=1$) is of the form
\cite{SV}
\beq
\phi_{a}(\vecx) = \eta(T)h(g\eta(T)r)\frac{x_{a}}{r},
\eeq
where $r$ now denotes radial distance, and $h$ is to be determined
 from the Euler-Lagrange equations.  We are less interested in
monopoles than strings for cosmological reasons.  Sufficient to say
that

a) the monopole diameter $a(T) = O(\xi (T)) = O(g^{-1}\eta^{-1}
(T)) = O(m_{H}^{-1})$, the Higgs correlation length

b) the monopole energy within a sphere of radius $r$ diverges
linearly with $r$.  Nonetheless, as with strings, the presence of
nearby defects cuts off the energy divergence.  Of greater interest,

c) monopoles with winding number $n>1$ are unstable

d) the net winding number of the monopoles in a volume  $v$
is
\beq
N_{v} = -\frac{\epsilon^{abc}\epsilon^{ijk}}{8\pi}
\int d^{3}x\;
%% FOLLOWING LINE CANNOT BE BROKEN BEFORE 80 CHAR
\frac{\partial_{i}\phi_{a}\partial_{j}\phi_{b}\partial_{k}\phi_{c}}{|\phi|^{3}}.
\eeq

In calculating the variance in winding number in a volume $v$ a
 similar analysis can be performed to that for strings. Instead
of $N_{v}$, we calculate the variance of its relative
\beq
{\bar N}_{v} = -\frac{\epsilon^{abc}\epsilon^{ijk}}{8\pi}
\int d^{3}x\;
\frac{\partial_{i}\phi_{a}\partial_{j}
\phi_{b}\partial_{k}\phi_{c}}{\eta^{3}(T)}.
\eeq
Then, in an approximation that ignores the connected correlation
functions i.e. replaces the six-point correlation
function by the product of three independent two-point functions, we
find \cite{AG}
\beq
<{\bar N_{v}}{\bar N_{v}}> = O((Rm_{H}(T))^{2})(g^{2}T/m(T))^{3},
\eeq
where $v = O(R^{3})$.  The $R^{2}$ behaviour, counting the number of
correlation patches on the surface of $v$ is generic, interpreted
as reflecting the random nature of the `angular' distribution of the
Goldstone modes. The remaining factor, derived here in an
approximation in which free-field propagators have been assumed
dominant, links the correlations of the Higgs field to those of the
Goldstone modes. As with strings, in the Ginzburg regime,
fluctuations are sufficiently
strong for field phase differences between Higgs correlation volumes
to be random and of order unity, the prerequisite for a large
monopole presence.

So far, everything accords  with the picture proposed by Kibble in
that the fluctuations at the Ginzburg temperature are large enough
to contain defects with the behaviour that we would have expected.
However, the approximations that we have been obliged to make are
more severe near the transition, where we most need reliable results.
In particular, the neglect of connected correlation functions means
that the variance $<NN>$ does not know whether defects are present.  We
conclude these observations on fluctuations by examining the
strings of the local $U(1)$ gauge theory.
Here the picture is complicated by gauge properties,
which have lead some authors \cite{RS} to doubt the Kibble mechanism as
presented so far.  However, as will be seen, in other ways they
permit more justifiable calculations.

%\subsection{ Thermal Equilibrium.}
%\subsection{ Out-Of-Equilibrium Behaviour.}

\subsection{ Local Strings}

The arguments about symmetry breaking and the existence of defects
relied only on the group structure of the vacuum manifold.  Gauging
the global symmetry does not effect these arguments as
the gauge fields are zero in the
vacuum.\tnote{Is it possible that
for larger number of strings present where one is expanding about
action minima which are not the lowest, is it possible for the
vacuum manifold of the Higgs is changed in a background of classical
sources/solutions? It might be the idea
that very high magnetic fields could help restore the symmetry.
This might be the same as saying that at high string densities, the
stings overlap and the symmetry is restored.}
The precise dynamics of string formation and interactions will of
course be rather different in the local case.
%So if the existence
%of strings is not in doubt in local theories, their relevance to
%physics and in particular to cosmology does have to be examined.
For instance, it might be that in gauge theories the density of strings
formed at a transition is far too low to be of much interest.

One obvious hurdle which comes
with the introduction of gauge symmetry is the fact that parameters
which were physical in the global case become gauge dependent and
hence unphysical in the case of local symmetry.  In our example of
strings in the O(2) theory, the different vacua are labeled by different
values of the phase $\alpha$ of the field (when thought of as a single
complex $U(1)$ field).
 However, $\alpha$ is gauge dependent.  Indeed in the unitary gauge
it is removed altogether!  It is the degree
of freedom which is eaten by the gauge fields as they become
massive.  The question is, does the move from global to local
symmetry effect the basic Kibble mechanism arguments?

One picture that might be used is to fix the gauge to be
something fairly straightforward (the so called unitary gauges are well
known to be difficult and so we exclude them).
%We know that this
%may entail the introduction of ghosts but at least the gauge problem
%has been dealt with.  This does not alter the fact that many
%parameters such as the Higgs phase in O(2) are unphysical even though
%their gauge dependence is not obvious any more.
With the
gauge dependence dealt with provisionally,
we can use unphysical fields as  convenient
intermediate tools provided we end up studying a physical quantity.

So the arguments now proceed as they did in the global case.
$N_{S}$ of (4.5) is equally valid for the local $U(1)$ gauge theory.
One can then consider following the phase
as one moves along a closed loop $\partial S$ in space,
keeping track of the number of
times the vacuum manifold is traversed.  The phase must return to the same
point on the manifold so that
$\oint \alpha = 2 \pi N_{S}$ where $N_{S}$ is the
winding number, an integer.  As the winding number is a physical
quantity it does make sense to discuss the line integral of the phase
even though the phase is not itself physical.  In moving round the
loop of length $R$, one passes through $R/\xi_{H}$ correlated patches in
the Higgs field, for which $\xi_{H} = m_{H}^{-1}$.  Provided the fixed,
but gauge-dependent, correlation length for the field phase $\alpha$
is, on average, scaled by $\xi_{H}$ and the phases are randomly
distributed in the domains, then from a random walk
analysis one expects that $\langle N_{S}N_{S} \rangle = R/\xi_{H}$, as before.

One might remain uneasy about relying so heavily on an unphysical
parameter, and there have been attempts to show that this picture is
wrong.  We therefore turn to a second approach which can be used in
the O(2) or U(1) case.  For this we to look at the classical
solutions which minimise the compactified action $S_{3}$ in which
all heavy modes have been integrated over and which correspond to strings.
These are rather more complicated than their global counterparts.
%Writing the field as
%$\phi = \rho \exp \{ i \alpha\}$, as before, the Euler Lagrange
%equations become
%\bea
%\Box \rho &=& \rho [ ( e A^\mu - \partial^\mu \alpha)^2 -
%\frac{g^{2}}{2}(\phi^2-\eta^2) ]
%\label{emodphi}
%\\
%\Box A^{\mu} &=&
%J^\mu - (1-\zeta)\partial^\mu \partial. A
%\label{eA}
%\\
%\partial^\mu J_\mu &=&
%\partial^\mu \left( \rho^2 ( e A^\mu - \partial^\mu \alpha)
%\right) = 0
%\label{ephase}
%\\
%J^\mu &=& \rho^2 ( e A^\mu - \partial^\mu \alpha)
%\label{J}
%\eea
Nonetheless, there are well known vortex
solutions to these equations\tnote{The solution to these
equations requires boundary conditions, how do they change if one
uses thermal boundary conditions?  See the Brazilian papers?}
\cite{NO} in which there is an infinite straight
(necessarily) static string with winding number $n$.

The key observation is that if the scalar field is to be in the
vacuum at distances large compared to the string size (set by
correlation lengths) then
%(\ref{emodphi}) tells us that
\beq
e A^\mu = \partial^\mu \alpha  \; \; .
\label{eAphase}
\eeq
Indeed on studying the exact\tnote{zero temperature} solutions we
find that in cylindrical coordinates $(r,\theta,z)$
with the string along the $z$
axis has as asymptotic solutions at large distances
\beq
A_{\theta} = \frac{n}{er} - c_V \frac{1}{\surd(m_{V}(T) r)} e^{-m_{V}(T) r}
\; , \; \;
\alpha = n\theta
\; , \; \;
\modphi = \eta (T) - c_{H} \frac{1}{\surd(m_{H}(T) r)} e^{-m_{H}(T) r}
\eeq
with other components of $A^\mu$ being zero.  The effective Higgs
mass $m_{H}(T)$ is $g\eta(T)$ as before, and the effective gauge
field mass is $m_{V} = e\eta(T)$. This
satisfies (\ref{eAphase}) up to small
corrections.\tnote{$\int dS.B$ = 0 is
the usual relation coming from no magnetic monoples.}
The idea then is to use the gauge field to track the Higgs phase and
hence count the strings passing through some loop.\tnote{But a gauge
field is just as gauge dependent as the phase!!!  Its only good
because we can use 2 and not 4 point functions.}
There is a critical value of $e/g$ below which strings with $n>1$
are unstable and we assume this to be the case,since it is
approximately correlated to the existence of a second-order
transition.

To calculate the fluctations in the numbers
of strings $N_{S}$ which pass through a surface $S$
we can now use (\ref{eAphase}).  The variance in $N_S$ is then expressed
completely in terms
of a two-point correlation of the gauge fields,
\bea
\langle N_S N_S \rangle &=&
\langle \left( \frac{1}{2\pi} \oint_{\partial S} d\vec{l} .
\vec{\partial}\alpha \right)^2 \rangle
\nnel
%&=& \langle \left(
%\oint_L d\vec{l}_i \frac{\phi {\stackrel{\leftrightarrow}{\nabla}}_i
%\phi}{2 \pi \modphi^2}
%\right)^2 \rangle
%\nnel
&=& \frac{e^2}{4\pi^2 } \langle \left( \oint_{\partial S} d\vec{l} .
\vec{A} \right)^2 \rangle
\nnel
&=& \frac{e^2}{4\pi^2 } \oint_{\partial S} d\vec{x}_i \oint_{\partial S}
d{\vecxdash}_j .
\langle A_i(t,\vec{x}) A_j(t,\vecxdash)  \rangle ,
\label{eNNAA}
\eea
rather than the four-point scalar field correlation function (4.12)
of the global theory.
Thus we just need to know the equal-time correlation function of the
$\vec{A}$ fields.
There are only two
symmetric tensors in the spatial indices which can be made in the covariant
and $R_\zeta$ gauges,
namely $\delta^{ij}, k^{i}k^j$.
The Ward or BRS identities and the Lorentz structure at finite temperature
then tell us that the most general form for the (time-dependent)
correlation functions in covariant and $R_\zeta$ gauges is
\bea
iG^{ij}(t-\tdash,\vecx-\vecxdash) &=&
\langle T A^i(t,\vecx) A^j(\tdash,\vecxdash) \rangle
\label{eprop}
\\
&=& \frac{i}{8 \pi^3 \beta} \sum_{n=-\infty}^{+\infty} \int d^3 \veck \;
e^{-i\{k_0(t-\tdash) -\vec{k}.(\vecx-\vecxdash) \}} iG^{ij}(k) \}
\\
-G^{ij}(k) &=& \frac{1}{k^2-e^2\eta^2-\Pi_T}
(\delta^{ij} -\frac{k^{i}k^j}{k^2})
\nnel && +
\left( \frac{1}{k^2-e^2\eta^2-\Pi_L} + f(k,M,\zeta)
\right) \frac{k^{i}k^j}{k^2}
\label{eprop2}
\eea
where the energy $k_0=2\pi i n / \beta$.
The $f$ function varies with the gauge chosen while the other
$\Pi_T$ and $\Pi_L$ terms correspond to the physical modes of the
photon, two transverse (magnetic) and a single
longitudinal (electric) mode.

In the case of global strings it was sufficient to take the $n=0$
equal-time term in the sum, which could be identified with the compactified
action $S_{3}$, in terms of which fluctuation probabilities were
given.  We can do this here, even though  $S_{3}$ is nonphysical,
but will not.  The reason is that, later, we need to know some
aspects of its time behaviour.

  From (\ref{eNNAA}), (\ref{eprop}), (\ref{eprop2})
we see that we require two sorts
of integration
\beq
\oint_{\partial S} dx_i\; e^{-i \veck.\vecx} , \; \; \oint_{\partial
S} dx_i\; e^{-i \veck.\vecx} k_i .
\eeq
Since the latter is zero we only pick up the term in the
propagator with the $\delta_{ij}$.  This is not surprising as we are
really looking at the magnetic field correlations. The transverse
correlation function with
$\delta_{ij}$ is associated with magnetic fields, its self-energy
$\Pi_T$ giving information about magnetic screening.
Thus we have that
\bea
\langle N_{S} N_{S} \rangle &=&
\frac{e^2}{4\pi^2 }
\frac{i}{8 \pi^3 \beta} \sum_{n=-\infty}^{+\infty} \int d^3 \veck \;
\frac{-i}{k^2-e^2\eta^2-\Pi_T}
\left| \oint_{\partial S} d\vec{x}_i\; e^{-i \veck.\vecx} \right|^2 .
\label{eNNAA2}
\eea
Note that this form is independent of the gauge chosen.

In order to restore a temperature dependence to $\eta$
we calculate $\Pi_{T}(k)$ by resumming an
infinite set of diagrams in the consistent manner of Braaten and
Pisarski.  The major effect will be that $\Pi_T$ provides the required
shift in the expectation value $\eta$.  However there are
further corrections due to the presence of strings in the background.
This means that $\Pi_T$ is a non-trivial function of
momentum leading to both cuts and poles in the propagator
%(\ref{emagprop})
in the complex energy plane.  However,
for a low density of strings, we know that the solution is almost
the same as being in the vacuum state so as a first guess we need
not consider corrections to $\Pi_T$ due to
the strings' presence.
%for now we
%shall ignore this effect and assume that the effective mass is a
%constant.
%We will provide further justification for this further on.
Thus we work with a photon mass of the form
\beq
e^2\eta^2 + \Pi_T(k) = e^2 \eta(T)^2 + m_{\rm mag}^2 = M^2 ,
\eeq
where we have explicitly included the self-energy effects which give
a temperature dependent $\eta$ and left other effects in the $m_{\rm
mag}^2$ term.  The mass $m_{mag}$ is the magnetic screening mass
that, in general, would be expected to be present. For the special
case of a $U(1)$ theory $m_{mag} = 0$ if the string
background is ignored.  We can now do the energy sum to give
\bea
\langle N_{S} N_{S} \rangle &=&
\frac{e^2}{32\pi^5 }
\int d^3 \veck \;
\frac{1}{\omega}\frac{1}{\exp \{ \beta \omega \} -1 }
\left| \oint_{\partial S} d\vec{x}_i\;
e^{-i \veck.\vecx} \right|^2 + (\mbox{T=0 terms})
\label{eNNAA3}
\eea
where $\omega^2 =\veck^2 + M^2$.

If we now limit ourselves to counting $N$ within a circular loop
$\partial S$ of radius $R$, we find that that in terms of the Bessel
function\tnote{GR 8.411.1 for $J_n$.}
\bea
\oint_{\partial S} d\vec{x}_i \oint_{\partial S} d\vecxdash_i\;
e^{-i \veck.(\vecx-\vecxdash)}
&=& R^2 \int_{-\pi}^{\pi} d\theta d \thetadash
\cos (\theta-\thetadash)
\exp \{-i R\kbar.(\cos(\theta) - \cos(\thetadash) ) \}
\nnel
&=& 4 \pi^2 R^2 | J_1(\kbar R) |^2
\label{eNNAA4}
\\
J_n(z) &=& \frac{e^{-i n \pi /2}}{2 \pi} \int_{-\pi}^{\pi} d\theta
\exp \{-i z.\cos(\theta) + i n \theta \}
\eea
and so we have that
\bea
\langle N_{S} N_{S} \rangle &=&
\frac{e^2 R^2 }{8 \pi^3 }
\int_{-\infty}^{\infty} dk_z
\int_0^\infty d\kbar \; 2 \pi \kbar\;
\frac{1}{\omega}\frac{1}{\exp \{\beta \omega \}- 1 }
|J_1(\kbar R) |^2
\label{eNNAA5}
\eea
where $k_z$ is the component of $\veck$ perpendicular to the loop
and $\kbar$  the size of the $\veck$ in the plane of the loop, so
that $\veck^2 = k_z^2 + \kbar^2$.

We now look at very large loops, and high temperatures  $T \gg m \gg
R^{-1}$.  We can then use the asymptotic expansion for
$J_1$\tnote{Abrom. and Stegun 9.2.1 pp.364}
\beq
(J_{1}(z))^2 \sim \frac{1}{\pi z} [1 + \cos (2z - 3 \pi/2).]
\eeq
Thus we find that the largest contribution comes from a region
where $m<\kbar<T$ so that\tnote{I need to multiply this by $4/\pi$
to get the answer in my notes!}
\bea
\langle N_{S} N_{S} \rangle &\simeq &
\frac{e^2 R }{4 \pi }
\int_{M}^{\infty} d\omega \;
\frac{1}{\exp \{\beta \omega \} - 1 }
\label{eNNAA6}
\eea
which using the high temperature limit gives\tnote{The factor of
${\gamma+\psi(1)}=1$ from the expansions of
the zeta and Gamma function respectively
in the high temperature
expansion.  Also the hope was that this all comes
from the infra-red behaviour small
$\kbar$ (so that we can use $m_{mag}$ instead of $\Pi_T(k)$).  This
contradicts the assumption used on the Bessel
function.  If one looks at that region of $\kbar$ though one ends up with
smaller contributions - oh no you don't!!!! Studying the plots of
the integrand on maple suggest that it is the small, well of order $T/10$.}
\bea
\langle N_{S} N_{S} \rangle \simeq
\frac{e^2 R T}{4 \pi } \ln \left( \frac{T }{M} \right)
\label{eNNres}
\eea
As we observed,
this result could have been obtained directly from the $n = 0$ term
alone in the mode sum, which corresponds to replacing
$\exp \{\beta \omega \} -
1$ by $\beta\omega$ in the denominators, and remembering that
compactification provides a cutoff at the thermal wavelength
$|\veck| = O(T)$.

%\bea
%\frac{1}{k^2-e^2 |\phi(k)|^2-\Pi_T(k)}&=&
%\frac{1}{k^2-e^2 \eta^2-\Pi_T(k)}
%- \frac{e^2 |\phi(2k)|^2-e^2\eta^2}{(k^2-e^2 \eta^2-\Pi_T(k))^2}
%+\ldots
%\label{emagprop}
%\eea

Superficially $<N_{S}N_{S}>$ of (4.47) looks very different from
its counterpart (4.23), apart form its linear dependence on the loop
perimeter $R$.  However, if we choose $\xi_{H}(T) = m_{H}^{-1}(T)$
as the natural length scale, then (4.47) can be written
\beq
<N_{S}N_{S}> = \left(
\frac{e^2}{g^2}\right)\left(\frac{g^{2}T}{m_{H}(T)}\right)
\left(\frac{Rm_{H}(T)}{4\pi}\right) ln\left(\frac{1}{eg} .
\frac{g^{2}(T)}{m_{H}(T)}\right) ,
\eeq
where we have set $m_{mag}$ to zero.  Assuming that $e/g\approx 1$,
compatible with a second-order transition, then (up to logarithms),
the scale of $<N_{S}N_{S}>$ is set by $g_{3} = g^{2}T/m_{H}$.  In
the Ginzburg regime, where $g_{3}\approx 1$, the variance in $N_{S}$
is $O(R/\xi_{H})$, with coefficient $O(ln(1/eg))$.  Qualitatively,
this is as before, provided $eg$ is not extremely small.
That is, once in the Ginzburg regime,
fluctuations are large enough to contain strings with high density.

However one question that remains unanswered is exactly how
does  $\langle N_{S} N_{S} \rangle$ know if there are
strings around or not?  The result is only sensitive to the presence
of strings through the $\modphi$ dependent mass term and we know
that, for dilute strings at least, the correction to due to strings being
present is small.
This can be made more rigorous by carefully calculating the photon
propagator in a nontrivial background which is a minimum of the
classical action.  If one leaves the terms cubic or higher in the
action as perturbations to be treated in the usual way,  it is easy
to see, in the first instance,
that the photon propagator has the same form as before but
where the $\modphi$ in the mass term is replaced by the classical
scalar field solution.  In the case of dilute strings this whole
procedure is justified as the deviations this produces from
calculations in a the usual flat background of the true vacuum  are
exponentially suppressed.  The end result is that, although the
fluctuations in the field are reduced, it seems that the result
(104) remains qualitativly correct.  The concerns that a local theory
would possess strings with a lower density than a global theory is
unfounded in this particular instance.

We should stress that the $O(N)$ theory, with its $Z$-strings, may be
special in some regards.  We have little idea as to the nature of
the symmetry-breaking at GUT scales, but it is equally likely that
the relevant homotopy group is $Z_{2}$, rather than $Z$. Since the
net winding number of $Z_{2}$ strings is $0,1$ the previous analysis
is inappropriate and we don't know where to begin.

Finally, the {\it local} $O(3)$ theory permits t'Hooft-Polyakov
\cite{tHP} monopoles.
 Unfortunately, their winding number variance cannot be expressed
simply in terms of gauge fields and we shall not consider them.
Since, in the early universe, monopoles are an embarrassment, this is
no great loss.

\section{ TIME SCALES}

In most of the calculations discussed hitherto, it is assumed that the
system is in thermal equilibrium, enabling us to use the
imaginary-time formalism. At best, in early universe calculations,
equilibrium can only be achieved
in some intermediate period.  Prior to this, as we noted in the
introduction, the universe is expanding too rapidly for the
particles to interact sufficiently to equilibriate.  Subsequent to
this particle species, or defects, will freeze out\tnote{See Kolb
and Turner \cite{KT} sec. 5.2. } for chemical or
thermal reasons.
%This is of course not going to be
%realistic for some of the most crucial eras of the early universe.
%For example, a necessary though not sufficient condition for the
%natural balance of the universe to be upset during cooling, be it a
%zero average baryon number or a zero average winding number density,
%is that the universe must go out of equilibrium while such
%quantities shift.  However, given that the universe is in some kind
%of thermal equilibrium in intermediate times, it is important to see
%if equilibrium fluctuations can wipe out such non-equilibrium
%effects.
It is worthwhile spending a few moments to see whether this
temporary lull can occur for a temperature range encompassing
the critical temperature $T_{c}$ appropriate for an $O(N)$ GUT
transition (or that of any other symmetry group).
If this is not possible our work of the previous sections
is irrelevant for the very early universe.

Let us assume equilibrium, for which it is relatively easy to
estimate the time scales.
We can then check for consistency.
The time scale associated with the rate at which the universe is
expanding $\tau_{\rm exp}$ is simple to
quantify.  It is the inverse Hubble constant\tnote{$m_{pl}=10^{19}$
GeV, $(8 \pi / 90)= (2.76)= (1.66)^2$, $g_{\ast} \simeq 100$
for standard model,}
\beq
\tau_{\rm exp} = \frac{1}{H} = \frac{a(t)}{ \partial_t a(t)} \simeq
\left(\frac{3}{8\pi G\rho}\right)^{\half}
= \left(\frac{90}{8\pi^3 g_{\ast}}\right)^{\half}
\frac{M_{\rm pl}}{{T^2}},
\eeq
where we have used the Friedmann equation to link the Hubble
constant to Newton's constant.  Here $g_{\ast}$ is the effective
number of degrees of freedom\tnote{Kolb and Turner sec. 3.5}
\cite{KT} (total number of relativistic degrees of freedom with a few
modifying factors).  It is around 100 for $T>1GeV$  in the
standard model.
The time scales for particle interactions are also easy to calculate.
Looking at the imaginary part of the self-energy tells us the
time taken for a particle of a given frequency to interact with any
possible other real-physical particles.  For particles stationary in
the heat-bath we find at high temperatures ($T$ larger than all
other particle physics scales)
\beq
\tau \simeq \frac{10}{N}\frac{1}{e^2 T }\; , \; \; \frac{1}{g^3 T}.
\label{edecaytime}
\eeq
The first expression is for a gauge boson in a pure SU(N) theory.  To
obtain the precise numerical factor requires
considerable effort but this is essential if one wants to have a
gauge invariant and therefore believable answer \cite{BP}.
In the early universe contributions from fermions and other matter will
decrease this expression further.  The decay rate for a fermion
in a gauge theory takes a similar form except that the $N$ is replaced by
group theory factors depending on the representation of the fermions
involved \cite{fdamp}.\tnote{See Baier, Kunstatter, Schiff Can. J.
Phys. 71 (1993) 208 for overview.}
One should note that there are unresolved problems
with the thermalisation rates in gauge theories of particles
moving with respect to the heat bath because of infrared
problems associated with the behaviour of magnetic fields in a
plasma \cite{Pi}.

The latter time scale\tnote{Elmfors
(\ref{El}) Banff, eqn 21, that its
$g^3$ as mass is $gT$.} in (\ref{edecaytime}) is for
self-interacting scalars \cite{El} with interaction
$g^2\phi^4$.  The reason for using this notation for the scalar
coupling is that for most quantities, it is when $e \simeq g$ that
the two types of interactions give similar-sized effects.  For
instance the thermal mass corrections are $\delta m \sim eT,gT$.
One of the exceptions to this simple rule are the thermalisation
rates quoted here.  This is also marks the boundary between
first and second order transitions as mentioned at the end of
section 2.

These time scales are an estimate of the time in which a particle
traveling through a plasma {\em starts} to lose its energy to the heat
bath and so becomes absorbed as just another part of the thermal
background \cite{We}.  They only tell us about the start of the energy loss
process as the result is to be interpreted through linear response
theory in which one considers only small or short time deviations from
equilibrium.  Essentially these results are telling us that the
calculations which produced them become invalid and inconsistent at
this time scale.  To follow the process of coming into equilibrium
properly requires that one goes beyond the usual linear response
analysis, which is very difficult to do.

A simpler approach is to consider a single interaction.  The
crossection goes as $\sigma \sim e^2/s, g^4/s$ where $s$ is the
characteristic centre of mass energy squared scale which here will be
$T^2$.  The many-body nature of the problem can then be taken
account.  First the density of particles per unit volume
in equilibrium is $n \sim T^3$ and second their
speed, $v$, is $O(1)$.  An estimate of interaction time is then
\beq
\tau = \frac{1}{n \sigma v} \simeq\frac{1}{e^2 T} \; ,\; \;
\frac{1}{g^4 T}.
\label{edecaytime2}
\eeq
Note that this simple argument does not give
the correct behaviour for the scalar case as given in
(\ref{edecaytime}).

%From this viewpoint the Ginzburg temperature reappears as that
%temperature for which the distance $r = \tau$ traversed by a
%relativistic particle in this time (essentially its mean free path)
%equals the correlation length $\xi = m^{-1}(T)$ of the field.

For equilibrium we require that
\beq
\tau_{\rm exp} \gg \tau \; \; \Rightarrow T \ll  g^3 M_{\rm pl} , \;
\; e^2 M_{\rm pl},
\eeq
where $M_{\rm pl}$ is the Plank mass.
Although qualitatively correct, the prefactors are important.
Assuming the Standard Model degrees of freedom,
the inequalities become
\bea
T & \ll & \frac{e^2 M_{\rm pl}}{1.6 g_{\ast}^{1/2}} \frac{N}{11.1}
= 5 N e^2 \times 10^{16} {\rm GeV},
\\
T &\ll &  2.5 g^3\times 10^{15}
{\rm GeV}
\label{eineq2}
\eea
Thus gauge particles are in equilibrium at GUT scales and below, at
least until the high temperature forms for the particle interactions
(\ref{edecaytime}) and (\ref{edecaytime2}) cease to be valid.
The second inequality (\ref{eineq2}) is the
result for a pure O(1) self-interacting scalar, suggesting that they are not
usually in equilibrium at GUT scales.
%In reality, $T$ as given here is better understood as the momentum
%scale for fluctuations that will decay in less than a Hubble
%time.\tnote{This I don't understand!  It just means that gauge
%singlets are indeed unlikely to be in equilibrium at the GUT scale.}
%Although an $O(1)$ theory is unrealistic (and we don't know the
%results for an $O(N)$ theory, equilibrium at the GUT scale can only
%be true for fluctuations significantly larger than a thermal
%wavelength.

In fact, these temperature upper bounds may be an underestimates.
There has been a suggestion that the
characteristic relaxation time of the plasma is actually \cite{Sm2}
$1/(e^4 T)$ rather than (\ref{edecaytime}).  We are unsure about this.
Apart from using more sophisticted formulae for particle interaction
times, valid over a larger range of temperatures,
the simple analysis given above does not take phase
transitions into account. Second order processes show critical slowing down.
Rather, the purpose of the calculation is to see whether thermal
boundary conditions are appropriate.

The above calculation shows that most species are in equilibrium
below GUT temperatures.  It is however possible for a species to fall
out of equilibrium, to `freeze' out at a later stage.  This process
is extremely relevant to defects.  However the principle is most
easily illustrated with the photons of the microwave background.

At recombination, the total crossection for all photon interactions falls
dramatically over a short time period.  This essentially leaves a background of
thermal non-interacting photons of temperature $T \sim O({\rm eV})$.
The number and energy distributions of the photons are not changed by
particle interactions to a very good approximation.  The photons
still interact gravitationally but this simply involves a dilution
of the photon density due to the usual stretching of the universe.
It is easy to check that the photons keep a black-body spectrum and
that the expansion of the universe causes their effective
temperature to fall from around $10^3$K to the present $2.7$K.

More generally, the freeze-out can be seen by using formulae for
particle interaction time scales appropriate at lower temperatures.
Essentially, the time scale goes from $e^{-2} T^{-1}$ to $ e^{-2} \Lambda^n
T^{-n-1}$ where $\Lambda$ is some other particle physics scale
relevant to the particular species.  If $n>1$ then we get a lower
bound on the temperatures until which a given species remains in
equilibrium
\bea
\frac{T}{\Lambda} & \gg &
\left(\frac{17 g_{\ast}^{1/2}\Lambda }{e^2 M_{\rm pl} N}
\right)^{1/(n-1)}
=
\left(\frac{\Lambda}{5 N e^2 \times 10^{16} {\rm GeV}}\right)^{1/(n-1)},
\label{eineq3}
\eea
This is the freeze-out temperature of the particle species.

This is a particularly simple example because the photons
become essentially non-interacting (except for gravity).  There are
however different shades of equilibrium and corresponding freeze out
as species fall out of `equilibrium'.  What we have in mind here is
the difference between thermal and chemical equilibrium.  So far we
have concentrated on thermal equilibrium.  By this we mean the
situation where the energy distribution of a species is being
maintained in a Bose-Einstein or Fermi-Dirac distribution.  This
sort of equilibrium will be maintained by interactions with {\em
any} other species already in equilibrium and this includes
self-interactions.  So to test for thermal equilibrium one need only
check that the time scale associated with the fastest interaction
process is shorter than the expansion rate of the universe.

Chemical equilibrium is concerned with the maintenance of
equilibrium particle numbers.  Each individual interactions mixes
only certain types of particle species together.  The quark-gluon
plasma provides an excellent example which has been studied
extensively becuase of the application to terrestial relativistic
heavy-ion collision experiments \cite{Mu}.  If one heats up
nucleons to obtain a quark-gluon plasma at a few hundred MeV, as is
done in relativistic heavy ion collisions, one starts with an
initial domination of up and down quarks plus gluons.
 The gluons thermalise about three times quicker than the quarks so
initially the plasma is pure glue.  While not strictly in
equilibrium, the gluons maintain a Bose-Einstein distribution.  The
up and down quarks then come into equilibrium in which there are
significant numbers of both up and down quarks and anti-quarks.
There is an imbalance between the quark and anti-quark numbers
reflecting the initial baryon number.  Strange quark-anti-quark
pairs can be created as easily as up and down quark pairs as at
these temperatures the mass differences are being negligible.  It
will, however,  take a considerable length of time for the imbalance
between quarks and anti-quarks to to be felt in the strange quark
densities as the stangeness changing processes are very slow.  The
time scale of such strangeness changing interactions, $\tau_{s}$ is
much longer than other quark interactions $\tau_{q}$.  In the
relativistic heavy-ion collision experiments this leads to a
situation where on time scales $t$ where $\tau_q < t < \tau_s$ the
plasma can have large patches where the energy density of the total
number of quarks follows a Fermi-Dirac distribution.  The net
baryon number is carried almost entirely by the up and down quarks
whereas in the long time limit it should be spread evenly between
all the quark flavours which are essentially massless at these
temperatures.   This can be modeled by having the number of up and
down quarks approximately conserved and the number of strange quarks
also conserved.  This means that one would have different chemical
potentials associated with the total number of up plus down quarks
and for strange quarks instead of one associated with the
total number of all types of quark.  These chemical potentials would
slowly change on time scales of order $\tau_s$.

In the present context of defects, the interesting question is not
if the kinetic energy in the defects follows a thermal distribution
but what is the density of defects plus anti-defects.  This is what
is important for the gravitational effects, for instance.   Thus we
are asking about the chemical equilibrium of defects, not the thermal
equilibrium.  The question then is what is the time scale for
processes which can change the net amount of string passing through
a loop or the net number of monopoles in a given volume.

In the usual picture the universe settles  down into equilibrium
after a phase transition and one then finds that the defects are
frozen in.  That is to say, it is assumed that fluctuations  are no
longer strong enough to change the numbers of zeros in
the field which represent positions of monopoles or strings in the
O(3) and O(2) cases considered here. To verify this we are therefore
interested in trying to calculate the fluctuations in the numbers of
defects, and more especially the time-scale over which such
fluctuations occur.  This is what we have to compare to the expansion
rate of the universe.

At first sight it might seem as if it is very unlikely that a
fluctuation could change the total string or monopole number.  They
are, after all, topological charges and to change the total winding
number of the phase at infinity for instance would require large
changes in the field at all points in space.  However, what is
relevant is the total defect density and not the defect minus anti-defect
density, which is the conserved number.  Thus we would like to try
and count how often fluctuations create loops of string or monopole
anti-monopole pairs for instance.

This is not an easy task.  One simple approach would be to take the
estimate of the fluctuations in the numbers of local strings
$\langle N N \rangle $ and now ask how does this change in time
i.e.\ to calculate
\beq
\langle [N(t)- N(0)]^2 \rangle  =
2 \langle N(0)N(0) \rangle  - 2 \langle N(t)N(0) \rangle  ,
\eeq
where $N(t)$ is the number of strings minus anti-strings passing
through the same loop in space but at different times $t$.  We have
assumed thermal equilibrium so that
$\langle N(t)N(t) \rangle = \langle N(0)N(0) \rangle$.   It is
easy to upgrade the calculation of $\langle N(0)N(0) \rangle$ made
above.  From (\ref{eNNAA}) we have
\bea
\langle N(t) N(0) \rangle &=&
\langle \frac{1}{2\pi} \oint_L dl_i . \partial_i\alpha(t,\vecx)
\frac{1}{2\pi} \oint_L d\dash{l}_i . \partial_i\alpha(0,\vecxdash)  \rangle
\nnel
&=& \frac{e^2}{4\pi^2 } \oint_L d\vec{x}_i \oint_L d{\vecxdash}_j .
\langle A_i(t,\vec{x}) A_j(0,\vecxdash)  \rangle
\label{eNNAAt}
\eea
This can be expressed in terms of the propagator as before at the
expense of one extra exponential in time
\bea
\langle N(t) N(0) \rangle &=&
\frac{e^2}{4\pi^2 }
\frac{i}{8 \pi^3 \beta} \sum_{n=-\infty}^{+\infty} \int d^3 \veck \;
\frac{-i}{k^2-e^2\eta^2-m^2_{\rm mag}}
\oint_L d\vec{x}_i d \vecxdash_i e^{i k_0 t-i \veck.(\vecx-\vecxdash)}
\nonumber\\ &&\mbox{ }
\label{eNNAA2t}
\eea
In doing the energy sum we can convert into a contour integral in
the usual way.  With the same approximations made on the propagator,
the energy sum leaves contributions coming from the residues of the
poles of the integrand which are at $k_0 = \pm \omega$ where
$\omega = (\veck + M^2)^{1/2}$ as before.  We merely pick up an
extra factor of $e^{it\omega}$ in the remaining integrals.

Putting all this together the generalisation of (\ref{eNNAA3}) is
\bea
\langle [N(t)- N(0)]^2 \rangle &=&
2\frac{e^2}{32\pi^5 }
\int d^3 \veck \;
\frac{1}{\omega}\frac{2 \sin^2(\omega t/2) }{\exp \{ \beta \omega \} -1 }
\left| \oint_L d\vec{x}_i\; e^{-i \veck.\vecx} \right|^2
\nnel
&& + (\mbox{T=0 terms})
\label{eNNAA3t}
\eea
Without going into details it is clear that this becomes large when
$t^{-1} \sim \omega$, where $\omega$ is effectively limited to values
between $M\sim eT$ and $T$.  Unfortunately, it is not clear in such a
simple calculation if this if the time scale for fluctuations due to
loops of string being created around the boundary of the loop (any
inside the loop will contribute nothing as the loop measures
string-anti-string numbers only).  It could just be picking up the
inevitable motion of the  strings in and out of the loop. In
short, it will take a more sophisticated approach that accomodates a
string background, to try and get a
measure of the time scale of defect number fluctuations.   Until
that time, the picture of defects frozen in shortly after the phase
transition can not be confirmed by analytic calculations.

\section{NONEQUILIBRIUM BEHAVIOUR}

We have seen that equilibrium methods are not powerful enough to
demonstrate how defects formed at the Ginzburg temperature can
freeze in after the transition to survive as topologically stable
(macroscopic) entities.  For this we need nonequilibrium methods.
As yet they too are poorly developed to deal with defect formation
at the Ginzburg temperature. However, in one sense this may not be
the issue.  One reason why so much thought has gone into fluctuations
at the Ginzburg temperature is that it was necessary to demonstrate
at least one circumstance in which
fluctuations could produce strings and monopoles in large numbers
for defects to be viable, in principle.
Now that we have done this we can move to
more compelling pictures of the way in which the initial
density of topological defects is fixed.
In practice, while the domain
mechanism outlined initially is almost certainly correct,
it ccould well be  that  the Ginzberg temperature is relevant
to nothing other than a thermally produced population of defects.

We saw earler that an expanding universe is driving the system very
fast.  As the system is
driven from some initial state
there comes a
point when the rate of change is too fast for
the evolution of the field to keep up \cite{Z}
[This is true even for
very slowly driven
transitions, but for appropriate couplings the out-of-equilibrium
behaviour occurs more closely to the transition, in time and
temperature, than the Ginzburg regime.  See Kibble, these proceedings].
The
transition may now be viewed as a quench and it is not clear that
either temperature or free energy mean anything at all. At this
point any
defects within the field are assumed to be frozen in until the transition is
complete. Upon completion, the field will try to establish thermal
equilibrium. At sufficiently low temperatures, however, the return to
equilibrium by thermal processes will be so slow that the evolution of
the initial defect density thus produced will be almost entirely by
interactions between the defects themselves.

Thus, in this scenario, the vortices cease
to be produced,
not at the Ginzberg temperature, but when the
scalar fields go out of equilibrium
%\footnote{Although thermal
%equilibrium is mode dependent, this does not matter for the crude argument
%repeated here.}.
The relevant scale which determines the
defect density is the coherence length, $\xi(t)$ at this time,
and for some time onwards, rather than the coherence length at the
Ginzberg temperature.
The initial number of vortices produced during a phase
transition would then be expected to be roughly the number when the
scalar field first goes out of equilibrium.
The next part of these notes will go some way in justifying this
expectation.

Gauge fields are too complicated to handle out of equilibrium.
We shall return to the
global $O(N)$ theory  in $D=3$ dimensions, with its
monopoles for $N=3$ and its strings for $N=2$.
 From the viewpoint above the
transition
is realised by the changing
environment inducing an explicit time-dependence in the field
parameters. Although we have the early universe in mind,
we remain as simple as possible, in flat space-time
with the $\phi$-field action of before:-
\beq
S[\phi] = \int  d^{D+1}x \biggl (
\frac{1}{2} \partial_{\mu} \phi_a \partial^{\mu} \phi_a - \frac{1}{2}
m^2(t) \phi_a^2 - \frac{1}{8} g^{2}(t) (\phi_a^2)^2
\biggr ).
\eeq
but for the t-dependence of $m^2(t)$ and $g^{2}(t)$ , which is assumed given.

We wish to calculate
the evolution of the defect density during the fall from
the false vacuum to the true vacuum after a rapid quench from an
initial state. The simplest
assumption, made here, is that the symmetry
breaking occurs at time $t=t_0$, with the sign of $m^2(t)$ changing
 from positive to negative at $t_0$. Further, after some short
period $\Delta t = t- t_0 >0$, $m^2 (t)$ and $\lambda (t)$ have relaxed to
their final values, denoted by $m^2$ and $g^{2}$ respectively.
The field begins
to respond to the symmetry-breaking at $t=t_0$ but we assume that
its response time is greater than $\Delta t$, again ignoring any mode
dependence.

To follow the evolution of the defect density during the fall off
the hill towards the true vvacuum involves two problems.
The first is how to follow the
evolution of the quantum field, the second is how to count the defects.  We
take these in turn.

\subsection{The Closed Timepath Approach to Non-Equilibrium}

During a rapid transition the dynamics of the
quantum field is intrinsically non-equilibrium. The normal techniques
of equilibrium thermal field theory are therefore inapplicable. Out of
equilibrium, one typically proceeds using a functional Schr\"{o}dinger
equation or using the closed time path formalism \cite{CTP}.
Here, we employ the latter,
following closely the work of Boyanovsky, de Vega and coauthors
\cite{B,boyanovsky}.

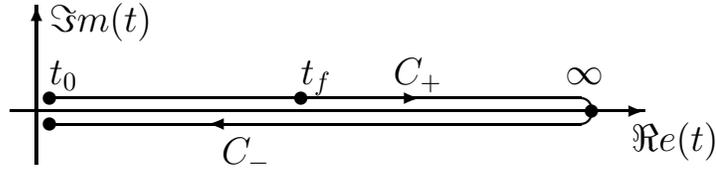
\begin{figure}[htb]
\begin{center}
\setlength{\unitlength}{5pt}
\begin{picture}(50,12)(1,64)
%\put( 7,56){\makebox(0,0)[lb]{\large $C_3$}}
\put(19,64){\makebox(0,0)[lb]{\large $C_{-}$}}
\put(6,74){\makebox(0,0)[lb]{\large $\Im m(t)$}}
\put(32,70){\makebox(0,0)[lb]{\large $C_{+}$}}
%\put(28,68){\makebox(0,0)[lb]{\large $0$}}
\put(50,65){\makebox(0,0)[lb]{\large $\Re e (t)$}}
\put(45,70){\makebox(0,0)[lb]{\large $\infty$}}
\put(25,70){\makebox(0,0)[lb]{\large $t_f$}}
\put( 6,70){\makebox(0,0)[lb]{\large $t_0$}}
%\put(7,49){\makebox(0,0)[lb]{\large $-i\beta$}}
\thicklines
\put(25,69){\circle*{1}}
\put( 6,69){\circle*{1}}
\put( 6,67){\circle*{1}}
%\put( 6,50){\circle*{1}}
\put(47,68){\circle*{1}}
\put(46,68){\oval(2,2)[r]}
\put( 3,68){\vector( 1, 0){48}}
\put( 6,69){\vector( 1, 0){28}}
\put(34,69){\line( 1, 0){12}}
\put(46,67){\vector(-1, 0){28}}
\put(18,67){\line(-1, 0){12}}
%\put( 6,67){\vector( 0,-1){11}}
%\put( 6,56){\line( 0,-1){ 6}}
%\put(4,50){\line( 1, 0){ 2}}
\put(5,64){\vector( 0, 1){12}}
\end{picture}
\end{center}
\caption{Closed Time Path}
\label{fctp}
\end{figure}

Take $t=t_0$ as our starting time. Suppose that, at this time, the
system is in a pure state, in which the measurement of $\phi$ would
give $\Phi_0(\vecx)$. That is:-
\beq
\hat{\phi}(t_0,\vecx) | \Phi_0,t_0 \rangle = \Phi_0 | \Phi_0,t_0 \rangle.
\eeq
The probability $p_{t_f}[\Phi_f]$ that, at time $t_f>t_0$, the
measurement of $\phi$ will give the value $\Phi_f$ is $p_{t_f}[\Phi_f]
= |c_{f0}|^2$, where:-
\beq
c_{f0} = \int_{\phi(t_0) = \Phi_0}^{\phi(t_f) = \Phi_f} {\cal D} \phi
\, \exp \biggl \{ i S[\phi] \biggr \},
\eeq
in which ${\cal D} \phi = \prod_{a=1}^N {\cal D} \phi_a$ and spatial
labels have been suppressed. It follows that $p_{t_f}[\Phi_f]$ can be
written in the closed time-path form:-
\beq
p_{t_f}[\Phi_f] = \int_{\phi_{\pm}(t_0) = \Phi_0}^{\phi_{\pm}(t_f) =
\Phi_f} {\cal D} \phi_+  {\cal D} \phi_-
\, \exp \biggl \{ i \biggl ( S[\phi_+]-S[\phi_-] \biggr ) \biggr \}.
\eeq
Instead of separately integrating $\phi_{\pm}$ along the time paths
$t_0 \leq t \leq t_f$, the integral can be interpreted as
time-ordering of a field $\phi$ along the closed path $C_+ \oplus
C_-$ as shown in figure \ref{fctp}
where $\phi =\phi_+$ on $C_+$ and $\phi= \phi_-$ on $C_-$.
It is convenient to extend the contour from $t_f$ to $t= \infty$.
% ********************************************
\typeout{figure: CTP2 }
Either $\phi_+$ or $\phi_-$ is an equally good candidate for the
physical field, but we choose $\phi_+$:-
With this choice and suitable normalisation, $p_{t_f}$ becomes:-
\beq
p_{t_f}[\Phi_f] = \int_{\phi_{\pm}(t_0) = \Phi_0}
{\cal D} \phi_+  {\cal D} \phi_- \, \delta [ \phi_+(t) - \Phi_f ]
\, \exp \biggl \{ i \biggl ( S[\phi_+]-S[\phi_-] \biggr ) \biggr \},
\eeq
where $\delta [ \phi_+(t) - \Phi_f ]$ is a delta functional, imposing
the constraint $\phi_+(t,\vecx) = \Phi_f (\vecx)$ for each $\vecx$.

The choice of a pure state at time $t_0$ is too simple to be of any
use. The one fixed condition is that we
begin in a symmetric state with $\langle \phi \rangle = 0$ at time
$t=t_0$. Otherwise, our ignorance is parametrised in the probability
distribution that at time $t_0$,
$\phi(t_0,\vecx) = \Phi(\vecx)$.
%Above, we took $P_{t_0}[\Phi]=
%\delta [ \Phi - \Phi_0 ]$.
If we allow for an initial probability
distribution $P_{t_0}[\Phi]$ \cite{SW}
then $p_{t_f}[\Phi_f]$ is generalised to:-
\beq
p_{t_f}[\Phi_f] = \int {\cal D} \Phi P_{t_0}[\Phi]
\int_{\phi_{\pm} (t_0) = \Phi} {\cal D} \phi_+  {\cal D}
\phi_- \, \delta [ \phi_+(t) - \Phi_f ] \, \exp \biggl \{ i \biggl (
S[\phi_+] - S[\phi_-] \biggr ) \biggr \}.
\eeq
At this stage, we have to begin to make approximations.
To make any progress, we shall find that it is necessary that
$p_{t_f}[\Phi_f]$ be Gaussian.  For this to be so
$P_{t_0}[\Phi]$ must be Gaussian also, with zero mean.  For all our
caveats about the difficulty of defining temperature we don't know
where to begin unless we assume that
$\Phi$ is Boltzmann distributed at time $t_0$ at an effective
temperature of $T_0 = \beta_0^{-1}$ according to a quadratic Hamiltonian
$H_0[\Phi]$. That is:-
\beq
P_{t_0}[\Phi] = \langle \Phi,t_0 | e^{- \beta H_0} | \Phi,t_0 \rangle
= \int_{\phi_3(t_0) = \Phi = \phi_3(t_0-i \beta_0)} {\cal D} \phi_3
\exp \biggl \{ i S_0 [\phi_3] \biggr \},
\eeq
for a corresponding action $S_0[\phi_3]$, in which $\phi_3$ is taken
to be periodic in imaginary time with period $\beta_0$. We take
$S_0[\phi_3]$ to be quadratic in the $O(N)$ vector $\phi_3$ as:-
\beq
S_0[\phi_3] = \int d^{D+1}x \biggl [
\frac{1}{2} (\partial_{\mu} \phi_{3 \, a} )(\partial^{\mu} \phi_{3 \,
a} )
- \frac{1}{2} m_0^2 \phi_{3 \, a}^2
\biggr ].
\eeq
We stress that $m_0$ and $\beta_0$ parametrise our uncertainty in the
initial conditions. The choice $\beta_0 \rightarrow \infty$ corresponds
to choosing the $p_t[\Phi]$ to be determined by the ground state
functional of $H_0$, for example. Whatever, the effect is to give an
action $S_{3}[\phi]$ in which we are in thermal equilibrium for $t<t_0$ during
which period the mass $m(t)$ takes the constant value $m_0$ and, by
virtue of choosing a Gaussian initial distribution, $g^{2}(t) = 0$
for $t<t_0$.
This may be less restrictive than it seems since the results are
essentially $\beta$-independent.

\begin{figure}[htb]
\begin{center}
\setlength{\unitlength}{5pt}
\begin{picture}(50,28)(1,48)
\put( 7,56){\makebox(0,0)[lb]{\large $C_3$}}
\put(19,64){\makebox(0,0)[lb]{\large $C_{-}$}}
\put(6,74){\makebox(0,0)[lb]{\large $\Im m(t)$}}
\put(32,70){\makebox(0,0)[lb]{\large $C_{+}$}}
%\put(28,68){\makebox(0,0)[lb]{\large $0$}}
\put(50,65){\makebox(0,0)[lb]{\large $\Re e (t)$}}
\put(45,70){\makebox(0,0)[lb]{\large $\infty$}}
\put(25,70){\makebox(0,0)[lb]{\large $t_f$}}
\put( 6,70){\makebox(0,0)[lb]{\large $t_0$}}
\put(7,49){\makebox(0,0)[lb]{\large $-i\beta$}}
\thicklines
\put(25,69){\circle*{1}}
\put( 6,69){\circle*{1}}
\put( 6,67){\circle*{1}}
\put( 6,50){\circle*{1}}
\put(47,68){\circle*{1}}
\put(46,68){\oval(2,2)[r]}
\put( 3,68){\vector( 1, 0){48}}
\put( 6,69){\vector( 1, 0){28}}
\put(34,69){\line( 1, 0){12}}
\put(46,67){\vector(-1, 0){28}}
\put(18,67){\line(-1, 0){12}}
\put( 6,67){\vector( 0,-1){11}}
\put( 6,56){\line( 0,-1){ 6}}
\put(4,50){\line( 1, 0){ 2}}
\put(5,48){\vector( 0, 1){28}}
\end{picture}
\end{center}
\caption{Closed Time Path with Initial Conditions}
\label{fctpinit}
\end{figure}
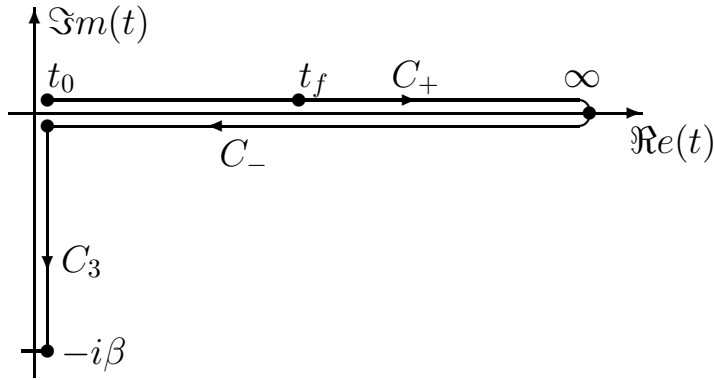

We now have the explicit form for $p_{t_f}[\Phi_f]$:-
\begin{eqnarray}
\lefteqn{p_{t_f}[\Phi_f] } \nonumber \\
&=& \int {\cal D} \Phi \int^{\phi_3(t_0) = \Phi}_
{\phi_3(t_0 - i \beta_0)= \Phi} {\cal D} \phi_3 \,  e^{i S_0[\phi_3]}
\int_{\phi_{\pm}(t_0) = \Phi}
 {\cal D} \phi_+ {\cal D} \phi_- \,
e^{i ( S[\phi_+] - S[\phi_-] ) } \delta [ \phi_+(t_f) - \Phi_f ]
\nonumber
\\
&=& \int_B {\cal D} \phi_3 {\cal D} \phi_+ {\cal D} \phi_- \, e^{
i S_0[\phi_3] + i ( S[\phi_+] - S[\phi_-]}) \,
\delta [ \phi_+(t_f) - \Phi_f ],
\end{eqnarray}
where the boundary condition $B$ is $\phi_{\pm}(t_0) = \phi_3(t_0) =
\phi_3(t_0- i \beta_0)$. This can be written as the time ordering of a
single field:-
\beq
p_{t_f} [ \Phi_f] = \int_B {\cal D} \phi \, e^{i S_C [\phi]} \, \delta [
\phi_+ (t_f) - \Phi_f ],
\eeq
along the contour $C=C_+ \oplus C_- \oplus C_3$, extended to include a
third imaginary leg, where $\phi$ takes the values $\phi_+$, $\phi_-$
and $\phi_3$ on $C_+$, $C_-$ and $C_3$ respectively, for which $S_C$
is $S[\phi_+]$, $S[\phi_-]$ and $S_0[\phi_3]$.
This
corresponds to working with a curve of the form shown in figure \ref{fctpinit}.
% ********************************************
\typeout{figure: CTP with Initial Conditions}
We stress again that
although $S_0[\phi]$ may look like the quadratic part of $S[\phi]$,
its role is solely to encode the initial distribution of configurations
$\Phi$ and need have nothing to do with the physical action.
Henceforth we drop the suffix $f$ on $\Phi_f$ and take the
origin in time from which the evolution begins as $t_0 =0$.

We perform one final manoeuvre with $p_t[\Phi]$ before resorting to
further approximation. This will enable us to avoid an ill-defined
inversion of a two-point function later on. Consider the generating
functional:-
\beq
Z[j_+,j_-,j_3] = \int_B {\cal D} \phi \, \exp \biggl \{ i S_C[\phi] +
i \int j \phi \biggr \},
\eeq
where $\int j \phi$ is a short notation for:-
\beq
\int j \phi \equiv \int_0^{\infty} dt \, \,  [ \, j_+(t) \phi_+(t) - j_-
\phi_-(t) \, ] \, + \int_0^{-i \beta} j_3(t) \phi_3(t) \, dt,
\eeq
omitting spatial arguments.
It may seem from the exponent $S_{0}[\phi_{3}] + S[\phi_{+}] -
S[\phi_{-}]$ that there is no communication between fields in
different branches of $C$.  This is not the case.  Suppose that
$t_{2} > t_{1} > t_{0}$.  Then the two-point function
$<T(\phi_{-}(t_{2}) \phi_{+}(t_{1}))>$ equals
$<\phi_{+}(t_{2})\phi_{+}(t_{1}))>$ since, if we remove the contour
for $t > t_{2}$, $\phi_{-}(t_{2}) = \phi_{+}(t_{2})$.
Thus, in fact , we have a nondiagonal matrix of propagators, of which the
$2\times 2$ submatrix between the $\phi_{+}, \phi_{-}$ fields is the
most important.  It is for this reason that the propagator (3) is
incomplete.

Then introducing $\alpha_a(\vecx)$ where
$a=1, \ldots , N$, we find:-
\begin{eqnarray}
p_{t_f} [\Phi] &=& \int {\cal D} \alpha \int_B {\cal D } \phi \,\,
\exp \biggl \{i
S_C[\phi] \biggr \} \,\,
\exp \biggl \{ i\int  d^4x \alpha_a(\vecx)
 [ \phi_+(t_f,\vecx) - \Phi(\vecx) ]_a \biggr \}
\\
&=& \int {\cal D} \alpha \, \, \exp \biggl \{ -i \int \alpha_a \Phi_a
\biggr \} \,Z[\overline{\alpha},0,0],
\end{eqnarray}
where $\overline{\alpha}$ is the source $\overline{\alpha}(t,\vecx) =
\alpha(\vecx) \delta (t-t_f)$. As with ${\cal D} \phi$, ${\cal D}
\alpha$ denotes $\prod_1^N {\cal D} \alpha_a$.

We have seen that analytic progress can only be made insofar as
$p_t[\Phi]$ is itself Gaussian, requiring in turn that
$Z[\overline{\alpha},0,0]$ be Gaussian in the source
$\overline{\alpha}$. In order to treat the fall from the false
into the true vacuum, at best this means adopting a self-consistent
or variational approach i.e. a Hartree approximation or a large
N-expansion.\tnote{In the latter case, given the relationship
between $N$ and
spatial dimension $D=N$, this corresponds to a large-dimension
expansion}.  However,
%we are only interested in the validity of the
%field approximation of Zurek, recounted earlier, then
if we limit ourselves to
small times $t$ then $p_t[\Phi]$ is genuinely Gaussian since the
field has not yet felt the upturn of the potential. That is, we may
treat the potential as an inverted parabola until the field begins to
probe beyond the spinodal point. The length of time
for which it is a good approximation to ignore the upturn of the
potential is greatest for weakly coupled theories which, for the
sake of calculation we assume, but physically, we
expect that if the defect counting approximation is going to fail, then
it will do so in the early part of the fall down the hill.

\section{ DEFECT PRODUCTION AT A QUENCH }

\subsection{ Counting The Defect Density}

To calculate the defect density
%we use the results of Halperin \cite{halperin},
requires knowledge of $p_t[\Phi]$ which, for the moment, we  write
as \cite{halperin}
\beq
p_t[\Phi] = {\cal N} \exp \biggl (
- \frac{1}{2} \int d^{D}x \, d^{D}y \, \, \Phi_a(\vecx)
K_{ab}(\vecx-\vec{y};t) \Phi_b( \vec{y})
\biggr ),
\eeq
with $K_{ab} = \delta_{ab} K$ and ${\cal N}$ a
normalisation.  The circumstances under which a Gaussian is valid
will be examined later.  For weakly coupled theories
we shall see  that, for short times after $t_0$ at least, a
Gaussian $p_t[\Phi]$ will occur.  If this is taken for granted it
has been shown by Halperin
\cite{halperin} how to calculate the number density of defects.
Postponing the calculation of $K$  until then, we quote those of
Halperin's results that are relevant to us.
%[In the absence of a Gaussian probability distribution, we can make
%little progress].

Suppose that the field $\phi(t,\vecx)$ takes the particular
value $\Phi(\vecx)$. We noted that strings are tubes of false
vacuum. We count the strings passing through a surface $S$
by identifying them with
its zeroes in $S$.  Similarly, monopoles are balls of false vacuum
in which the field vanishes.  The only way for a zero to occur with significant
probability is at the centre of a topological defect so, but for a
set of measure zero, all zeroes are topological defects.
%\footnote{This counting procedure alone gives no
%information about the length  distribution of the defects.}

Consider the $O(D)$ theory in $D$ spatial dimensions,
with global {\it monopoles}. Although less relevant than strings for
the early universe they are slightly easier to perform calculations
for. Almost everywhere, monopoles occur
at the zeroes of $\Phi(\vecx)$,
labelled $\vec{x}_i, i=1,2,\ldots$, at which the orientation
$\Phi(\vecx)/ |\Phi(\vecx) |$ is ill-defined.
A topological winding
number $n_i=\pm 1$ can be associated with each zero $\vec{x}_i$ by the
rule:-
\beq
n_i = {\rm sgn} \det ( \partial_a \Phi_b) \biggr | _{ \vecx=\vec{x}_i}.
\eeq
Monopoles with higher winding number are understood as multiple zeroes
of $\Phi(\vecx)$ at which the $n_i$ are additive.
The {\it net} monopole density is then given by:-
\beq
\rho_{{\rm net}} ( \vecx) = \sum_i n_i  \, \delta (\vecx - \vec{x}_i).
\eeq
The volume integral of this gives the number of monopoles minus the
number of antimonopoles. The correlations of $\rho_{{\rm net}}$ give
us information on monopole-(anti)monopole correlations but, in the
first instance, we are interested in the cruder grand totals.
The quantity of greater relevance to us,
is the {\it total} monopole density:-
\beq
\rho(\vecx) = \sum_i \delta(\vecx - \vec{x}_i),
\eeq
whose volume integral gives the total number of monopoles {\it plus}
antimonopoles. in the volume of integration.

Now consider an ensemble of systems in which the fields $\Phi$ are
distributed according to $p_t[\Phi]$ at time $t$. Then, on average,
the total monopole density is:-
\beq
\rho_{m}(t) =\langle \rho(\vecx) \rangle_{t} = \biggl \langle \, \sum_i
\delta(\vecx - \vec{x}_i) \, \biggr
\rangle _t,
\eeq
where the triangular brackets denote averaging with respect to
$p_t[\Phi]$ and the suffix $m$ denotes $monopoles$. That is:-
\beq
\langle F[\Phi] \rangle_t = \int {\cal D} \Phi \, F[\Phi] \, p_t[\Phi],
\eeq
with $p_t[\Phi]$ normalised so that $\int {\cal D} \Phi \, p_t[\Phi]
=1$. The translational invariance of the Gaussian kernel of the
probabililty density ensures that $\rho(t)$ is translationally invariant.

In terms of the fields $\Phi_a$, vanishing at $\vec{x}_{i}$, $\rho_{m}(t)$
can be re-expressed as:-
\beq
\rho_{m}(t) = \langle \, \delta^D [\Phi_c(\vecx)] \, \,  |
\det \, (\partial_a \Phi_b(\vecx) ) \, | \, \, \rangle_t.
\eeq
The second term in the brackets is just the Jacobian of the
transformation from $\vecx$ to $\Phi(\vecx)$.
%Similarly, although
%we have less need for it, the {\it net} monopole density at any time is:-
%\beq

%\langle \rho_{{\rm net}} (t) \rangle = \langle  \, \delta^D [\Phi({\bf
%x})] \, \det (\partial_a \Phi_b(\vecx) ) \rangle_t.
%\eeq

It follows from the Gaussian form of the probability density that
the $\Phi_a$ are individually and independently Gaussian
distributed with zero mean, as
\beq
\langle \, \Phi_a(\vecx) \, \Phi_b (\vecx) \, \rangle _t =
\delta_{ab} W(|\vecx - \vec{y}| ;t),
\eeq
where $W(|\vecx - \vec{y}| ;t) = K^{-1}(\vecx - \vec{y} ;t)$.
So also are the first derivatives of
the field $\partial_a \Phi_b$, which are independent of the field:-
\beq
\langle \Phi_c(\vecx) \partial_a \Phi_b(\vec{y}) \rangle_t = 0
\eeq
due to the fact that
$W$ is dependent only on the magnitude of $\vecx - \vec{y}$.

Thus, the total defect density may be separated into two independent
parts:-
\beq
\rho_{m}(t) = \langle \, \delta^D [\Phi (\vecx) ] \,
\rangle_t \, \, \langle \, | \det ( \partial_a \Phi_b ) | \, \rangle _t.
\eeq
The first factor is easy to calculate, the second less so.

Consider first the delta-distribution factor:-
\begin{eqnarray}
\langle \delta[ \Phi (\vec{x}_0) ] \rangle_{t} &=& \int {\cal D} \Phi \,
\delta[ \Phi (\vec{x}_0) ] \, \exp \biggl \{ - \half\int
d^{D}\vecx d^{D}\vec{y}
\Phi(\vecx) K(\vecx - \vec{y};t) \Phi(\vec{y}) \,
% {\bf d^4x} {\bf d^4y}
\biggr \}
\\
&=& \int d \! \! \! / \alpha \int  {\cal D} \Phi \,
\exp \biggl \{  i \alpha \Phi (\vec{x}_0) - \half\int
d^{D}\vecx d^{D}\vec{y}
\Phi(\vecx) K(\vecx - \vec{y};t) \Phi(\vec{y}) \,
% {\bf d^4x} {\bf d^4y}
\biggr \}
\end{eqnarray}
where $O(N)$ indices and integrals over spatial variables have been
suppressed and $d \! \! \! / \alpha = d  \alpha / 2 \pi$. On defining
$\overline{\alpha} = \delta (\vecx-\vec{x}_0) \alpha$, we find:-
\begin{eqnarray}
\lefteqn{\langle \delta[ \Phi (\vec{x}_0) ] \rangle_{t}} \nonumber
\\
 &=& \int  d \! \! \! /
\alpha \int {\cal D} \Phi
\, \exp \biggl \{ \int d^{D}\vecx\, \biggl (i \overline{\alpha}(\vecx)
\Phi(\vecx)  - \half\int
d^{D}\vecx d^{D}\vec{y}  \Phi(\vecx) K(\vecx - \vec{y};t) \Phi(\vec{y})
\biggr ) \,
\biggr \}
\nonumber
\\
&=& \int  d \! \! \! / \alpha \exp \biggl \{ - \half\int
d^{D}\vecx d^{D}\vec{y}\;
\overline{\alpha}(\vecx) W(|\vecx-\vec{y}|;t)
\overline{\alpha}(\vec{y}) \biggr \} \,
%\delta(\vecx-{\bf x_0})
\\
&=& \int  d \! \! \! / \alpha \exp \biggl \{
 - \half\,\int
d^{D}\vecx d^{D}\vec{y}\;  \delta(\vecx-\vec{x}_0) \alpha
W(|\vecx-\vec{x}'|;t) \alpha \delta(\vec{x}'-\vec{x}_0)
\, \,
%{\bf d^3x}{\bf d^3x'}
\biggr \}
\\
&=& \frac{1}{2 \pi} \biggl ( \frac{1}{\sqrt{K^{-1}}} \biggr ) ^2 =
\frac{1}{2 \pi \langle \Phi \Phi \rangle}
=\frac{1}{2 \pi W(0;t)}
\end{eqnarray}

Consider now the second factor. Writing out the determinant
explicitly for $N=D=2$, and exploiting the fact that the field is
Gaussian, we have:-
\begin{eqnarray}
\langle \,  | \, \det ( \partial_a \phi_b(\vecx)) \, | \,
\rangle_{t} ^2 &=& \biggl \langle \, \biggl [ \, \det ( \partial_a
\phi_b(\vecx) ) \, \biggr ]^2 \, \biggr \rangle_{t}
\\
&=& \biggl \langle ( \partial_1 \phi_1 \partial_2 \phi_2
)^2 +(\partial_1 \phi_2 \partial_2 \phi_1)^2 - 2 \partial_1 \phi_2
\partial_2 \phi_1 \partial_1 \phi_1 \partial_2 \phi_2 \biggr
\rangle_{t}  .
\end{eqnarray}
The first term may be factorised into a product of two Gaussian variables
and calculated as follows:-
\begin{eqnarray}
\langle \, ( \partial_1 \phi_1 \partial_2 \phi_2 )^2 \, \rangle_{t} &=&
\langle \, (\partial_1 \phi_1 )^2 \, \rangle_{t} \, \langle \, (\partial_2
\phi_2 )^2 \,  \rangle_{t}
\\
&=& [ - \delta_{11} \partial_1 \partial_1 W(|\vecx|;t) ] \,  [ -
\delta_{22} \partial_2 \partial_2 W(|\vecx|;t) ] = [ \partial_1
\partial_1 W(|\vecx|;t) ]^2
\end{eqnarray}
where $W(|\vecx|;t)$ is as before.
%defined by:-
%\beq
%\langle \phi_a (\vecx) \, \phi_b ({\bf x'}) \rangle = \delta_{ab}
%W(\vecx - {\bf x'}).
%\eeq
Fourier transforming the two-point function, we find:-
\begin{eqnarray}
\langle \, ( \partial_1 \phi_1 \partial_2 \phi_2 )^2 \, \rangle_t &=&
\biggl ( \partial_1 \partial_1 \int e^{i\veck . \vecx }
\tilde{W}(\veck;t) \,  d \! \! \! / ^3 k \biggr )
\biggl ( \partial_2 \partial_2 \int e^{i \veck . \vecx }
\tilde{W}(\veck;t) \,   d \! \! \! / ^3 k \biggr )
\\
&=& \biggl (
\int k^2 \cos ^2(\theta) W(\veck;t) \,   d \! \! \! / ^3 k
\biggr )^2
\\
&=& [ \nabla^2 W(0;t) ]^2
\end{eqnarray}
A similar result applies for the second term whereas
the third term vanishes.
%Thus, in terms of the two point
%function $\langle \Phi_a(\vecx) \Phi_b({\bf x'}) \rangle =
%\delta_{ab} W(\vecx - {\bf x'})$,
The final result is:-
\beq
\rho_{m}(t) = C_N \Biggl | \frac{W''(0;t)} {W(0;t )},
\Biggr | ^{N/2}
\eeq
where the second derivative in the numerator is with
respect to $x=|\vecx|$. $C_N$
is $1/ \pi^2$ for $N=D=3$ and $1/ 2 \pi$ for $N=D=2$ (had we
performed the calculation in $D=2$ dimensions), the difference
coming entirely from the determinant factor \cite{halperin}.

Let us now consider the case of global strings in $D=3$ dimensions
that arise from the $O(2)$ theory. Strings are identified with
lines of zeroes of $\phi(t,\vecx)= \Phi(\vecx)$ and the net vortex
density (vortices minus antivortices) in a plane
perpendicular to the $i$-direction is:-
\beq
\rho_{{\rm net},i}(\vecx) = \delta^2[\Phi(\vecx) ] \, \epsilon_{ijk}
(\partial_j \Phi_1) ( \partial_k \Phi_2 ),
\eeq
with obvious generalisations to $N=D-1$, for all $N$ in terms of the
Levi-Cevita symbol $\epsilon_{i_1,i_2, \ldots ,i_D}$.
As before, the total vortex density is of more immediate use.
On a surface perpendicular to
the $i$-direction this is:-
\beq
\rho_{i}(\vecx) = \delta^2[\Phi(\vecx) ]
\, | \epsilon_{ijk} (\partial_j \Phi_1) ( \partial_k \Phi_2 ) |.
\eeq
in analogy with the monopole case.
The expectation value of this total density, when calculated as before
reproduces the same expression (where $s$ denotes $strings$):-
\beq
\rho_{s,i}(t) = <\rho_{i}(\vecx)> = C_N \Biggl |
\frac{W''(0;t)} {W(0;t )}
\Biggr | ^{N/2},
\eeq
independent of $i$, but for $N=D-1$.
Thus,
whether we are concerned about global
monopole or global string density, once we have calculated
$W(|\vecx|;t)$ we can find the total string density.
Similar results apply for the correlations between net densities,
which are important in determining the subsequent evolution of the
defect network.

\subsection{ Evolution Of The Defect Density}

The onset of the phase transition at time $t=0$ is characterised by
the instabilities of long wavelength fluctuations permitting the growth of
correlations. Although the initial
value of $\langle \phi \rangle$ over any volume is zero,
the resulting phase separation or spinodal decomposition will lead to
domains of constant $\langle \phi \rangle$ whose boundaries will
evolve so that ultimately, the average value of $\phi$ in some finite
volume, will be non-zero. That is, the relativistic system has a
non-conserved order parameter. In this sense, the model considered
here is similar to those describing the $g^{2}$ transition in liquid
helium or transitions in a superconductors.

Consider small amplitude fluctuations of $\phi_a$, at the top of the
parabolic potential hill described by $V(\phi) = \frac{1}{2} m^2 (t)
\phi_a^2$. At $t<0$, \,
$m^2(t) > 0 $ and, for $t>0$,\, $m^2(t)<0$. However, by $t \approx \Delta
t$, \,$m^2(t)$ and $g^{2}$ have achieved their final values, namely
$-\mu^2$ and $g^{2}$. Long wavelength fluctuations, for which $|\veck|^2
< - m^2(t)$, begin to grow exponentially. If their growth rate
$\Gamma_k \approx \sqrt{-m^2(t) - |\veck|^2}$ is much slower than
the rate of change of the environment which is causing the quench,
then those long wavelength modes are unable to track the quench. For
the case in point, this requires $m \Delta t \ll 1$.
We take this to be the case.
%That is, we assume that the static field
%approximation is good at least until the field begins to fall down the
%hill with the final values of $\mu^2$ and $g^{2}$. Here, we are
%concerned only with the general physics rather than with the details
%of any particular system. Therefore,
To
exemplify the growth of domains and the attendant dispersal of
defects, it is sufficient to take the idealised case, $\Delta t =0$ in
which the change of parameters at $t=0$ is instantaneous. That
is, $m^2(t)$ satisfies
\beq
\[
m^2(t) = \left \{
\begin{array}{ll}
m_0^2 >0 & \, \, \mbox{if $t<0$,} \\
 - \mu^2 <0 & \, \, \mbox{if $ t>0$}
\end{array} \right .
\]
\eeq
where for $t<0$, the field is in thermal equilibrium at inverse
temperature $\beta_0$. As for $g^{2}(t)$, for $t<0$, it has already
been set to zero, so that $p_{t_0} [\Phi] $ be Gaussian. For small t,
when the amplitude of the field fluctuations is small, the field has
yet to experience the upturn of the potential and we can take
$g^{2}(t) = 0$ then as well. At best, this can be valid until the
exponential growth $|\phi| \approx \mu e^{\mu t}$ in the amplitude
reaches the point of inflection $|\phi| \approx \mu / \sqrt{g^{2}}$,
that is $\mu t \approx O(\ln (1/ g^{2}) )$. The smaller the coupling
then, the longer this approximation is valid.
%However long the
%approximation holds, however, it seems physically reasonable that any
%failure in the static field approximation will occur while it is still good.
As noted earlier, it should be possible to perform more sophisticated
calculations with the aim of evolving the defect density right through
the transition. For our present purposes, however, the small time or Gaussian
approximation is adequate.

We are now in a position to evaluate $p_t[\Phi]$, identify $K$ and
calculate the defect density accordingly. $S_C[\phi]$ becomes
$S_0[\phi_3]$ on segment $C_3$ so setting the boundary condition
$\phi_+(0,\vecx) = \phi_3(0,\vecx ) = \phi_3(-i \beta_0, \vecx)$
and we have
\beq
S[\phi_+] = \int d^{4}x \, \, \biggl [ \, \frac{1}{2}
(\partial_\mu \phi_{+a})
(\partial^\mu \phi_{+a}) + \frac{1}{2} \mu^2 \phi_{+a}^2 \, \biggr ],
\eeq
on $C_+$. The Gaussian integrals can now be performed to give
\beq
p_t[\Phi] = \int {\cal D} \alpha \, \exp \biggl \{-i \int  d^3x
\, \alpha_a  \Phi_a \biggr \}
\exp \biggl \{ \frac{i}{2} \int d^3x \, d^3y \, \alpha_a(\vecx)
G(\vecx - \vec{y};t,t) \alpha_b(\vec{y}) \biggr \},
\eeq
where  $G(\vecx - \vec{y};t,t)$ is the equal time correlation, or
Wightman, function
with thermal boundary conditions. Because of the time evolution
there is no time translation invariance in the double time label.
As this is not simply
invertible, we leave the $\alpha$ integration unperformed. The form
is then a mnemonic reminding us that $K^{-1} = G$.

In fact, there is no need to integrate the $\alpha$s since from the
previous equation it follows that the characteristic functional
$\bigl \langle \exp \bigl \{ i \int J_a \Phi_a \bigr \}
\bigr \rangle_t$ is directly
calculable as
\begin{eqnarray}
\biggl \langle \exp \biggl \{ i \int j_a \Phi_a \biggr \}
\biggr \rangle_t &=& \int {\cal D} \Phi
\, p_t[\Phi] \, \exp \biggl \{ i \int j_a \Phi_a \biggr \}
\\
&=& \exp \biggl \{ \frac{1}{2} \int d^3x \,  d^3y \, j_a(\vecx) G(\vecx -
\vec{y} ; t,t ) j_a(\vec{y}) \biggr \}.
\end{eqnarray}
Thus for example, the first factor in the monopole density $\rho_{m}(t)$
is
\begin{eqnarray}
\langle \, \delta ^D [ \Phi(\vecx) ] \, \rangle_t &=& \Biggl \langle
\int dj \, \exp \bigl ( i \Phi_a(\vecx) j_a \bigr ) \Biggr \rangle_{t}
\\
&=& \int dj \, \exp \biggl \{ \frac {1}{2} j_a^2 G(\vec{0} ;t,t) \biggr
\} = [ -iG(\vec{0};t,t)]^{-D/2},
\end{eqnarray}
with suitable normalisation, without having to invert $G(\vec{0};t)$.
Thus, on identifying $-iG(\vecx;t,t)$ with $W(\vecx,t)$
as defined earlier, $\rho_{m}(t)$  becomes \cite{GR}
\beq
\rho_{m}(t) = C_N \Biggl |
\frac { -iG''(\vec{0};t,t) } {-iG(\vec{0};t,t)}
\Biggr | ^{N/2},
\eeq
where $-iG(\vecx;t,t)$ has to be calculated from the equations of
motion, subject to the initial condtition.

Details are given by Boyanovsky et al.,
\cite{B} and we quote their results, which give $-iG(\vecx;t,t)$
as the real, positive quantity
\begin{eqnarray}
\lefteqn{ -iG(\vecx;t,t) =
\int \frac { d \! \! \! / ^D k }{2 \omega_<(k)}
\, e^{i \veck . \vecx } \coth ( \beta_0 \omega_<(k) / 2) \times}
\\
&&
\Biggl \{
\biggl [ 1+ A_k(\cosh(2W(k)t) - 1 ) \biggr ] \theta(\mu^2 - |\veck|^2)
+ \biggl [ 1+ \alpha_k ( \cos(2\omega_>(k)t)-1 \biggr ] \theta (
|\veck|^2 - \mu^2)
\Biggr \}
\nonumber
\end{eqnarray}
with
\begin{eqnarray}
\omega_<^2(k) = |\veck|^2 + m_0^2
\\
\omega_>^2(k) = |\veck|^2 - \mu^2
\\
W_<^2(k) = \mu^2 - |\veck|^2
\\
A_k = \frac{1}{2} \Biggl (
1+ \frac{\omega_< ^2(k) }{W^2(k)} \Biggr )
\\
\alpha_k = \frac{1}{2} \Biggl (
1 - \frac{\omega_<^2(k) }{\omega_>^2(k)} \Biggr ).
\end{eqnarray}
The first term is the contribution of the unstable long wavelength
modes, which relax most quickly;
the second is that of the short
wavelength stable modes which provide the noise. The first term will
dominate for large times and even though the approximation is only
valid for small times, there is a regime, for small couplings, in which
$t$ is large enough for $\cosh(2 \mu t) \approx \frac{1}{2} exp(2 \mu
t)$ and yet $\mu t$ is still smaller than the time $O(\ln 1/g^{2})$ at
which the fluctuations sample the deviation from a parabolic hill. In
these circumstances the integral at time t is dominated by a
peak in the integrand $k^{D-1} e^{2W(k)t}$ at $k$ around $k_c$, where
\beq
t k_c^2 = \frac{(D-1)}{2} \mu \biggl ( 1 + O \biggl ( \frac {1}{\mu t}
\biggr ) \biggr ).
\eeq
The effect of changing $\beta_0$ is only visible in the $O(1/ \mu t)$
term. In the region $|\vecx| < \sqrt{ t/ \mu} $ the integral is
dominated by the saddle-point at $k_c$, to give
\beq
- i G( \vecx ; t,t ) = W(x;t ) \approx W(0;t) \, \exp \Biggl
( \frac{- \mu x^2 }{ 8t} \Biggr ) \, {\rm sinc }
\Biggl ( \frac{x}{ \sqrt{t/\mu} } \Biggr ),
\eeq
for $D=3$, where
\beq
W(0;t) \approx C \frac {e^{2 \mu t} } {( \mu t) ^{3/2} },
\eeq
for some C, which we don't need to know. The exponential growth
of $G({\bf 0 };t)$ in $t$ reflects the way the field amplitudes fall
off the hill $\langle \Phi \rangle = 0$. It is sufficient for our
purposes to retain $D=3$ only.

After symmetry breaking to $O(N-1)$ the mass of the Higgs is $m_H =
\sqrt{2} \mu$ with cold correlation length $\xi(0) = m_H^{-1}$. On
identifying $e^{- \mu x^2 / 8t}$ as $e^{-x^2 / \xi^2(t)}$ we interpret
\beq
\xi(t) = (8 \sqrt{2}) ^{1/2} \sqrt {\, t \, \xi(0) },
\eeq
as the size of Higgs field domains. This $t^{1/2}$ growth behaviour at
early times is characteristic of relativistic systems (with a double
time derivative) with a
non-conserved order parameter.

To calculate the number density of defects at early times we have to
insert this expression for $-iG$ or $W$ into the equations derived
earlier.
Expanding $W(x;t)$ as
\beq
W(x;t) = W(0;t) \, \exp \biggl ( \frac{ - x^2}{\xi^2(t)}
\biggr ) \, \biggl ( \, 1 - \frac{4}{3} \frac{x^2}{\xi^2(t)} +
O \biggl ( \frac{x^2}{\xi^2(t)} \biggr ) \, \biggr ),
\eeq
gives \cite{GR}
\beq
\rho_{m}(t) = \frac{1}{\pi^2} \Biggl (
\frac{ \sqrt{14/3} } {\xi(t)} \Biggr )^3 \approx \frac{1.02}{\xi^3},
\eeq
for an $O(3)$ theory with monopoles in three dimensions and
\beq
\rho_{s,i}(t) = \frac{1}{2 \pi} \Biggl (
\frac{ \sqrt{14/3} } {\xi(t)} \Biggr )^2 \approx \frac{0.74}{\xi^2},
\eeq
for an $O(2)$ theory with strings in three dimensions.
The first observation is that the dependence of the density on time
$t$ is only through the correlation length $\xi (t)$.  As the
domains of coherent field form and expand, the interdefect distance
grows accordingly.  This we would interpret as
the domains carrying the defects
along with them on their boundaries.  Secondly, as we signalled in
the introduction, there is roughly
one defect per coherence size,  a long held belief for whatever
mechanism.  However, in this case the density is exactly calculable.
It is also possible to use Halperin's results to calculate
defect-defect correlation functions.
Details will be given elsewhere.

Finally, we mentioned in our introduction that superfliud $He$ is,
if anything, more likely to provide an environment in which these
mechanisms can be tested.  One way to convert the relativistic
formalism above into that for a non-relativistic field theory is to
include a chemical potential of value $m$ equal to the rest mass of
the quanta.  Provided $\beta m\gg 1$ antiparticles are eliminated
and effective particle energy is non-relativistic.  Otherwise
everything goes through as before.  Results will be presented elsewhere.
%For example, the monopole-monopole
%correlation function on scales larger than a coherence length is found
%to be
%\beq
%\langle \rho_{{\rm net}} (\vecx) \, \rho_{{\rm net}} ({\bf 0})
%\rangle_t = \rho_{m}(t) \delta(\vecx) + g(\vecx)_t,
%\eeq
%where $g(\vecx)_t$ is a measure of the screening of a monopole at
%the origin,satisfying
%\beq
%\int {\bf d^{D}x} \, g(\vecx)_{t} = -\rho_{m}(t)
%\eeq
%Explicit calculation yields
%\beq
%g(\vecx)_t = - \frac
%{ 3 \sqrt{2} \exp ( - 3 x^2 / \xi^2 ) \sin^3 (2 \sqrt{2} x / \xi )}
%{ 8 \pi^3 x \xi^5}\Biggl ( 1 + O\Biggl (\frac{\xi}{x}\Biggr )\Biggr ).
%\eeq
%This alternation in the sign of the screening on scale $\xi$ is
%compatible with the density result presented earlier.
%Similar calculations can be performed for the correlations of the
%string densities $\rho_{net,i}(\vecx)$ introduced earlier.  At the
%moment this is more information than we need.

\section{ LATER EVOLUTION}

Once field fluctuations are large enough to reach the points of
inflexion of the classical potential $V(\phi) = g^{2}(\phi^{2} -
\eta^{2})^{2}/8$, the simple Gaussian approximation breaks down, if
not before.  For a while the evolution of domains is controlled by
the dissipation of the field energy in particle production.
However, because of their topological stability some defects will
survive.

One way to proceed is to try to develop Landau-like Langevin
equations for the evolution of coarse-grained fields in $V(\phi)$.
 From our previous considerations the relevant scale would be $\xi
(t_{i})$, where $t_{i} = O(ln(1/g^2))$ is the time at which the
field fluctuations begin to feel the points of inflexion of
$V(\phi)$.
This is essentially the approach of Davis and Martin \cite{Anne},
although they
choose $\xi_{G}$, the Ginzburg correlation length as their starting
point.  From the point of view of the equations this does not
matter, and this is all that concerns us here.  As yet we have no
new results to report, but we conclude these lectures with a few
comments on dissipation that might help in developing a programme
along the lines of \cite{Anne}.  In large part, we are just paraphrasing
the most recent work of Boyanovsky et al.\ \cite{boyanovsky}.

It is a truism to say that coarse-graining introduces dissipation
into the otherwise time-reversal invariant evolution equation for
the fields.  To see how this comes about, we begin with the simplest
example of coarse-graining, the effective potential.

\subsection{ The Effective Potential}

We have discussed the effective potential earlier since, as the
energy density, it determines the ground state of the system, and
hence its phase.  However, we avoided a detailed calculation of it.
For exemplary purposes it is sufficient to consider a single scalar
field $\phi$ at {\it zero} temperature.

The effective potential $V_{eff}(\bar{\phi})$
is obtained most succinctly by summing over
field histories in the partition function $Z$,
subject to the constraint
\beq
\frac{1}{\Omega}\int d^{4}x\; \phi (t,\vecx) = \bar{\phi},
\eeq
where $\Omega$ is the large space-time volume of the system.  This
total coarse-graining of $\phi$, both in space {\it and} time,
precludes any evolution equation, leading just to
\beq
\frac{dV_{eff}}{d\bar{\phi}}(\bar{\phi}) = 0
\eeq
as the definition of the ground state of the theory.
Nonetheless, eq.(179) is the simple counterpart of the Langevin
equation that we shall attempt later.

To see how (179) occurs it is most useful to make the separation
\beq
\phi (t,\vecx) = \bar{\phi} + \eta (t,\vecx),
\eeq
where $\eta$ contains no fluctuations with $k = 0$. That is,
\beq
\int d^{4}x\; \eta(t,\vecx) = 0.
\eeq
To calculate $V_{eff}$ we begin with the generating functional
$Z[j]$, in which $\phi$ is coupled to a {\it constant} source $j$ as
\beq
Z[j] = \int {\cal D}\phi\; e^{iS[\phi] + ij\int\phi}
\eeq
where $S[\phi]$ is given in (4).   For the sake of argument $m^{2}$
is taken positive.
Upon the decomposition (180) $Z[j]$ becomes
\beq
Z[j] = \int d\bar{\phi}\int{\cal D}\eta\; \delta\left(\int\eta\right)
e^{iS[\bar{\phi} + \eta] + ij\Omega\bar{\phi}}.
\eeq
A series expansion of $S[\bar{\phi} + \eta]$ in powers of $\eta$
leads to
\beq
Z[j] = \int d\bar{\phi} e^{ij\Omega\bar{\phi} -i\Omega V(\bar{\phi})}
\int {\cal D}\eta\;\delta\left(\int\eta\right)e^{iS_{eff}[\eta]},
\eeq
where
\beq
S_{eff}[\eta] = \int d^{4}x\left(\half\partial_{\mu}
\eta\partial^{\mu}\eta - \half
(m^2 + 3g^{2}\bar{\phi}^{2}/2)\eta^{2} + ...\right).
\eeq
The constraint (181) eliminates all terms in $S_{eff}$ linear in
$\eta$.
To one-loop it is only necessary to
\begin{enumerate}
\item retain quadratic terms in $\eta$ in $S_{eff}$
\item ignore the $\delta$-function.
\end{enumerate}
On performing the $\eta$ integration we find \cite{RR2}
\beq
Z[j]\simeq \int d\bar{\phi}\; e^{i\Omega(j\bar{\phi} -
V_{eff}(\bar{\phi}))},
\eeq
where
\beq
V_{eff}(\bar{\phi}) = V(\bar{\phi}) + \half i\int d^{4}k\; ln(-k^{2} +
m^{2} + 3g^{2}\bar{\phi}^{2}/2).
\eeq
is the effective potential. In the $\Omega\rightarrow\infty$ limit the
integral is dominated by a stationary phase for $\phi$ satisfying
(179), which thereby determines the ground state of the theory.

We conclude this introductory section by making some simple
observations  on approximations.
The one-loop approximation is correct to $O(\hbar)$ and all orders
in $g^2$.  Its $g^{2}$-expansion, to second order, is
\beq
V_{eff}(\bar{\phi}) = V(\bar{\phi}) +
\frac{1}{2}\left(\frac{3}{2}g^{2}\bar{\phi}^{2}\right)\Sigma +
\frac{1}{2}\left(\frac{3}{2}g^{2}\bar{\phi}^{2}\right)
I\left(\frac{3}{2}g^{2}\bar{\phi}^{2}\right)
+ O(g^{6}),
\eeq
where $\Sigma$ is the one-loop self-mass and $I$ is the one-loop
two-vertex function.

If we had only been given (188) as a first approximation to
$V_{eff}$ we would know that it would need to be elevated to the
one-loop potential (187).  Further, in many circumstances even this
is not sufficient.  Any attempt to incorporate non-perturbative
effects in $V_{eff}$ would require that, at least, we extend the
one-loop potential (187) to a self-consistent one-loop potential,
using the Hartree approximation or a large-N calculation of the
$O(N)$ theory. For the case in hand, when $N$ is directly related to
spatial dimension $D$ as $N=D$ or $N=D-1$ the latter is not useful.

When we attempt a Langevin equation it is the counterpart of (188)
that we can most simply derive.  We shall do this in the knowledge
that both the steps listed above need to be implemented before we
can have believable results.

\subsection{ Spatial Averaging}

As the next step we determine the evolution equation for the spatial
average
\beq
\Phi (t) = \lim_{v\rightarrow\infty} \phi_{v}(t)
\eeq
for $\phi_{v}(t)$ of (43).
Let us return to the closed time-path of (127) where, as there, we
have assumed an initial thermal distribution of states at inverse
temperature $\beta_{0}$. As with the effective potential we wish to
separate the $\phi$ field into those modes that describe the
`system', and those that describe the `environment' that will be
integrated over.  In this case these latter are the $\veck\neq\vec{0}$
modes, permitting the decomposition
\beq
\phi_{\pm}(t,\vecx)=\Phi_{\pm}(t) + \eta_{\pm}(t,\vecx),
\eeq
where
\beq
\int d^{3}x\; \eta_{\pm}(t,\vecx) = 0
\eeq
at all times $t$.  Although $\Phi_{+}(t)\neq\Phi_{-}(t)$,
$<\Phi_{+}(t)> = <\Phi_{-}(t)>$ since each is equally acceptable as the
physical field.  They only differ by the $\veck = \vec{0}$ fluctuations,
\beq
\sigma (t) = \Phi_{+}(t) - \Phi_{-}(t).
\eeq
The effective coarse-grained field is
\beq
\Phi (t) = \half (\Phi_{+}(t) +\Phi_{-}(t))
\eeq
with common expectation value.

The generating functional for the effective field $\Phi$ and its
correlations is
\beq
Z[j] = \int_C {\cal D} \phi_3 {\cal D} \phi_+ {\cal D} \phi_- \, e^{
i S_0[\phi_3] + i ( S[\phi_+] - S[\phi_-])} \,e^{iv\int j\Phi},
\eeq
where $S_{0}$ sets the initial conditions and, from above
\beq
\phi_{\pm}(t,\vecx) = \Phi (t)\pm\half\sigma(t) +
\eta_{\pm}(t,\vecx).
\eeq
On substituting in $Z$, the constraint (191) guarantees that there
are no linear terms in $\eta_{\pm}$, just as no terms in $\eta$
survived in (185).  Thereafter, at one loop we can ignore the
constraint in the path integral.  Further, to lowest nontrivial
order it is sufficient to neglect terms $O(\eta_{\pm}^3)$ in
$\eta_{\pm}$ and terms $O(\sigma ^3)$ in $\sigma$.  As a result
\bea
\lefteqn{S[\Phi + \half\sigma +\eta_{+}] - S[\Phi - \half\sigma + \eta_{-}] =}
\\
&&\int dt d\vecx\left(\sigma(t)\frac{\partial {\cal L}(\Phi)}{\partial
\Phi (t)} + [\half (\partial\eta_{+})^{2} -\half
M_{+}^{2}\eta_{+}^{2}] -  [\half (\partial\eta_{-})^{2} -\half
M_{-}^{2}\eta_{-}^{2}]\right)
\eea
where
\beq
M_{\pm}^{2} = m^{2} + \frac{3}{2}g^{2}(\Phi\pm\half\sigma)^{2}
\eeq
and
\beq
{\cal L}(\Phi) = \frac{1}{2}\dot{\Phi}^{2} -
\frac{1}{2}m^{2}\Phi^{2} -\frac{1}{8}g^{2}\Phi^{4}.
\eeq
We can now integrate out the $\eta_{\pm}$, remembering that they are
coupled and expand the resultant trace of the logarithm (cf. (187)) in
powers of $g^{2}$.  At second order in $g^{2}$ the effect is to give
$Z[j]$ as
\beq
Z[j] = \int{\cal D}\Phi{\cal D}\sigma\;  e^{iv[S_{eff}[\Phi,\sigma]
+\int j\Phi]},
\eeq
where $S_{eff}$ is
\beq
S_{eff} = -\int dt\; \sigma (t)L(\Phi (t)) +\half i\int dt dt'\; \sigma
(t)K(t,t';\Phi)\sigma (t')
\eeq
for $L$ and $K$ {\it real}.  $L(\Phi)$ is the dissipative form
\bea
\lefteqn{L(\Phi(t)) =
\ddot{\Phi}(t) + \left(m^{2} + \frac{3}{2}g^{2}\Sigma (t)\right)\Phi (t)
+\left(\frac{3}{2}g^{2}\right)\Phi^{3}(t)
} \nonumber\\
& &
- \left(\frac{3}{2}g^{2}\Phi(t)\right)\int_{-\infty}^{t}
dt' I(t,t')\left(\frac{3}{2}g^{2}\Phi^{2}(t')\right)
\nonumber
\\
&& \mbox{ }
\eea
the counterpart to $dV_{eff}/d\phi$ for $V_{eff}$ of (188), in which
$\Sigma (t)$ is the `tab' diagram counterpart to $\Sigma$ .$I(t,t')$,
real {\it and} retarded, is built from one-loop two-vertex
diagrams, the counterpart to $I$ of (188).
With propagators $G_{\pm\pm}$ for $\eta_{\pm}$, more one-loop
diagrams are possible than for the effective potential.
Details are given in Boyanovsky et al. but see also \cite{GlR}.
At one-loop $I(t,t')$
provides dissipation.  Meanwhile $K(t,t')$, of the form $(3
g^{2}\Phi(t)/2)R(t,t')(3g^{2}\Phi(t')/2)$, is also real, where $R$ is
the other combination of one-loop two-vertex diagrams, but gives a
relative imaginary part to $S_{eff}$. $K$ describes the `noise' of
the environment, most easily seen by using the identity
\beq
e^{-\half v\int\sigma K\sigma} = \int {\cal D}\xi\; e^{-\half v\int\xi
K^{-1}\xi + iv\int\xi\sigma}
\eeq
to define $P[\xi,\Phi]$ by
\beq
P[\xi,\Phi] = e^{-\half v\int\xi K^{-1}\xi}.
\eeq
Then Z[j] becomes
\beq
Z[j] = \int{\cal D}\Phi{\cal D}\xi{\cal D}\sigma\; P[\xi,\Phi]
e^{-iv\int\sigma (L - \xi)}
e^{iv\int j\Phi}.
\eeq
The $\sigma$-integration is trivial, giving
\beq
Z[j] = \int{\cal D}\Phi{\cal D}\xi\; P[\xi,\Phi]\delta [L(\Phi) -\xi]
e^{iv\int j\Phi},
\eeq
where the square brackets denote a $\delta$-functional, valid at
each time.  That is, for each noise function $\xi (t)$,
$\Phi (t)$ satisfies the
dissipative Langevin equation
\beq
L(\Phi (t)) = \xi (t)
\eeq
for $L$ of (201).  Both $\Sigma (t)$ and $I(t,t')$ depend on the
initial conditions and describe the shift in mass and the
dissipative effect of the coarsegraining from a particular
distribution of initial states.
As it stands, if the natural frequencies (masses) of the initial distribution
do not match the frequencies (masses) of the potential there will be
transient shocks.  However, since the expression (201) can only be
valid for short times, at best, `transient' effects may be
important.
The noise has distribution $P[\xi,\Phi[\xi]]$, depending on $\xi$
through $\Phi$ but, in the Gaussian approximation, $\bar{\Phi} (t) =
<\Phi (t)>$ satisfies the purely dissipative equation
\beq
L(\bar{\Phi} (t)) = 0.
\eeq
Equation (207), an Ehrenfest equation for field theory, can be
obtained directly form the functional Schroedinger equation, if
required.

As we noted with the effective potential, this approximation has to
be elevated to  a full one-loop equation and then to a consistent
one-loop approximation before it begins to be believable.
Nonetheless, this shows how dissipation and noise arise in quantum
field theory.  [It should be remembered that, in an expanding
universe, the evolution of the metric provides its own dissipation,
$3H\dot{\Phi}$].

As yet we are not in a position to build upon the understanding of
domain formation that we have developed in the previous sections.
Complete spatial averaging removes any reference to defects in the
same way that the space-time averaging of the effective potential
obliterates any information of
finite-size fluctuations. [This
can be seen very easily in the context of equilibrium theory, where
the effective potential essentially describes the (logarithm of)
the probability that a particular field average is achieved over
{\it all} space-time \cite{JL}].

In order to describe defects it is necessary
 to coarse-grain on correlation-length scales.
That is, we separate fields $\phi (t,\vecx)$ into `system' and
`environment' as
\beq
\phi (t,\vecx) = \phi_{f}(t,\vecx) + \phi_{1-f}(t,\vecx),
\eeq
where
\beq
\phi_{f}(t,\vecx) = \int \dbar k^{3} e^{i\veck . \vecx}\phi_{\veck}(t)
f(\veck)
\eeq
contains only wavelengths with with $|\veck|<\xi^{-1}$, for some
$\xi$.
That is,
\beq
f(\veck) = \theta (k_{c} - |\veck|),
\eeq
where $k_{c} = \xi ^{-1}$.
We can now try to integrate out the short-wavelength modes, in the
first instance in a one-loop approximation at  $O(g^4)$.  There will
be dissipation and noise as before, arising from diagrams
with similar ultraviolet
behaviour, but different infra-red properties.  The resulting
Langevin equation should permit us to follow the evolution of
defects, but serious sums have yet to be done.  This seems a good
place to stop.

\section{CONCLUSIONS}

In these notes we have made several attempts to understand defects
produced at a phase transition.  We will only recapitulate our main
result.  This is that defect density at formation is controlled by the field
correlation lengths essentially in the way anticipated by Kibble and
others.

It is one thing to count defects in fluctuations, another
to demonstrate how they freeze out as the system becomes cold.  For
quasi-equilibrium in the Ginzburg regime defects appear copiously,
but we are unable to show how they can survive as temperature is
lowered. More hopefully, defects produced at a quench by phase
separation do track domain boundaries and have every likelihood of
freezing-in in numbers determined by these domains, but as yet we do
not have the machinery to follow them through.  Much more work
remains to be done.

% ------------------------------------------------------------------
\typeout{--- references ---}


\begin{thebibliography}{99}

\bibitem{TK}T.W.B.\ Kibble, J.Phys. {\bf A9} (1976) 1387;
J.Phys.Rep.\ {\bf 67} (1980) 183.

\bibitem{HLM} P.C.\ Hendry, N.S.\ Lawson, R.A.M.\ Lee,
P.V.E.\ McClintoch and C.D.H.\ Williams, Nature {\bf 368},
 (1994) 315.

\bibitem{Z} W.H. Zurek, Acta Physica Polonica, {\bf B24} (1993) 1301;
Nature {\bf 368} (1994) 292.

\bibitem{HR} M.B.\ Hindmarsh and R.J.\ Rivers,
Nucl.Phys.\ {\bf B417} (1994) 506.

\bibitem{RR} R.J.\ Rivers, \ttitle{Fluctuations at Phase Transitions} in
proceedings of  Nato Advanced Research Workshop on
Electroweak Physics and the Early
Universe, Sintra (Portugal), March 1994 (Plenum press), Imperial College
preprint Imperial/TP/93-94/39.

\bibitem{GR} A.J.\ Gill and R.J.\ Rivers, \ttitle{The dynamics of
vortex and monopole production by quench induced phase separation},
Imperial College preprint Imperial/TP/93-94/55, {\tt hep-th/9410159}.

\bibitem{B}D.\ Boyanovsky and H.J.\ de Vega, Phys.Rev.\ {\bf D47}
(1993) 2343;
D.\ Boyanovsky, D.-S.\ Lee and A.\ Singh, Phys.Rev.\ {\bf D48} (1993) 800.

\bibitem{boyanovsky} D. Boyanovsky, H.J. de Vega, R. Holman. D-S.
Lee and A. Singh, preprint DOR-ER/40682-77.

\bibitem{halperin} B.I. Halperin, Les Houches, Session XXXV 1980 NATO
ASI, editors Balian, Kl\'{e}man and Poirier.

\bibitem{Banff} F.C.\ Khanna, R.\ Kobes, G.\ Kunstatter and H.\
Umezawa (ed.s), \ttitle{Proceedings of the Banff/CAP Workshop on Thermal Field
Theory} (World Scientific, 1994).

\bibitem{TFT}  { A. Fetter and J. Walecka}, \ttitle{Quantum Theory of
Many-Particle Systems} (McGraw-Hill, New York, 1971);
 { A.A. Abrikosov, L.P. Gor'kov and I.Ye. Dzyaloshinskii},
\ttitle{Quantum Field Theoretical Methods in Statistical Physics}
(Pergamon Press, Oxford, 1965);
 { J.I. Kapusta and P.V. Landshoff},
{J.Phys.G.}  {\bf 15} (1989) 267;
{ R.J. Rivers}, \ttitle{Path Integral Methods in
Quantum Field Theory} (Cambridge University Press, Cambridge, 1987);
{ N.P. Landsman and Ch.G. van Weert},
{Phys. Rep.}  {\bf 145} (1987) 141.

\bibitem{Ka} {J.I. Kapusta}, \ttitle{Finite Temperature Field Theory}
(Cambridge University Press, Cambridge, 1989).

\bibitem{DJ} L.\ Dolan and R.\ Jackiw, Phys.Rev.\ {\bf D9} (1974) 3320.

\bibitem{KL}D.\ A.\ Kirzhnits and A.\ D.\ Linde, Phys.Lett.\ {\bf
42B} (1972) 471.

\bibitem{W} S.\ Weinberg, Phys.Rev.\ {\bf D9} (1974) 3357.

\bibitem{LL}M.\ Dine, R.G.\ Leigh, P.\ Huet, A.\ Linde and D.\
Linde, Phys.Rev.\ {\bf D46} (1992) 550.

\bibitem{KK} K.\ Farakos, K.\ Kajantie, K.\ Rummukainen and M.\
Shaposhnikov, \ttitle{3D physics and the electroweak phase
transition: perturbation theory}, CERN preprint CERN-TH.6373/94,
{\tt hep-ph/9404201}.

\bibitem{G} V.I.\ Ginzburg, Fiz.Tverd.Tela {\bf 2} (1960) 2031;
[Sov.Phys.Solid State {\bf 2} (1961) 1826].

\bibitem{GK}M.\ Gleiser, Phys.Rev.\ {\bf D42} (1990) 3350,
M.\ Gleiser, E.W.\ Kolb and R.\ Watkins, Nucl.Phys.\ {\bf B364}
 (1991).411.

\bibitem{Chris} I.D.\ Lawrie, J.Phys.\ {\bf C11} (1978) 3857;
D.\ O'Connor, C.R.\ Stephens and F.\ Freire,
Mod.Phys.Lett.\ {\bf A25} (1993) 1779; N.\ Tetradis and C. Wetterich,
Nucl.Phys.\ {\bf B398} (1993) 659; M.A.\ van Eijck and C.R.\ Stephens
in Banff proceedings \cite{Banff} p.55; I.D.\ Lawrie, in Banff
proceedings \cite{Banff}, p.352; N.\ Tetradis and C. Wetterich,
Int.J.Mod.Phys.\ {\bf A9} (1994) 4029.

\bibitem{SV} A.\ Vilenkin and E.P.S.\ Shellard, \ttitle{Cosmic Strings
and Other Topological Defects}, (Cambridge University Press, 1994).

\bibitem{CCER} G.J.\ Cheetham, E.J.\ Copeland, T.S.\ Evans and
R.J.\ Rivers,  Phys.Rev. {\bf D47} (1993) 5316.

\bibitem{AG}A.\ Gill, Imperial College preprint, in preparation.

\bibitem{RS} S.\ Rudaz and A.M.\ Srivastava, Mod.Phys.Lett.\
{\bf A8} (1993) 1443.

\bibitem{NO} H.\ Nielsen and P.\ Olesen, Nucl.Phys.\ {\bf B61} (1973) 45.

\bibitem{tHP} G.\ t'Hooft, Nucl.Phys.\ {\bf B79} (1974) 276; A.M.\
Polyakov, JETP Letters {\bf 20} (1974) 194.

\bibitem{KT} E.W.\ Kolb and M.S.\ Turner, \ttitle{The Early Universe}
(Addison-Wesey, 1988).

\bibitem{BP} E.\ Braaten, R.D.\ Pisarski, Phys.Rev.\ {\bf D42} (1990) 2156.

\bibitem{fdamp} E.\ Braaten and R.D.\ Pisarski,
Phys.Rev.\ {\bf D46} (1992) 1829;
R.\ Kobes, G.\ Kunstatter, K.\ Mak, Phys.Rev. {\bf D45} (1992) 4632.

\bibitem{Pi} R.D.\ Pisarski, Phys.Rev.\ {\bf D47} (1993) 5589.

\bibitem{El} P.\ Elmfors, in Banff proceedings \cite{Banff}, pp.34;

\bibitem{We} H.A.\ Weldon, Phys.Rev.\ {\bf D28} (1983) 2007.

\bibitem{Sm2} A.V.\ Smilga, Can.J.Phys.\ {\bf 71} (1993) 295.

\bibitem{Mu} B.\ Muller,  \ttitle{Physics and Signatures of the
Quark-Gluon Plasma}, to appear in Rep.Prog.Phys., {\tt nucl-th/9410005}.

\bibitem{CTP} J.\ Schwinger, J.Math.Phys. {\bf 2} (1961) 407 ;
%K.T.\ Mahanthappa and P.M.\ Bakshi, J.Math.Phys. {\bf 4}, 1; ibid, 12 (1963);
L.V.\ Keldysh, Sov.Phys.\ JETP {\bf 20} (1965) 1018;
K-C.\ Chou, Z-B.\ Su, B-L.\ Hao and L.\ Yu, Phys.Rep.\ {\bf 118} (1985) 1.

\bibitem{SW} G.W.\ Semenoff and N.\ Weiss, Phys.Rev.\  {\bf D31} (1985)
689; {\em ibid}  {\bf D31} (1985) 699;
E.\ Calzetta and B.L.\ Hu, Phys.Rev.\ {\bf D35} (1988) 495;
{\em ibid} {\bf D37} (1988) 2838;
I.D.\ Lawrie, Phys.Rev.\  {\bf D40} (1989) 3330; J. Phys.
 {\bf A25} (1992) 2493.

\bibitem{Anne} A. Martin and A.-C.\ Davis, \ttitle{Evolution of fields
of a second order phase transition}, preprint DAMTP 94-87,
{\tt hep-ph/9410374}.

\bibitem{RR2} R.J.\ Rivers, Z.Phys.\ {\bf C22} (1984) 137.

\bibitem{GlR}  B.L.\ Hu, in Banff proceedings \cite{Banff}, p.309 (1994);
M.\ Gleiser and R.O.\ Ramos, Phys.Rev.\ {\bf D50} (1994)
2441; F.\ Cooper, S.\ Habib, Y.\ Kluger, E.\ Mottola, J.P.\ Paz and
P.\ Anderson, Phys.Rev.\ {\bf D50} (1994) 2848.

%\bibitem{KE}K.\ Enqvist, Magnetic fields of electroweak origin,
%these proceedings.
%K.\ Enqvist and P.\ Oleson, Phys.\ Lett.\ {\bf B319} (1993) 178.

\bibitem{JL}G.\ Jona-Lasinio, in  \ttitle{Scaling and Self-Similarity
in Physics}, ed. J.\ Frolich, Progress in Physics, Vol.7, 11
(Birkhauser Press, 1983).

\end{thebibliography}
\end{document}